\documentclass[journal]{IEEEtran}
\usepackage{cite}
\usepackage{color}
\usepackage{amsmath,amssymb,amsfonts}
\usepackage{algorithmic}
\usepackage{graphicx}
\usepackage{textcomp}
\usepackage{underscore}
\usepackage{leftidx}
\usepackage{graphicx}
\usepackage{subfigure}
\usepackage{multirow}
\usepackage{multicol}
\usepackage{booktabs}
\usepackage{float}
\newcommand{\forcond}{$t=1$ \KwTo $T$}

\usepackage[ruled]{algorithm2e}

\begin{document}
%
\title{Feedback Refined Local-Global Network for Super-Resolution of Hyperspectral Imagery}
%
%
%

\author{Zhenjie Tang,
        Qing Xu,
        Zhenwei Shi,
        Bin Pan
\thanks{This work was supported by the National Key R\&D Program of China under the Grant 2019YFC1510905, the National Natural Science Foundation of China under the Grant 62001251 and 62001252 and the
Scientific and Technological Research Project of Hebei Province
Universities and Colleges under the Grant ZD2021311. (Corresponding
author: Bin Pan.)}
\thanks{Zhenjie Tang is with the School of Civil Engineering, Tianjin University, Tianjin 300350, China (e-mail:tangzhenjie.hebut@gmail.com).}
\thanks{Qing Xu is with the College of Intelligence and Computing, Tianjin University, Tianjin 300350, China (e-mail:qingxu@tju.edu.cn). Zhenjie Tang and Qing Xu contributed equally to this manuscript.} 
\thanks{Zhenwei Shi is with the Image Processing Center, School of Astronautics, Beihang University, Beijing 100191, China (e-mail: shizhenwei@buaa.edu.cn).}
\thanks{Bin Pan (Corresponding author) is with the School of Statistics and Data Science, Nankai University, Tianjin 300071, China (e-mail: panbin@nankai.edu.cn).}
}

%
%

\markboth{}
{Shell \MakeLowercase{\textit{et al.}}: Bare Demo of IEEEtran.cls
for IEEE Journals}
%



\maketitle

\begin{abstract}
With the development of deep learning technology, multi-spectral image super-resolution methods based
on convolutional neural network have recently achieved great progress.
However, the single hyperspectral image super-resolution remains a
challenging problem due to the high-dimensional and complex spectral
characteristics of hyperspectral data, which make it difficult to simultaneously capture spatial and spectral information. To deal with this issue, we propose a novel
Feedback Refined Local-Global Network (FRLGN) for the super-resolution of
hyperspectral image. To be specific, we develop a new
Feedback Structure and a Local-Global Spectral Block to alleviate the difficulty in spatial and spectral feature extraction. 
The Feedback Structure can transfer the high-level information to guide the generation process of low-level feature, 
which is achieved by a recurrent structure with finite unfoldings.
Furthermore, in order to effectively use the high-level information passed back, a Local-Global Spectral Block is constructed 
to handle the feedback connections. The Local-Global Spectral Block utilizes the feedback high-level information to correct 
the low-level feature from local spectral bands and generates powerful high-level representations among global spectral bands.
By incorporating the Feedback Structure and Local-Global Spectral Block, the FRLGN can fully exploit spatial-spectral correlations among spectral bands and gradually reconstruct high-resolution hyperspectral images. The
source code of FRLGN is available at https://github.com/tangzhenjie/FRLGN.

\end{abstract}


%
\IEEEpeerreviewmaketitle

\section{Introduction}
%
%
%
%
\IEEEPARstart{H}{yperspectral} imaging sensors collect and process
information across different bands of the entire electromagnetic
spectrum. Compared with multi-spectral image, the resulting
hyperspectral image (HSI) contains richer spectral information and
has been applied to resource management, target
detection and land cover detection
\cite{7175025,6734720,8418840,yan2016band}, etc. However, because of the
limitation of imagery system, it is difficult to acquire an
HSI with high spatial resolution. Therefore, how to
obtain a reliable high-resolution HSI is sill a very
challenging problem.

Recently, HSI super-resolution approaches have been
intensively studied in remote sensing \cite{9334383}. Based on the number of input images, the HSI super-resolution methods can be roughly divided into fusion-based HSI
super-resolution \cite{9508188,9363338,chen2021hyperspectral} and single HSI super-resolution \cite{fu2021bidirectional,9329109,wang2022dilated}. The
fusion-based HSI super-resolution methods improves the spatial resolution by combining the
observed low-resolution HSI with high-resolution
multispectral image or panchromatic. For example, Wei \emph{et al}.
\cite{7010915} introduced a variational-based approach to merge a
high-resolution multispectral image with a low-resolution
HSI. By considering the HSI as a 3D
tensor, Wan \emph{et al}. \cite{9082892} designed a nonlocal 4-D tensor dictionary learning-based fusion approach.
More recently, deep learning-based fusion methods have achieved excellent performance with the powerful representation capability of convolution neural network. For instance, Wei \emph{et al}. \cite{9292466} suggested using the deep neural network to capture plenty of HSI statistics and then putting these priors to regularize the super-resolution procedure of HSIs. Wei \emph{et al}. \cite{9162463} recently further designed a deep recursive residual network to probe the deep statistical prior information.
Most fusion-based methods assume that the high-resolution auxiliary
image is well co-registered with the low-resolution HSI. In real applications, it is difficult to obtain these co-registered auxiliary images, which hinders the progress of such technique.

By contrast, the single
HSI super-resolution approaches do not need any auxiliary
information and have better feasibility in practice, which only reconstruct the high-resolution HSI from a low-resolution HSI. To explore the spatial-spectral prior information of HSIs, some single HSI super-resolution methods based on dictionary learning, sparse representation and low-rank approximation have been proposed. For instance, Huang
\emph{et al}. \cite{6854256} designed a noise-insensitive super-resolution mapping method based on multi-dictionary sparse representation. Wang \emph{et al}. \cite{2017Hyperspectral} introduced a new
tensor-based approach to solve the HSI super-resolution problem by
modeling three intrinsic characteristics of hyperspectral data. However, the
hand-crafted priors can only reflect one aspect of the hyperspectral
data, which make the reconstruction effect obvious only for the specified HSIs.
In recent years, due to the success of deep learning technology in many fields, it has been applied to the single hyperspectral super-resolution task, and achieved satisfying super-resolution results \cite{pan2021structure}. For example, to alleviate spectral distortion, Hu \emph{et al}. \cite{hu2017hyperspectral} designed a spectral difference
convectional network. Besides, Mei \emph{et al}.
\cite{mei2017hyperspectral} constructed a 3D super-resolution network to extract the prior information. Although the spectral correlation can be well exploited by 3D convolution operator, the amount of computation required by the model is very large. To solve the problem of high computation of 3D convolution, Jiang \emph{et al}. \cite{jiang2020learning} introduced a group convolution to explore the spatial information and the correlation among the spectral bands. Recently, Wang \emph{et al}. \cite{9380508} further designed a recurrent structure to investigate the spectral correlation among groups. Nonetheless, because of the high dimension and complex spectral patterns of hyperspectral data, it is hard to simultaneously explore the joint spatial and spectral information between continuous bands.

In this paper, in order to alleviate the difficulty of extracting spatial-spectral information from hyperspectral data, we propose a novel network for the single HSI super-resolution task, namely Feedback Refined Local-Global Network (FRLGN). FRLGN is motivated by the feedback mechanism \cite{hupe1998cortical}, which can make the network transmit high-level semantic information back to the previous layers and refine these low-level feature representations. Recently, some researchers have adopted this feedback mechanism to design the network architecture for various vision tasks \cite{cao2015look,carreira2016human,li2019feedback}. For instance, Han \emph{et al} \cite{han2018image} designed a two-state recurrent neural network, in which the information flows between two hidden states are exchanged in both directions. Taking advantage of the feedback mechanism to enhance the super-resolution results of HSIs, we designed a Feedback Structure (FS) and a Local-Global Spectral Block (LGSB) in FRLGN. To be specific, the Feedback Structure allows to use the feedback high-level information to correct the low-level representations through feedback connections. The FS is achieved by a recurrent structure with finite unfoldings. Furthermore, we construct a Local-Global Spectral Block to take full advantage of the feedback high-level information. The LGSB is composed of local and global spectral feature extraction layers, which can adjust the local spectral low-level representation input using the feedback high-level information and create a powerful high-level global spectral representation. The FRLGN is essentially a recurrent neural network with a Local-Global Spectral Block, which is specifically designed to explore the spatial and spectral prior of hyperspectral data. Experimental results indicate that LGSB is more suitable for HSI super-resolution task. Besides, in order to
make the feedback high-level feature contains the high-resolution HSI information, we aggregate the losses of each iteration to optimize the network model. On the whole, the principle of the feedback mechanism is that the information of a coarse reconstructed HSI facilitates the low-resolution HSI to generate a better super-resolution HSI.

The main contributions of our work can be  summarized as follows:
\begin{itemize}
\item A novel Feedback Refined Local-Global Network is proposed for the single HSI super-resolution task, which can effectively explore spatial-spectral priors between spectral bands.
\item We construct a new Feedback Structure to correct the low-level representation using feedback high-level semantic information.
\item We design a Local-Global Spectral Block to refine the local spectral low-level representations using the feedback information, and then generate a more powerful global spectral high-level representation.
\end{itemize}

 \begin{figure*}[!ht]
 \centering
 \includegraphics[width=17.5cm]{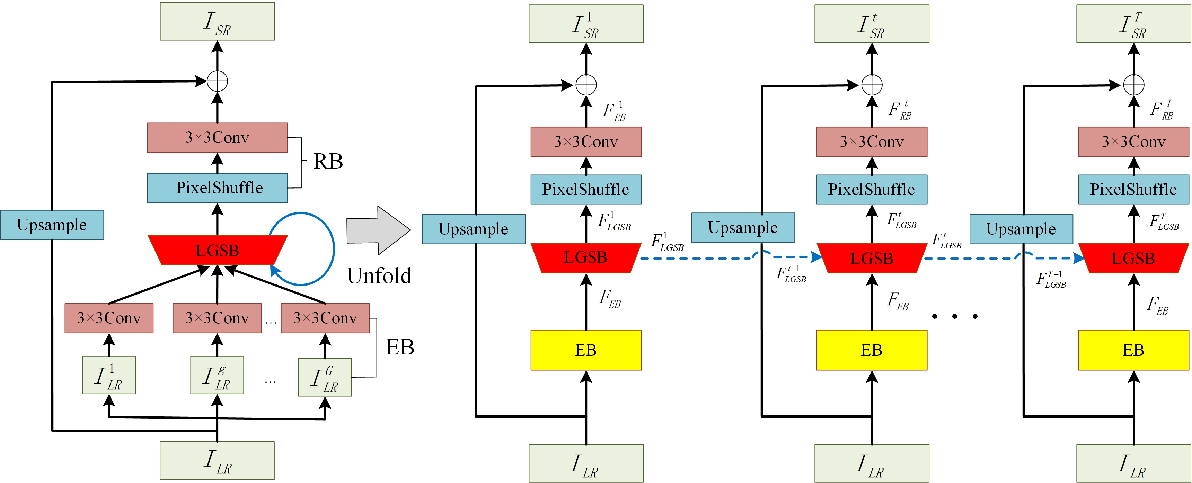}
 \caption{The overview of our proposed Feedback Refined Local-Global Network (FRLGN). The blue arrows are the feedback connections and the Local-Global Spectral Block (LGSB) represented by the trapezoidal block is specifically designed for the task of super-resolution of hyperspectral images.}
 \label{fig:figure2}
 \end{figure*}

\section{PROPOSED METHOD}
In this section, the detail description of FRLGN is first presented, and then the proposed Feedback Structure and Local-Global Spectral Block are introduced. At last, we interpret the loss function used in FRLGN approach.
\subsection{Network Architecture}
As shown in Fig.~\ref{fig:figure2}, the FRLGN is unfolded to $T$ iterations, where the order of each iteration $t$ is from 1 to $T$. We link the losses of each iteration together so that the hidden state in FRLGN can contain the information of the high-resolution HSI. We describe the specific details of the loss function in the next Loss Function part. The sub-network in each iteration $t$ consists
of three Blocks: the Embedding Block, the Local-Global Spectral Block and the Reconstruction Block. Each iteration shares the weights of each block. For each iteration $t$, we also design a global skip connection that transmits an up-sampled HSI to the final output. Therefore, each iteration $t$ of the sub-network is used to recover a residual image when a low-resolution HSI is input.

\subsubsection{The Embedding Block}
Different from the previous method of treating the HSI as a whole or multiple single-channel images, we divide the entire input low-resolution HSI into several groups. With this strategy, we can not only explore the correlation between adjacent spectral bands of the input HSI more easily, but also reduce the spectral dimension of the HSI. Specifically, the input low-resolution
HSI $I_{LR}$ is divided into $G$ groups.
More details are discussed in the experiment section. As shown in
Fig.~\ref{fig:figure2}, for each group $I_{LR}^{g}$, we use one convolution operation to extract its shallow feature
$F_{EB}^{g}$,
\begin{equation}
I_{LR} = [I_{LR}^{1}, I_{LR}^{2}, I_{LR}^{g}, \cdot\cdot\cdot ,I_{LR}^{G}]
\end{equation}
\begin{equation}
F_{EB}^{g} = f_{EB}(I_{LR}^{g})
\end{equation}
\begin{equation}
F_{EB}=[F_{EB}^{1}, F_{EB}^{2}, F_{EB}^{g}, \cdot\cdot\cdot, F_{EB}^{G}]
\end{equation}
where the $f_{EB}$ denotes the operations of Embedding Block, eg.,
feature extraction layer for all groups. The $[]$ represents a cascading function. After that, $F_{EB}$ is used as input to
the Local-Global Spectral Block.

\subsubsection{The Local-Global Spectral Block}
For $t$-th iteration, Local-Global Spectral Block receives the the shallow feature $F_{EB}$ and hidden state
from past iteration $F_{LGSB}^{t-1}$ through a feedback
connection. $F_{LGSB}^{t}$ denotes
the result of LGSB. The mathematical formula of LGSB is as follows:
\begin{equation}
F_{LGSB}^{t} = f_{LGSB}(F_{LGSB}^{t-1}, F_{EB})
\end{equation}
where $f_{LGSB}$ denotes the operations of the LGSB. More details
of the LGSB can be found in Local-Global Spectral Block part.

\subsubsection{The Reconstruction Block}
The reconstruction block firstly uses PixelShuffle
\cite{shi2016real} to upscale the feature $F_{LGSB}^{t}$ to high-resolution one,
and then a 3 $\times$ 3 convolution operation is applied to create
the residual image $I_{Res}^{t}$. The formula for reconstruction block is defined as:
\begin{equation}
I_{Res}^{t} = f_{RB}(F_{LGSB}^{t})
\end{equation}
where $f_{RB}$ is the operation of the reconstruction block.

For the $t$-th iteration, the output super-resolution image $I_{SR}^{t}$ is obtained
by:
\begin{equation}
I_{SR}^{t} = I_{Res}^{t} + f_{UP}(I_{LR})
\end{equation}
where $f_{UP}$ represents an upsampling operation. The choice of upsampling method is arbitrary. In this paper, we apply a
Bicubic upsample approach. After $T$ iterations, we will generate $T$ super-resolution
images $(I_{SR}^{1}, I_{SR}^{2}, \cdot\cdot\cdot, I_{SR}^{T})$.

 \begin{figure*}[!t]
 \centering
 \includegraphics[width=14cm]{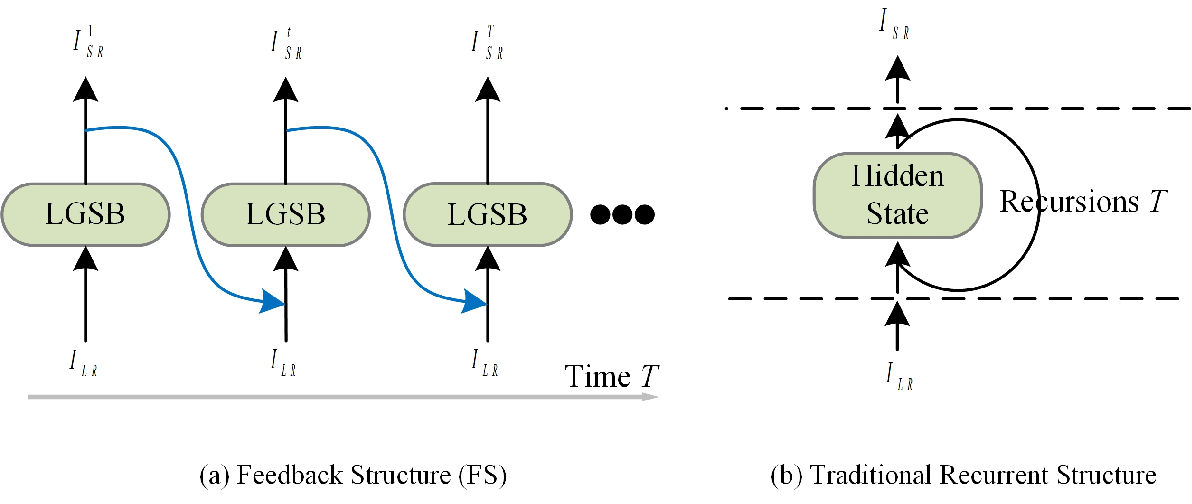}
 \caption{(a) The illustration of the Feedback Structure (FS) in the proposed FRLGN network and the blue arrow is the feedback connection. (b) The architecture of Traditional Recurrent Structure.}
 \label{fig:figure1}
 \end{figure*}
 
\subsection{Feedback Structure}
In HSI super-resolution task, some researchers \cite{li2018single,9162463,9380508} have made an effort to introduce the recurrent structure to improve super resolution results. However, in their network frameworks, the information flow from the low-resolution HSI to final super-resolution HSI is still feed-forward. As can be seen from Fig.~\ref{fig:figure1}(b), the recurrent structure adopted by these methods can be abstracted into a single-state recurrent network. These methods improve the feature representation of the model by running recursively on a specially designed network structure.

In this work, we design a Feedback Structure to reroute the output of the HSI super-resolution system to correct the input in each iteration. Fig.~\ref{fig:figure1}(a) illustrates the Feedback Structure of FRLGN. Specifically, the Local-Global Spectral Block receives the information of input low-resolution HSI and feedback high-level information from last iteration, then generates coarse super-resolution result and high-level semantic guidance information for next iteration. The Feedback Structure can be characterized by:
\begin{equation}
I_{SR}^{t} = f_{FS}(I_{LR}, I_{SR}^{t-1})
\end{equation}
where the $f_{FS}$ denotes the function of Feedback Structure.

 \begin{figure*}[!ht]
 \centering
 \includegraphics[width=14cm]{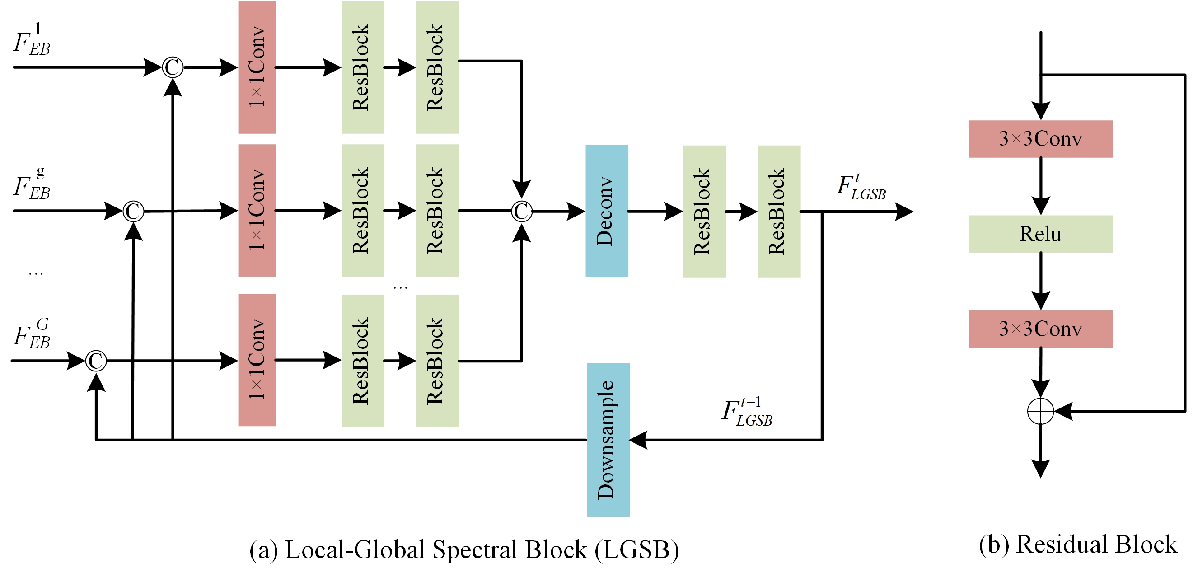}
 \caption{The network architecture of the Local-Global Spectral Block (LGSB): (a) the Local-Global Spectral Block, (b) the Residual Block}
 \label{fig:figure3}
 \end{figure*}

\subsection{Local-Global Spectral Block}
As an ill-posed problem, image super-resolution requires additional prior knowledge to regularize the reconstruction process. Traditional super-resolution methods usually make an effort to construct the regular terms of the super-resolution model, such as low-rank \cite{dian2019hyperspectral},
total variation \cite{he2016super} and sparse \cite{xu2019nonlocal,9417623}. Whether the designed prior knowledge can characterize the observed HSI data directly determines the performance of the super-resolution method. Therefore, for the HSI
super-resolution task, it is also essential to study the inherent
characteristics of hyperspectral data, \emph{e.g.}, the spatial
non-local self-similarity and the high-correlation among spectral bands \cite{wang2021contextual}.
However, the manually designed constraints are not enough to achieve
accurate restoration of HSIs.

In this work, a novel Local-Global Spectral Block is introduced to exploit the
spatial-spectral prior with the help of feedback high-level semantic information from hidden state. As can be seen in Fig.~\ref{fig:figure3}(a), for iteration \emph{t}, the LGSB inputs the feedback global
spectral high-level information $F_{LGSB}^{t-1}$ to correct the $G$
groups local spectral low-level representations, $F_{EB} =
[F_{EB}^{1}, F_{EB}^{2}, F_{EB}^{g}, \cdot\cdot\cdot ,F_{EB}^{G}]$, and then
creates more effective high-level feature $F_{LGSB}^{t}$ for
the next iteration and the reconstruction block. The LGSB contains
$G$ groups local spectral feature extraction layers and one global
spectral feature extraction layer. For simplicity, we use $Conv(k)$ and
$Deconv(k)$ to denote a convolution operation and a deconvolutional operation, where the $k$ represents the size of convolution kernel.

At the beginning of the LGSB, the downsampled $F_{LGSB}^{t-1}$ and
each group $F_{EB}^{g}$ are concatenated and compressed by one
$Conv(1)$ operation to refine the input each group feature
$F_{EB}^{g}$ by feedback information $F_{LGSB}^{t-1}$, producing the
refined group feature $F_{g}^{t}$.
\begin{equation}
F_{g}^{t} = f_{Com}([f_{Down}(F_{LGSB}^{t-1}), F_{EB}^{g}])
\end{equation}
where $f_{Down}$ refers to downsample operation using average
pooling with a kernel of 2 and stride of 2. The
$[f_{Down}(F_{LGSB}^{t-1}), F_{EB}^{g}]$ refers to the concatenation
of $f_{Down}(F_{LGSB}^{t-1})$ and $F_{EB}^{g}$. The $f_{Com}$ denotes
the initial compression operation.

After obtaining the refined group feature $F_{g}^{t}$, we add a
local spectral feature extraction layer to explore the local
spectral correlation, which consists of two residual blocks as shown
in Fig.~\ref{fig:figure3}(b). Let $L_{g}^{t}$ be the $g$-th group
local spectral LR feature map. $L_{g}^{t}$ can be obtained by:
\begin{equation}
L_{g}^{t} = f_{Local}(F_{g}^{t})
\end{equation}
where the $f_{Local}$ denotes local spectral feature extraction
layer.

After that, we pass all the local spectral LR feature maps to the
global spectral feature extraction layer, which contains one
upsample $Deconv(2)$ operation and two residual blocks. Note that we
propose a strategy of progressive super-resolution reconstruction to
stabilize the training process. Particularly, in addition to the
reconstruction block, we also add an upsampling operation in the
global spectral feature extraction layer. At last, the global
spectral high-level feature $F_{LGSB}^{t}$ can be obtained by:
\begin{equation}
F_{LGSB}^{t} = f_{Global}([L_{1}^{t}, L_{2}^{t}, \cdot\cdot\cdot,
L_{G}^{t}])
\end{equation}
where the $f_{Global}$ denotes the the global spectral feature
extraction layer.

\subsection{Loss Function}
To optimize the FRLGN,  we choose the most commonly used $L1$
loss function to measure the HSI reconstruction performance. Finally, the output result of FRLGN is the weighted average of all intermediate super-resolution results:
\begin{equation}
I_{SR} = \frac{1}{T} \sum_{t=1}^{T} I_{Res}^{t} + f_{UP}(I_{LR})
\end{equation}
The loss function of FRLGN is determined by:
\begin{equation}
L(\Theta) = \left \| I_{HR} - I_{SR} \right \|_{1}
\end{equation}
where the $\Theta$ represents the parameters of our proposed FRLGN and the $I_{HR}$ is the corresponding target high-resolution HSI. The training procedure of FRLGN is shown in Algorithm 1.

\begin{algorithm}
    \caption{Training Process of FRLGN}
    \label{alg:Algorithm}
    \SetKwInOut{Input}{input}\SetKwInOut{Output}{output}
    \Input{Low-resolution HSI $I_{LR}$; \\
     High-resolution HSI $I_{HR}$;\\
     The number of iterations $T$;\\
     The number of local spectral groups in LGSB $G$.}
    \Output{Super-resolution HSI $I_{SR}$}
    \Repeat{convergence}
    {
        Initialization:\\
        $I_{LR} = \{I_{LR}^{1}, I_{LR}^{2}, \cdot\cdot\cdot
        ,I_{LR}^{G}\};$\\
        $F_{LGSB}^{0}=None;$\\
        Shallow feature extraction by the Embedding Block:\\
        $F_{EB} = f_{EB}(I_{LR});$\\
        $T$ intermediate prediction result generation by the Local-Global Spectral Block and the Reconstruction Block:\\
        \For{\forcond}{
            $F_{LGSB}^{t} = f_{LGSB}(F_{LGSB}^{t-1}, F_{EB});$\\
            $I_{Res}^{t} = f_{RB}(F_{LGSB}^{t})$
        }
        Final output:\\
        $I_{SR} = \frac{1}{T} \sum_{t=1}^{T} I_{Res}^{t} +
        f_{UP}(I_{LR})$\\
        Update the FRLGN network parameters by minimizing the loss
        between the reconstructed $I_{SR}$ and the corresponding
        label $I_{HR}$
    }
\end{algorithm}

\section{EXPERIMENTS AND RESULTS}

\subsection{Datasets}

\subsubsection{CAVE dataset}
The CAVE dataset \cite{yasuma2010generalized} is a HSI dataset of real-world materials and objects, which are captured by a Cooled CCD camera. The hyperspectral camera collects information from the 400nm-700nm spectral range in 10 nm steps. This dataset consists of 32 HSIs with a size of $512 \times 512 \times 31$ pixels, which are further divided into 5 groups, namely food and drinks, skin and hair, paints, real and fake, and stuff.

\subsubsection{Harvard dataset}
The Harvard dataset \cite{chakrabarti2011statistics} contains 77 HSIs of $1040 \times 1392 \times 31$ size from outdoor and indoor scenes. These HSIs are captured by a commercial hyperspectral camera, which collects the spectral data in 10 nm steps over the wavelength range of 400 nm to 700 nm. 

\subsubsection{Chikusei dataset}
The Chikusei dataset \cite{NYokoya2016} consists of $2517 \times 2335$ pixels with a spatial resolution of 2.5 m. The dataset was taken by an airborne hyperspectral imaging sensor in the agricultural and urban areas of Chikusai, Japan. This dataset captures 128 spectral bands from the 363 nm to 1018 nm. Since the lack of edge information, we first cut the original HSI to generate an image of 2304×2048×128 pixels and then the generated image is further split into a training set and a test set. In particular, we first extract the top region of the generated image to create the test set, which consists of four HSIs with a pixel size of $512 \times 512 \times 128$ that do not overlap each other. And the remaining region of the generated image is used as training data.

\subsection{Implementation Details}
Since HSIs are collected by different hyperspectral imaging sensors, HSI datasets tend to have different numbers of spectral channels. Therefore, we need to learn a super-resolution HSI model separately for each HSI dataset. In the next experiments, 80\% of samples in the dataset are used to train the super-resolution models and the remaining samples are utilized for testing.

During training, 12 randomly selected patches are fed to the FRLGN network. To obtain low-resolution HSIs, we down-sample these patches to $32 \times 32 \times L$ pixels based on the scale factor $s$. Furthermore,we use the bicubic interpolation function to down-sample these patches. In our network, the convolution operators with a kernel 3 adopt a zero-padding strategy to ensure that the intermediate features have the same spatial size. We up-sample the resulting features by a factor of 2 using a deconvolution with a kernel 2 and a stride 2. The ADAM \cite{kingma2014adam} with an initial learning rate of 2e-4 is used to optimize the FRLGN network.

At the testing stage, in order to improve testing efficiency,  we use only the $512 \times 512$ area in the upper left corner of test HSIs for evaluation. In this work, the Pytorch library is used to implement and train our proposed FRLGN network.

\subsection{Evaluation Metrics}
In this section, we choose six commonly used quantitative metrics to evaluate the performance of FRLGN, i.e.,  cross correlation
(CC) \cite{loncan2015hyperspectral}, spectral angle mapper (SAM)
\cite{yuhas1992discrimination}, root mean squared error (RMSE), the
erreur relative globale adimensionnelle de synthese (ERGAS)
\cite{wald2002data}, peak signal-to-noise ratio (PSNR) and structure
similarity (SSIM) \cite{wang2004image}. As the CC, RMSE, PSNR and SSIM are
widely used quantitative metrics in HSI super-resolution tasks, we omit their detailed description here. In addition, ERGAS performs a global statistical measure on the reconstructed HSIs, which is calculated by
\begin{equation}
ERGAS(I_{HR}, I_{SR}) = 100s \sqrt{\frac{1}{L} \sum_{l=1}^{L}
\left(\frac{RMSE_{l}}{\mu_{l}}\right)^{2}}
\end{equation}
in which $RMSE_{l}=(\|I_{SR}^{l}-I_{HR}^{l}\|_{F} / \sqrt{n})$.
Here, $n$ and $\mu_{l}$ represent the number of spatial pixels and mean of the $l$th band from the ground truth $I_{HR}$, respectively. The $I_{SR}^{l}$ and $I_{HR}^{l}$ denote the $l$th band of
$I_{SR}$ and $I_{HR}$, respectively. SAM is used to evaluate the preservation of spectral band information for each spatial location of the HSI. SAM is obtained by calculating the angle between two spectral vectors from the same spatial position of $I_{SR}$ and $I_{HR}$. The formula of SAM is presented as
\begin{equation}
SAM(\textbf{x}, \hat{\textbf{x}}) = \arccos \left(\frac{\langle
\textbf{x},
\hat{\textbf{x}}\rangle}{\|\textbf{x}\|_{2}\|\hat{\textbf{x}}\|_{2}}\right)
\end{equation}
in which $\hat{\textbf{x}}$ and $\textbf{x}$ denote the two spectral vectors from $I_{SR}$ and $I_{HR}$, respectively. And the $\langle\cdot,\cdot\rangle$ is the dot product of two
vectors, $\|\textbf{x}\|_{2}$ represent the $l_{2}$ regularization operation of a vector.
For PSNR and SSIM, we present the average metric values of all spectral bands. The best values for CC, SAM, RMSE, ERGAS, PSNR, SSIM are 1, 0, 0, 0, $+\infty$, and 1, respectively.

 \begin{table}
\renewcommand{\arraystretch}{1.3} 
\begin{center}
\caption{Convergence analysis of $T$ when
$G$=8}\label{tab:Ablation1}
\setlength{\tabcolsep}{3pt}  
\begin{tabular}{cccccccc}
 \toprule[1pt]
  $T$  &  1 &  2  &  3 &   4   & 5  &  6   \\
\hline
PSNR(dB) &   37.0815   &   37.5073  &   37.7574  &  37.8131  &   37.8639  &   37.8905   \\
\bottomrule[1pt]
\end{tabular}
\normalsize
\end{center}
\end{table}

\begin{table}
\renewcommand{\arraystretch}{1.3} 
\begin{center}
\caption{Convergence analysis of $G$ when
$T$=6}\label{tab:Ablation2}
\setlength{\tabcolsep}{3pt}  
\begin{tabular}{cccccccc}
 \toprule[1pt]
  $G$  &  1 &  2  & 4 &   8   \\
\hline
SAM &   3.5977   &   3.5276  &   3.4792  &  3.4332   \\
\bottomrule[1pt]
\end{tabular}
\normalsize
\end{center}
\end{table}

\subsection{Study of T and G}
In this part, we discussed the effect of 
iterations (denoted as $T$) and local spectral groups (denoted as $G$) in the Local-Global Spectral Block on the FRLGN performance on the CAVE dataset. In
subsequent experiments, we set the base number of filters to 256. By fixing $G$ to 8, we first explore the influence of $T$ on HSI reconstruction. Table \ref{tab:Ablation1} shows that the super-resolution performance is improved with the help of feedback connections
compared to the network without feedback connections ($T$=1).
Moreover, the quality of reconstruction has been further improved as the increasing iteration $T$ . On the other hand, it also indicates that our proposed Local-Global Spectral Block would certainly benefit from cross-time feedback information. After that, we also discuss the influence of $G$ by fixing the $T$ to 6. From Table \ref{tab:Ablation2}, we can observe that with the help of local-spectral grouping strategy, the spectrum reconstruction performance is enhanced compared to the network without the grouping strategy ($G$=1). 
In addition, with the increase of $G$, the spectral representation of FRLGN becomes more powerful and the spectral reconstruction quality is also improved. In a word, choosing larger $T$ or $G$ can obtain better
super-resolution results. In next experiments, we set $T$=6, $G$=8 for CAVE dataset and Harvard dataset, and $T$=6, $G$=12 for Chikusei dataset.

\begin{table}
\renewcommand{\arraystretch}{1.25} 
\begin{center}
\caption{Quantitative analysis of seven different
comparison methods on CAVE test dataset involving six
metrics.}\label{tab:CAVE}
\setlength{\tabcolsep}{3pt}  
\begin{tabular}{cccccccc}

  \hline
    &   $s$ &   CC$\uparrow$    &   SAM$\downarrow$ &   RMSE$\downarrow$    &   ERGAS$\downarrow$   &   PSNR$\uparrow$  &   SSIM$\uparrow$  \\
\hline \hline
Bicubic &   4   &   0.9846  &   5.1832  &   0.0224  &   7.7384  &   34.5069     &   0.9472  \\
\hline
VDSR \cite{kim2016accurate} &   4   &   0.9896  &   4.3622  &   0.0188  &   6.3067  &   36.1348     &   0.9612  \\
RCAN \cite{zhang2018image}  &   4   &   0.9913  &   4.3058  &   0.0172  &   5.7796  &   36.7979 &   0.9657  \\
\hline
3DCNN \cite{mei2017hyperspectral}   &   4   &   0.9862  &   4.2297  &   0.0212  &   7.3182  &   34.9853     &   0.9549  \\
GDRRN \cite{li2018single}   &   4   &   0.9891  &   4.2970   &   0.0192  &   6.5087  &   35.8465     &   0.9594  \\
SSPSR \cite{jiang2020learning}    &   4   &   0.9915  &   3.7384  &   0.0168  &   5.7527  &   37.0479     &  0.9682  \\
\hline
FRLGN   &   4   &   \textbf{0.9930} &   \textbf{3.4332} &   \textbf{0.0152} &   \textbf{5.1599} &   \textbf{37.8905}    &   \textbf{0.9737} \\
    \hline
    \hline
Bicubic &   8   &   0.9564  &   7.3210  &   0.0385  &   12.8323  &   29.5763     &   0.8741  \\
\hline
VDSR \cite{kim2016accurate} &   8   &   0.9615  &   5.8692  &   0.0369  &   12.0527  &   30.0080     &   0.8999  \\
RCAN \cite{zhang2018image}  &   8   &   0.9671  &   5.9008 &   0.0340  &    11.1373 &   30.7372 &   0.9061  \\
\hline
3DCNN \cite{mei2017hyperspectral}   &   8   &   0.9594  &   5.6079  &   0.0370  &   12.3341  &   29.8880     &   0.8961  \\
GDRRN \cite{li2018single}   &   8   &   0.9611  &   5.8864  &   0.0368  &   12.0684  &   30.0042     &   0.8966  \\
SSPSR \cite{jiang2020learning}    &   8   &   0.9675  &   5.6617  &   0.0341  &   11.0506  &   30.7976     &   0.9098  \\
\hline
FRLGN   &   8   &   \textbf{0.9712} &   \textbf{5.0550} &   \textbf{0.0323} &   \textbf{10.3982} &   \textbf{31.4007}    &   \textbf{0.9159} \\
  \hline
\end{tabular}
\normalsize
\end{center}
\end{table}

\begin{table}
\renewcommand{\arraystretch}{1.25} 
\begin{center}
\caption{Quantitative analysis of seven different
comparison methods on Harvard test dataset involving six
metrics.}\label{tab:Harvard}
\setlength{\tabcolsep}{3pt}  
\begin{tabular}{cccccccc}

  \hline
    &   $s$ &   CC$\uparrow$    &   SAM$\downarrow$ &   RMSE$\downarrow$    &   ERGAS$\downarrow$   &   PSNR$\uparrow$  &   SSIM$\uparrow$  \\
\hline \hline
Bicubic &   4   &   0.9606   &   2.5671  &   0.0101  &   3.0957  &   43.9037     &   0.9582  \\
\hline
VDSR \cite{kim2016accurate} &   4   &   0.9640  &   2.5709  &   0.0090  &   2.8602  &   44.6486     &   0.9634  \\
RCAN \cite{zhang2018image}  &   4   &   0.9671  &   2.4097  &   0.0086  &   2.7537  &   45.1204 &   0.9663  \\
\hline
3DCNN \cite{mei2017hyperspectral}   &   4   &   0.9614  &   2.3917  &   0.0098  &   3.0324  &   44.1815     &   0.9600  \\
GDRRN \cite{li2018single}   &   4   &   0.9630  &   2.4924   &   0.0093  &   2.9276  &   44.4577     &   0.9620  \\
SSPSR \cite{jiang2020learning}    &   4   &  0.9704  &  2.2766  &  0.0082  &   2.5893  &   45.5460     &   0.9684  \\
\hline
FRLGN   &   4   &   \textbf{0.9722} &   \textbf{2.2496} &   \textbf{0.0074} &   \textbf{2.4463} &   \textbf{46.1866}    &   \textbf{0.9730} \\
    \hline
    \hline
Bicubic &   8   &   0.9098  &   3.0165  &   0.0179  &   5.0694  &   39.6681     &   0.9131  \\
\hline
VDSR \cite{kim2016accurate} &   8   &   0.9185  &   3.0093  &   0.0165  &   4.7369  &   40.2490     &   0.9223  \\
RCAN \cite{zhang2018image}  &   8   &   0.9312  &   2.7808 &    0.0150  &    4.3438 &   40.9853 &   \underline{0.9313}  \\
\hline
3DCNN \cite{mei2017hyperspectral}   &   8   &   0.9128  &   2.7853  &   0.0172  &   4.9422  &   39.9615     &   0.9175  \\
GDRRN \cite{li2018single}   &   8   &   0.9175  &   2.8669  &   0.0166  &   4.7946  &   40.1831      &   0.9214  \\
SSPSR \cite{jiang2020learning}    &   8   &   0.9338  &   \textbf{2.6202}  &   0.0149  &   4.2458  &   41.1869     &  0.9313  \\
\hline
FRLGN   &   8   &   \textbf{0.9373} &   2.7665 &   \textbf{0.0139} &   \textbf{4.0316} &   \textbf{41.6320}    &   \textbf{0.9374} \\
  \hline
\end{tabular}
\normalsize
\end{center}
\end{table}

\begin{table}
\renewcommand{\arraystretch}{1.25} 
\begin{center}
\caption{Quantitative analysis of seven different
comparison methods on Chikusei test dataset involving six
metrics.}\label{tab:Foster}
\setlength{\tabcolsep}{3pt}  
\begin{tabular}{cccccccc}

  \hline
    &   $s$ &   CC$\uparrow$    &   SAM$\downarrow$ &   RMSE$\downarrow$    &   ERGAS$\downarrow$   &   PSNR$\uparrow$  &   SSIM$\uparrow$  \\
\hline \hline
Bicubic &   4   &   0.8987   &   3.7666  &   0.0176  &   7.6532  &   36.5603     &   0.8882  \\
\hline
VDSR \cite{kim2016accurate} &   4   &   0.9176  &   3.1003  &   0.0155  &   6.9534  &   37.5648     &   0.9113  \\
RCAN \cite{zhang2018image}  &   4   &   0.9142  &   3.0936  &   0.0156  &   7.1099  &   37.4313 &   0.9104  \\
\hline
3DCNN \cite{mei2017hyperspectral}   &   4   &   0.9047  &   3.4808  &   0.0169  &   7.3419  &   36.9090     &   0.8931  \\
GDRRN \cite{li2018single}   &   4   &   0.9144  &   3.2178   &   0.0159  &   7.0426  &   37.3754     &   0.9060  \\
SSPSR \cite{jiang2020learning}    &   4   &   0.9250  &   2.8281  &   0.0148  &   6.6082  &   37.9698     &   0.9193  \\
\hline
FRLGN   &   4   &   \textbf{0.9283} &   \textbf{2.7580} &   \textbf{0.0143} &   \textbf{6.4953} &   \textbf{38.2085}    &   \textbf{0.9240} \\
    \hline
    \hline
Bicubic &   8   &   0.7546  &   5.9617  &   0.0274  &   11.9665  &   32.7047     &   0.7829  \\
\hline
VDSR \cite{kim2016accurate} &   8   &  0.7840   &  5.3103   &   0.0250  &  10.9097   &   33.4964     & 0.8069    \\
RCAN \cite{zhang2018image}  &   8   &  0.7630  &   6.5447 &   0.0258  &    11.9078 &   33.0475 &   0.7946  \\
\hline
3DCNN \cite{mei2017hyperspectral}   &   8   &   0.7723  &   5.5506  &   0.0257  &   11.0971  &   33.3107     &   0.7955  \\
GDRRN \cite{li2018single}   &   8   &   0.7842  &   5.3033  &   0.0249  &   10.9107  &   33.5236      &   0.8062  \\
SSPSR \cite{jiang2020learning}    &   8   &   0.7880  &   5.2415  &   0.0247  &   \textbf{10.7863}  &   33.6194     &   0.8106  \\
\hline
FRLGN   &   8   &   \textbf{0.7887} &   \textbf{5.2122} &   \textbf{0.0246} &   10.8033 &   \textbf{33.6332}    &   \textbf{0.8145} \\
  \hline
\end{tabular}
\normalsize
\end{center}
\end{table}


\begin{figure*}[htbp]
    \centering
    \subfigure{
        \begin{minipage}[t]{0.135\textwidth}
            \includegraphics[width=1\textwidth]{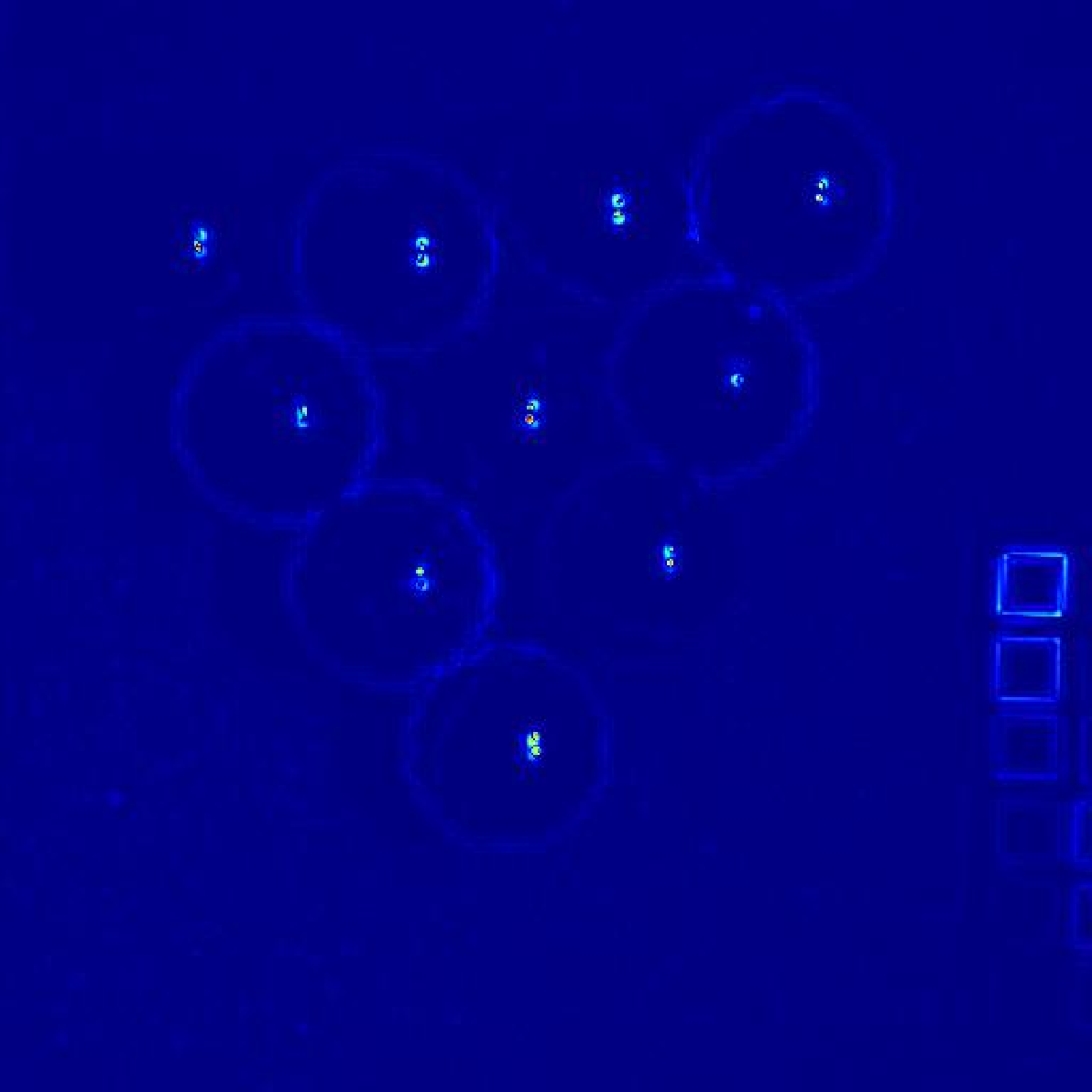}
            \centerline{Bicubic}
        \end{minipage}
    }\hspace{-2.7mm}
    \subfigure{
        \begin{minipage}[t]{0.135\textwidth}
            \includegraphics[width=1\textwidth]{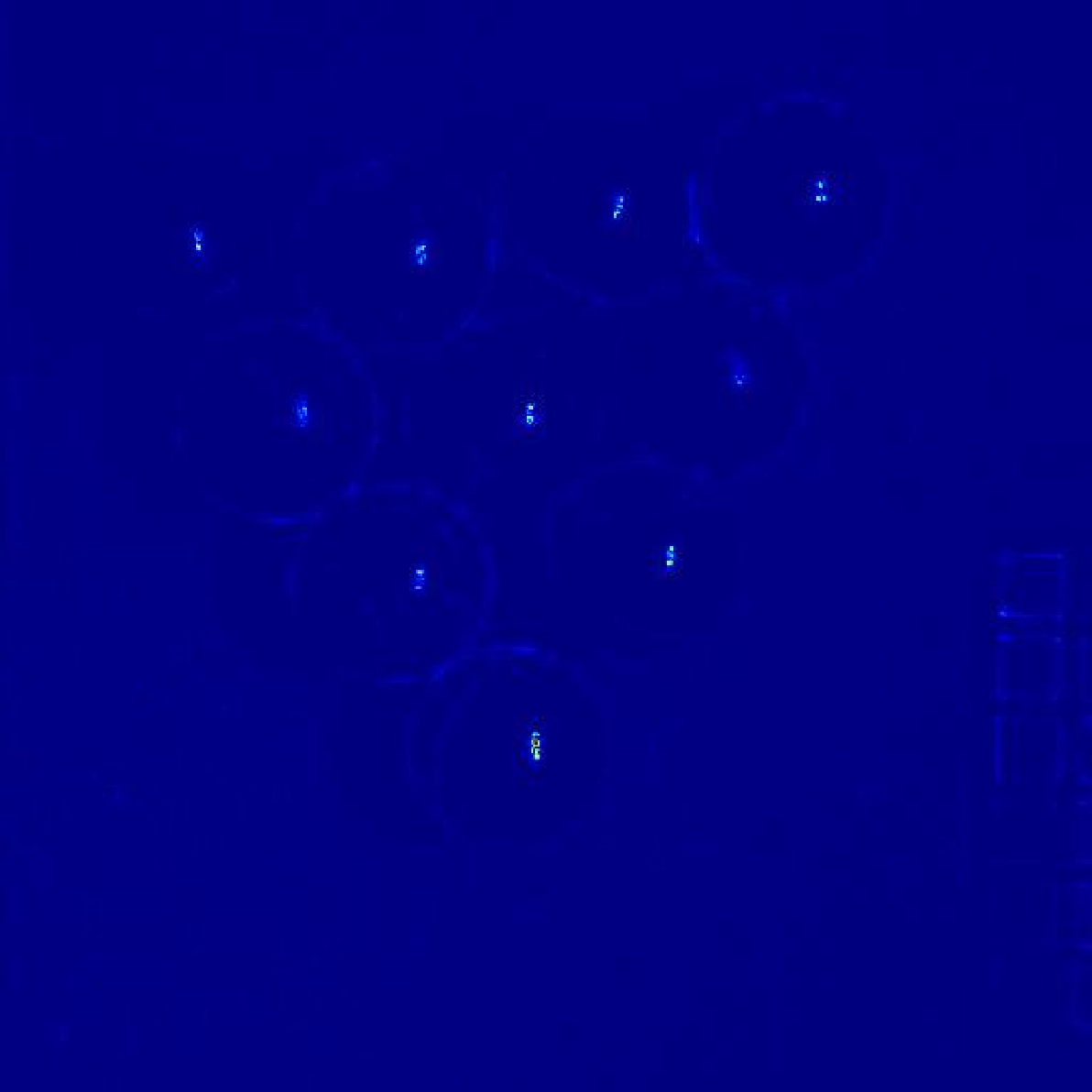}
            \centerline{VDSR}
        \end{minipage}
    }\hspace{-2.7mm}
    \subfigure{
        \begin{minipage}[t]{0.135\textwidth}
            \includegraphics[width=1\textwidth]{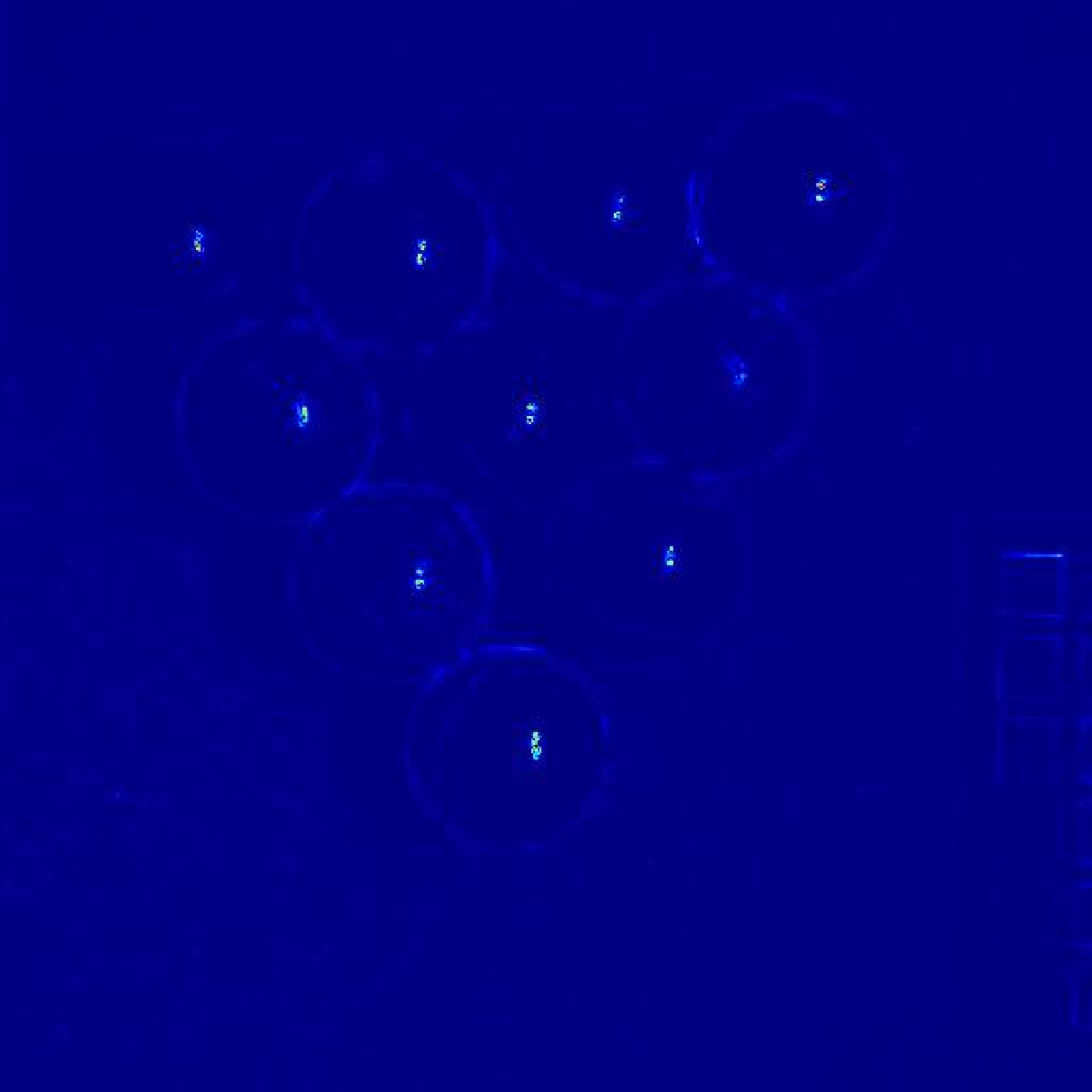}
            \centerline{RCAN}
        \end{minipage}
    }\hspace{-2.7mm}
    \subfigure{
        \begin{minipage}[t]{0.135\textwidth}
            \includegraphics[width=1\textwidth]{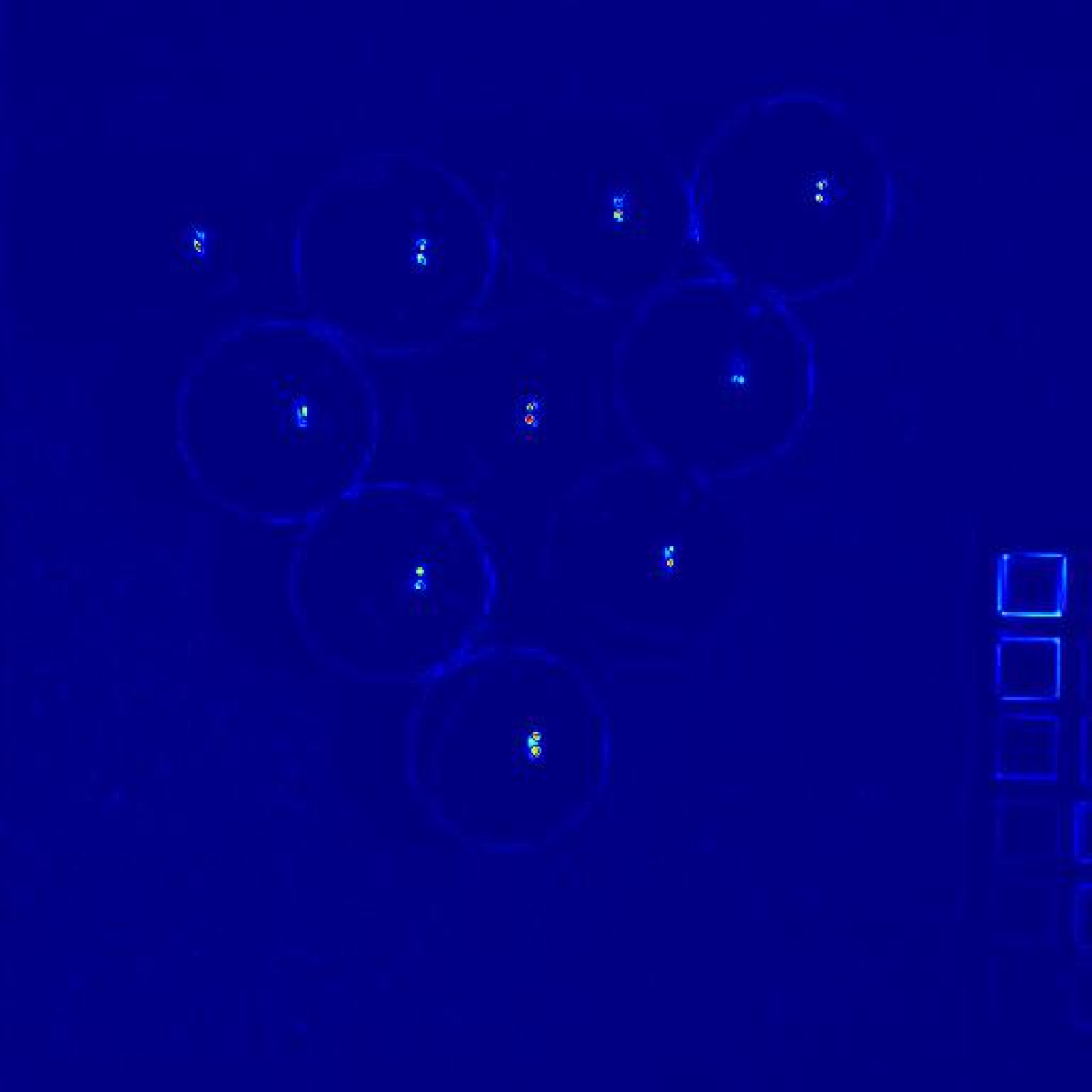}
            \centerline{3DCNN}
        \end{minipage}
    }\hspace{-2.7mm}
    \subfigure{
        \begin{minipage}[t]{0.135\textwidth}
            \includegraphics[width=1\textwidth]{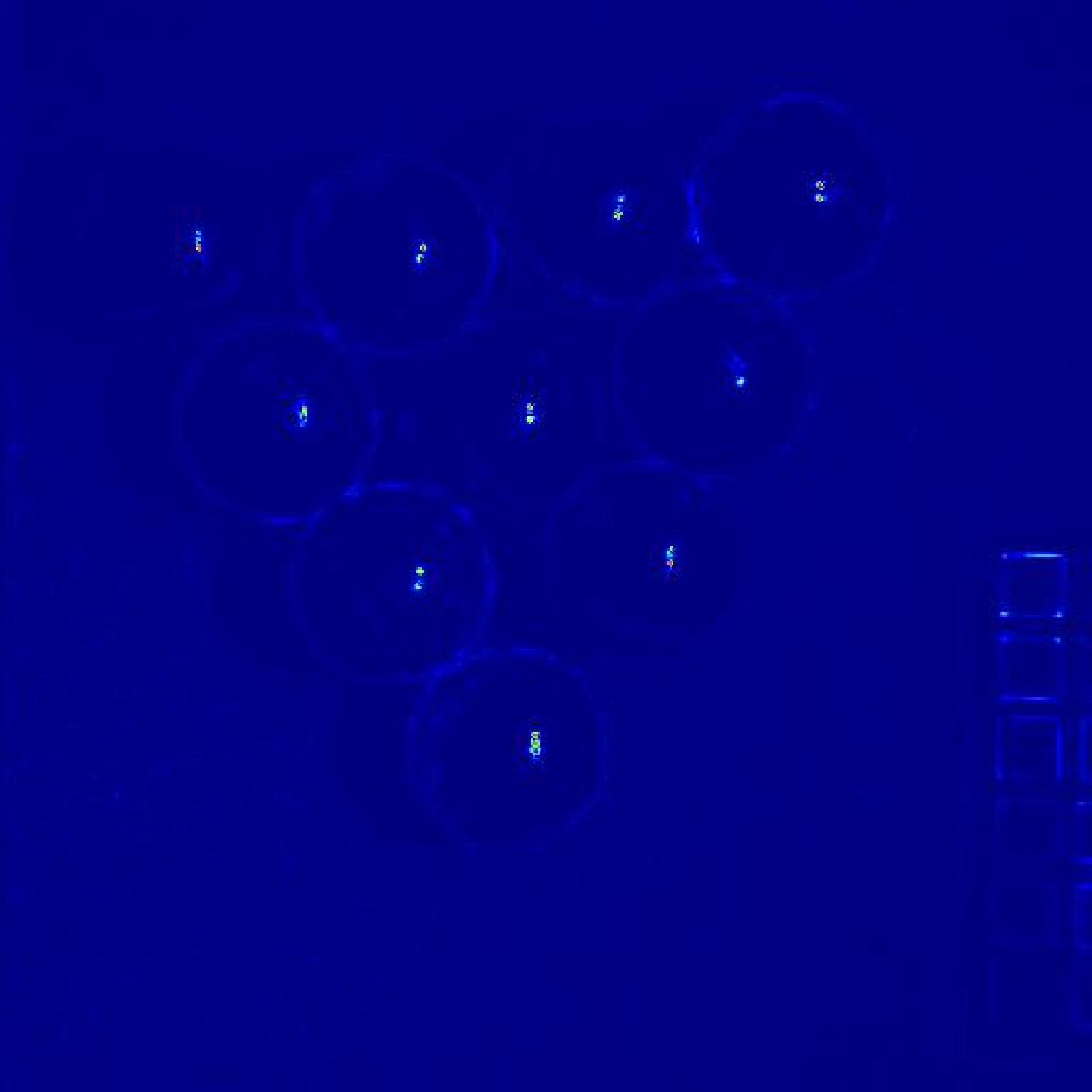}
            \centerline{GDRRN}
        \end{minipage}
    }\hspace{-2.7mm}
    \subfigure{
        \begin{minipage}[t]{0.135\textwidth}
            \includegraphics[width=1\textwidth]{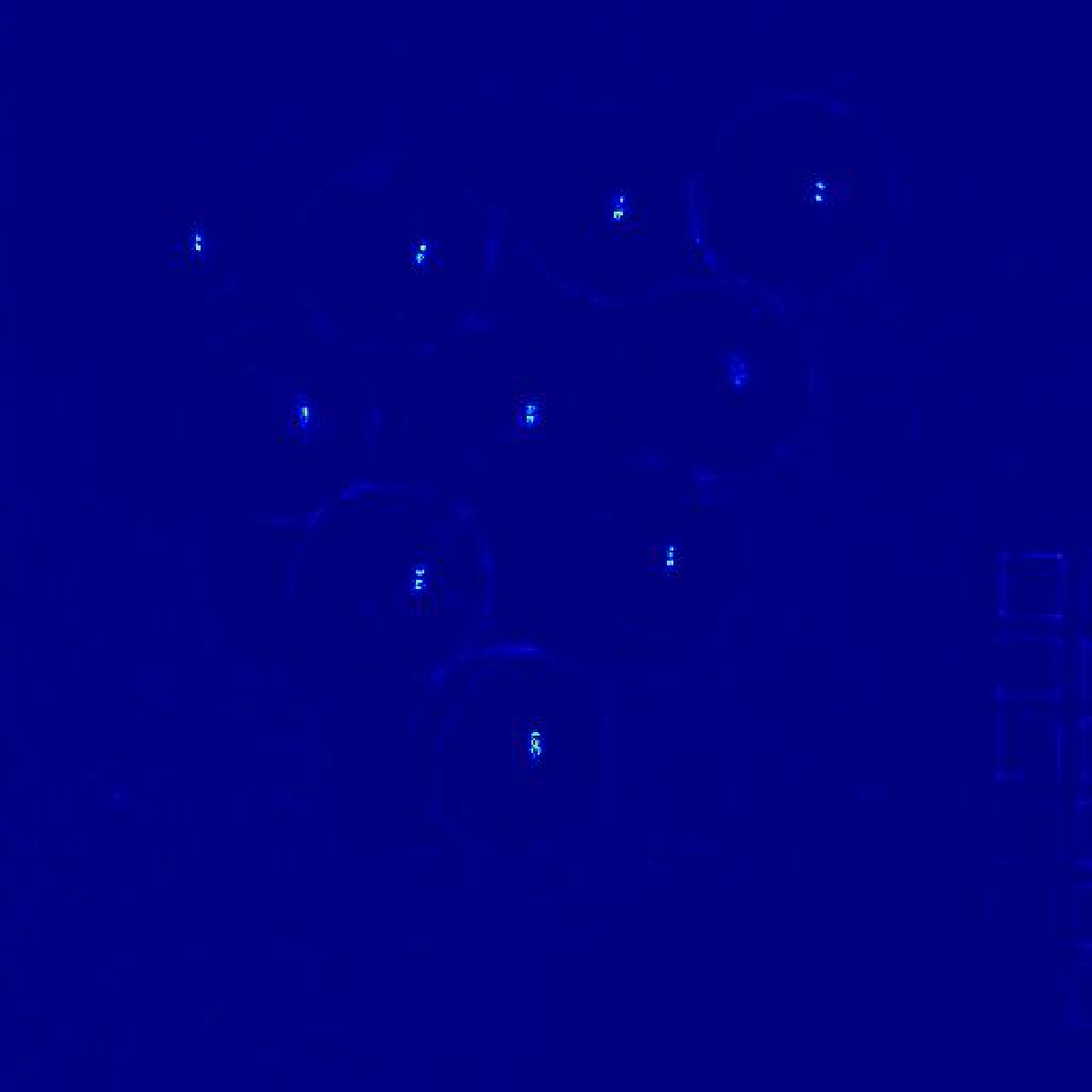}
            \centerline{SSPSR}
        \end{minipage}
    }\hspace{-2.7mm}
    \subfigure{
        \begin{minipage}[t]{0.135\textwidth}
            \includegraphics[width=1\textwidth]{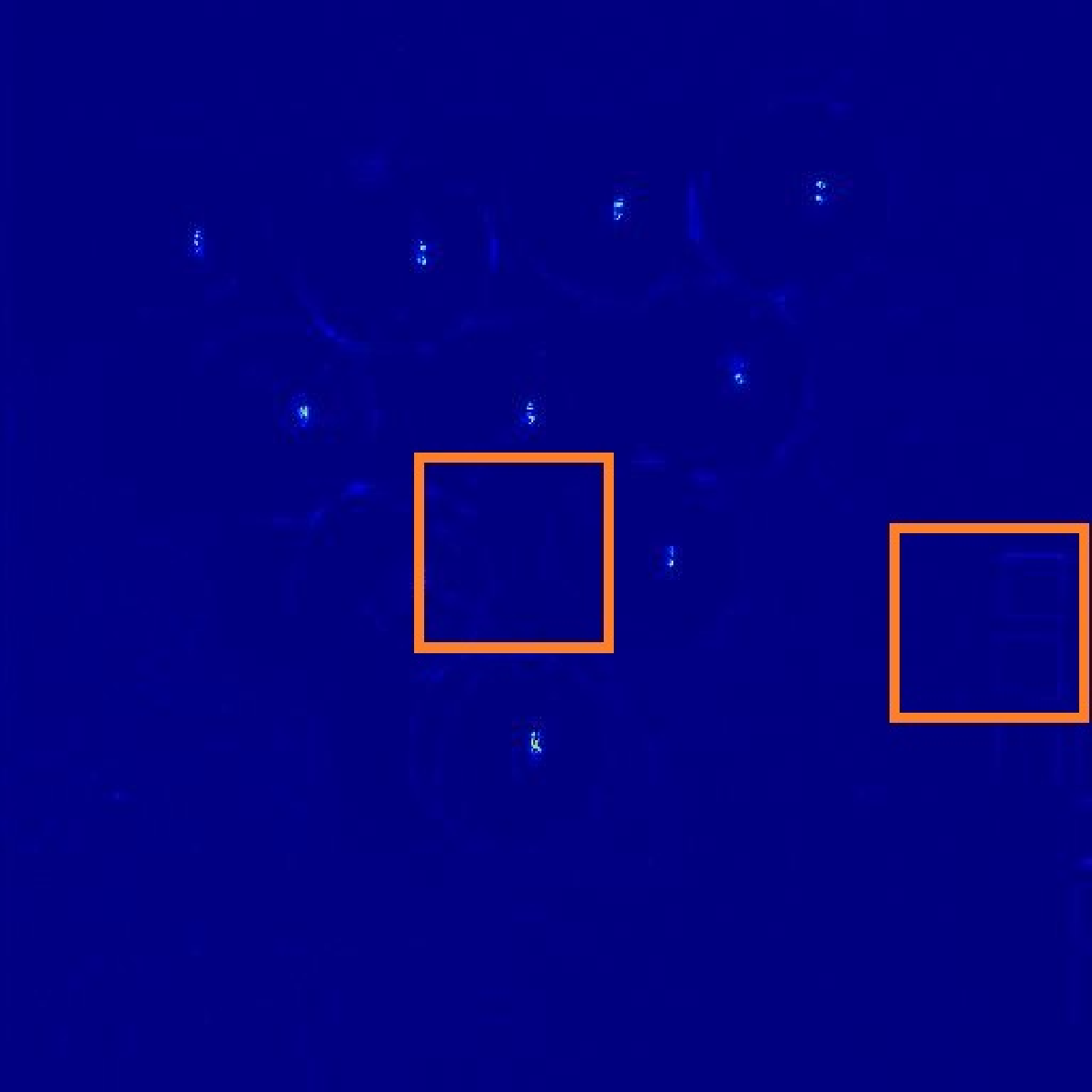}
            \centerline{FRLGN}
        \end{minipage}
    }
    \\
    \subfigure{
        \begin{minipage}[t]{0.135\textwidth}
            \includegraphics[width=1\textwidth]{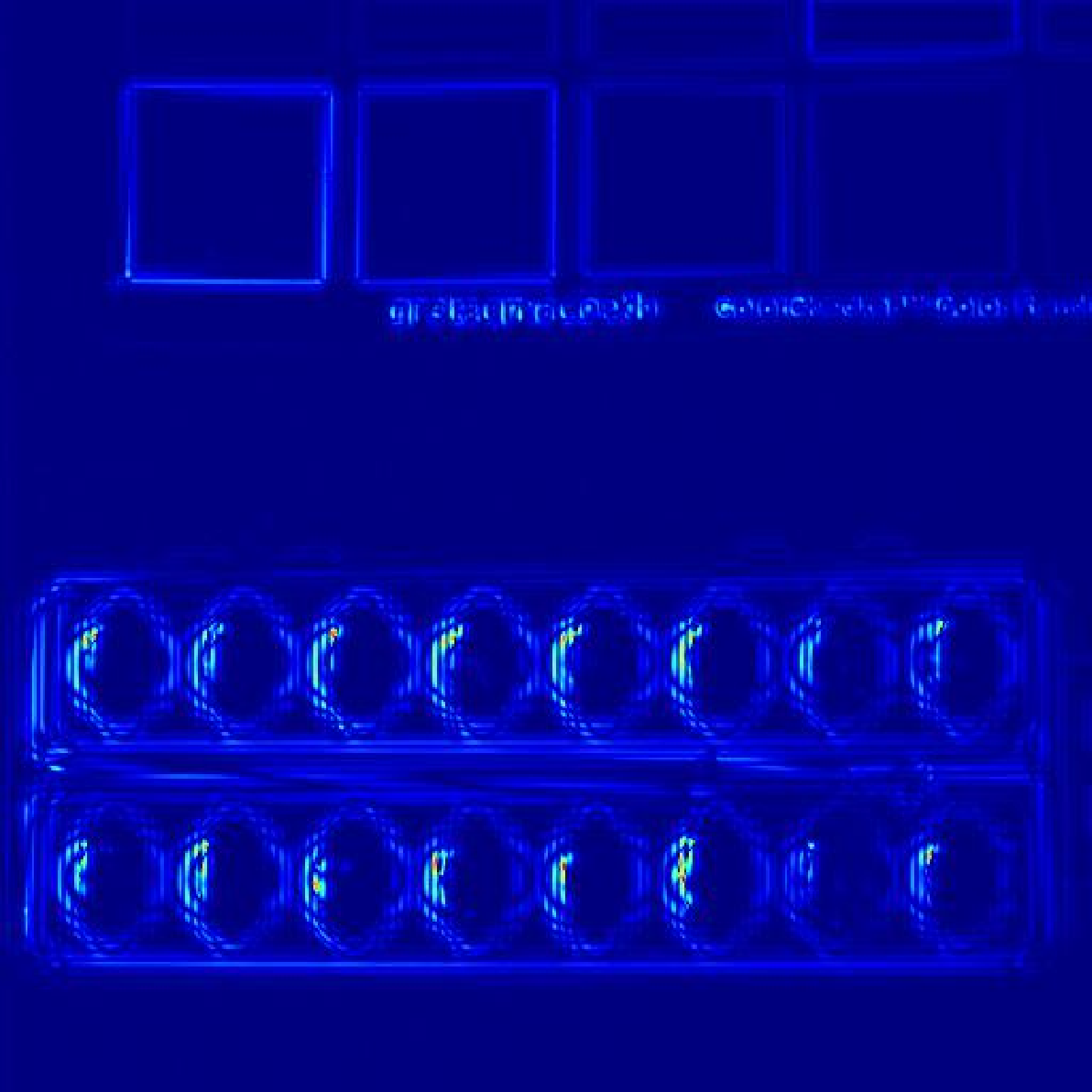}
            \centerline{Bicubic}
        \end{minipage}
    }\hspace{-2.7mm}
    \subfigure{
        \begin{minipage}[t]{0.135\textwidth}
            \includegraphics[width=1\textwidth]{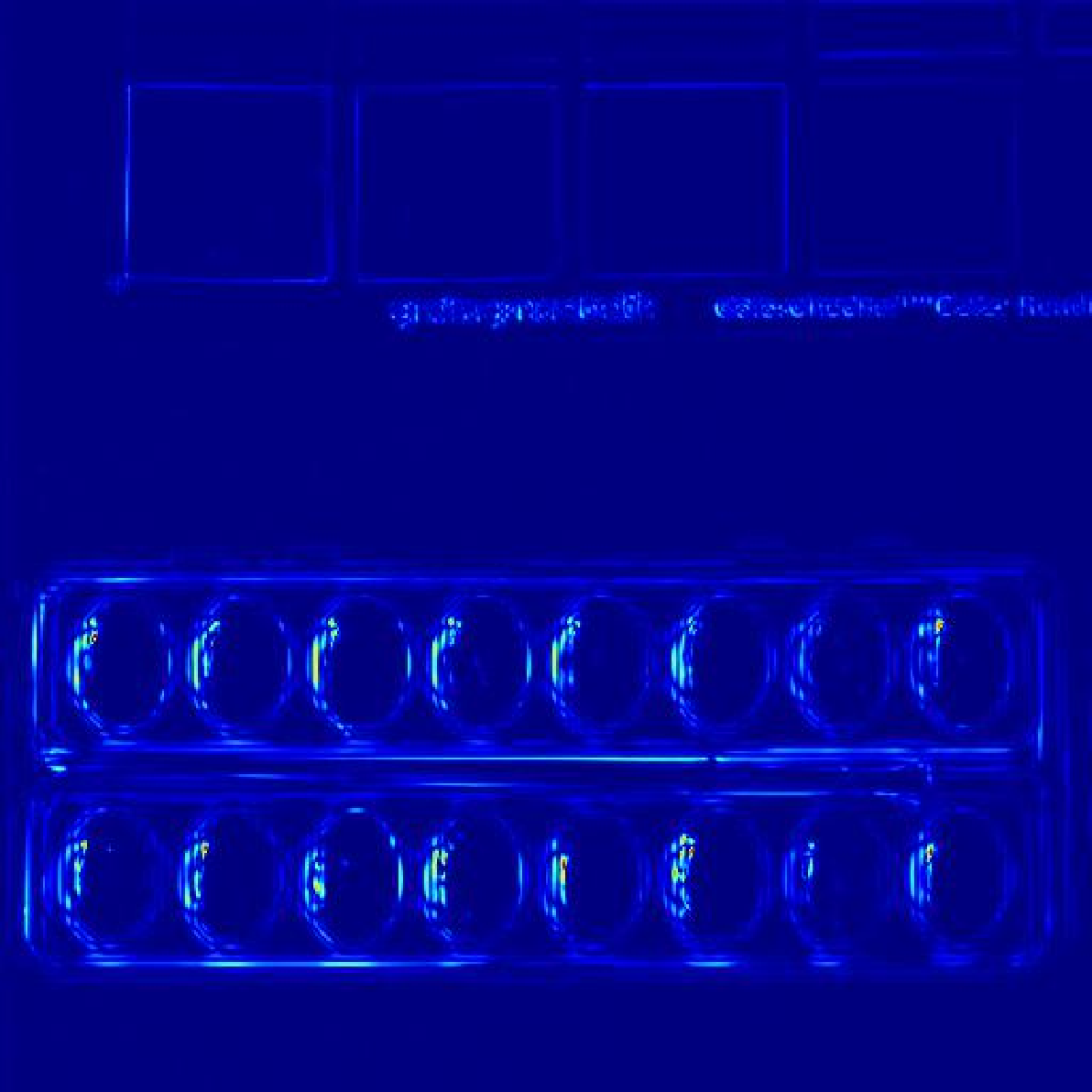}
            \centerline{VDSR}
        \end{minipage}
    }\hspace{-2.7mm}
    \subfigure{
        \begin{minipage}[t]{0.135\textwidth}
            \includegraphics[width=1\textwidth]{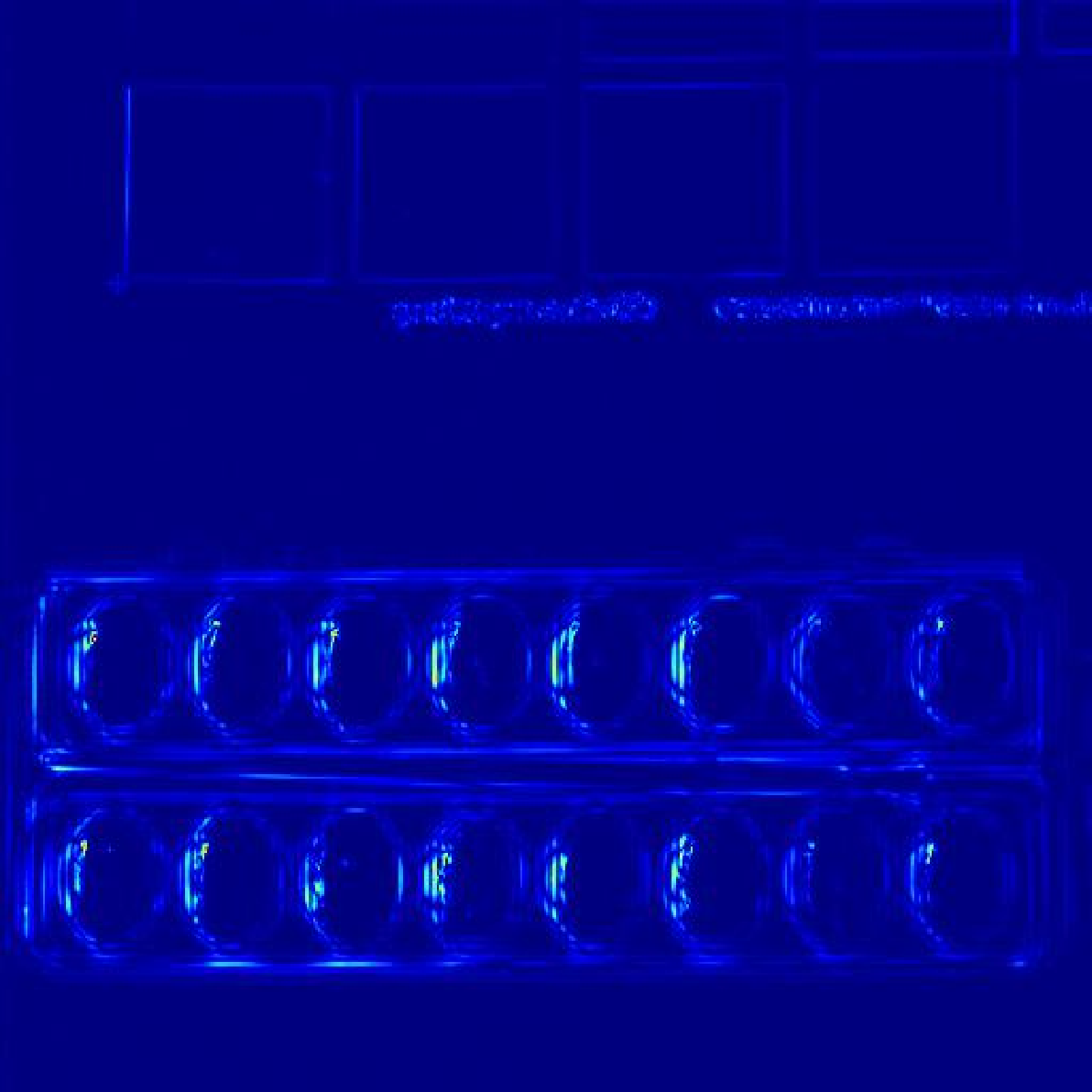}
            \centerline{RCAN}
        \end{minipage}
    }\hspace{-2.7mm}
    \subfigure{
        \begin{minipage}[t]{0.135\textwidth}
            \includegraphics[width=1\textwidth]{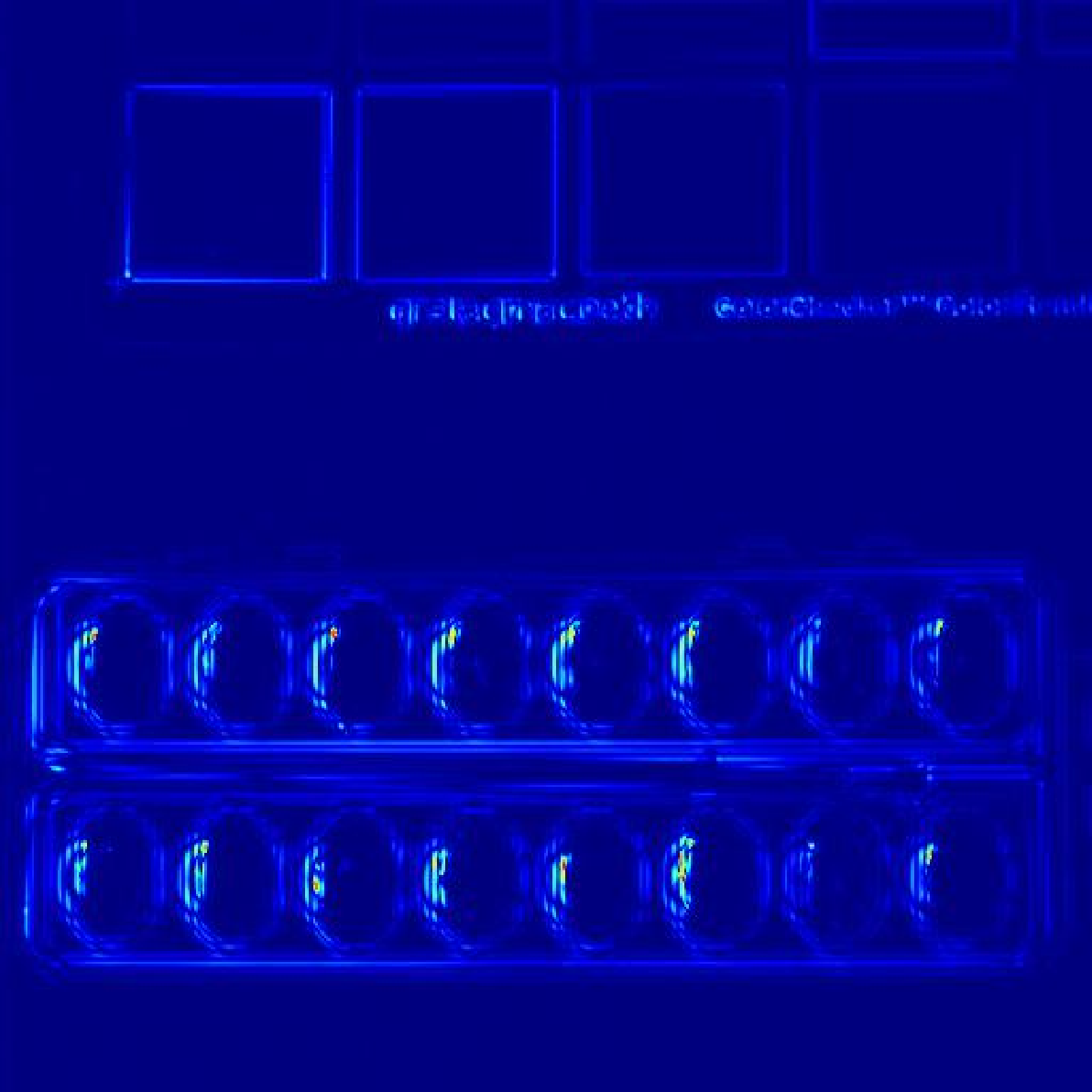}
            \centerline{3DCNN}
        \end{minipage}
    }\hspace{-2.7mm}
    \subfigure{
        \begin{minipage}[t]{0.135\textwidth}
            \includegraphics[width=1\textwidth]{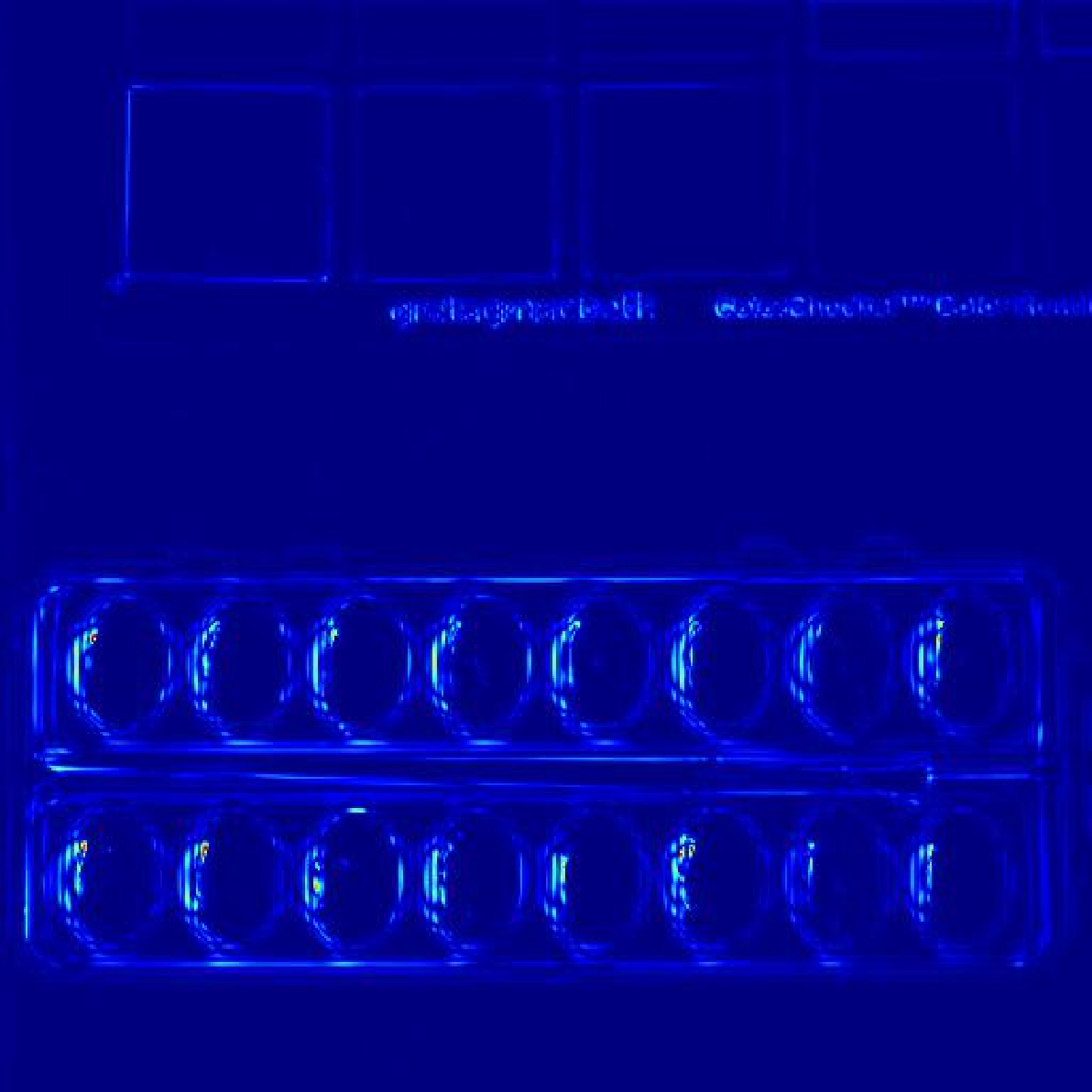}
            \centerline{GDRRN}
        \end{minipage}
    }\hspace{-2.7mm}
    \subfigure{
        \begin{minipage}[t]{0.135\textwidth}
            \includegraphics[width=1\textwidth]{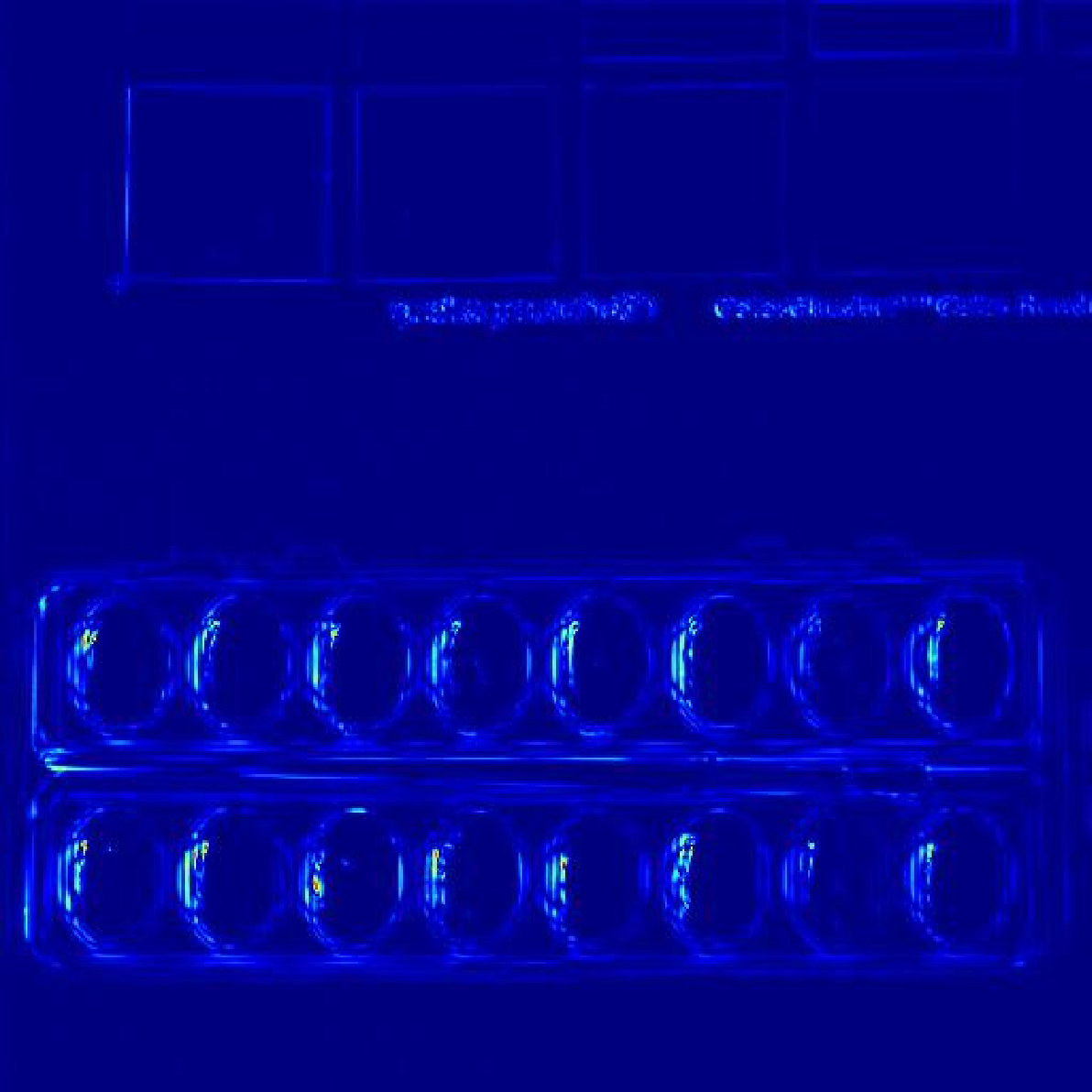}
            \centerline{SSPSR}
        \end{minipage}
    }\hspace{-2.7mm}
    \subfigure{
        \begin{minipage}[t]{0.135\textwidth}
            \includegraphics[width=1\textwidth]{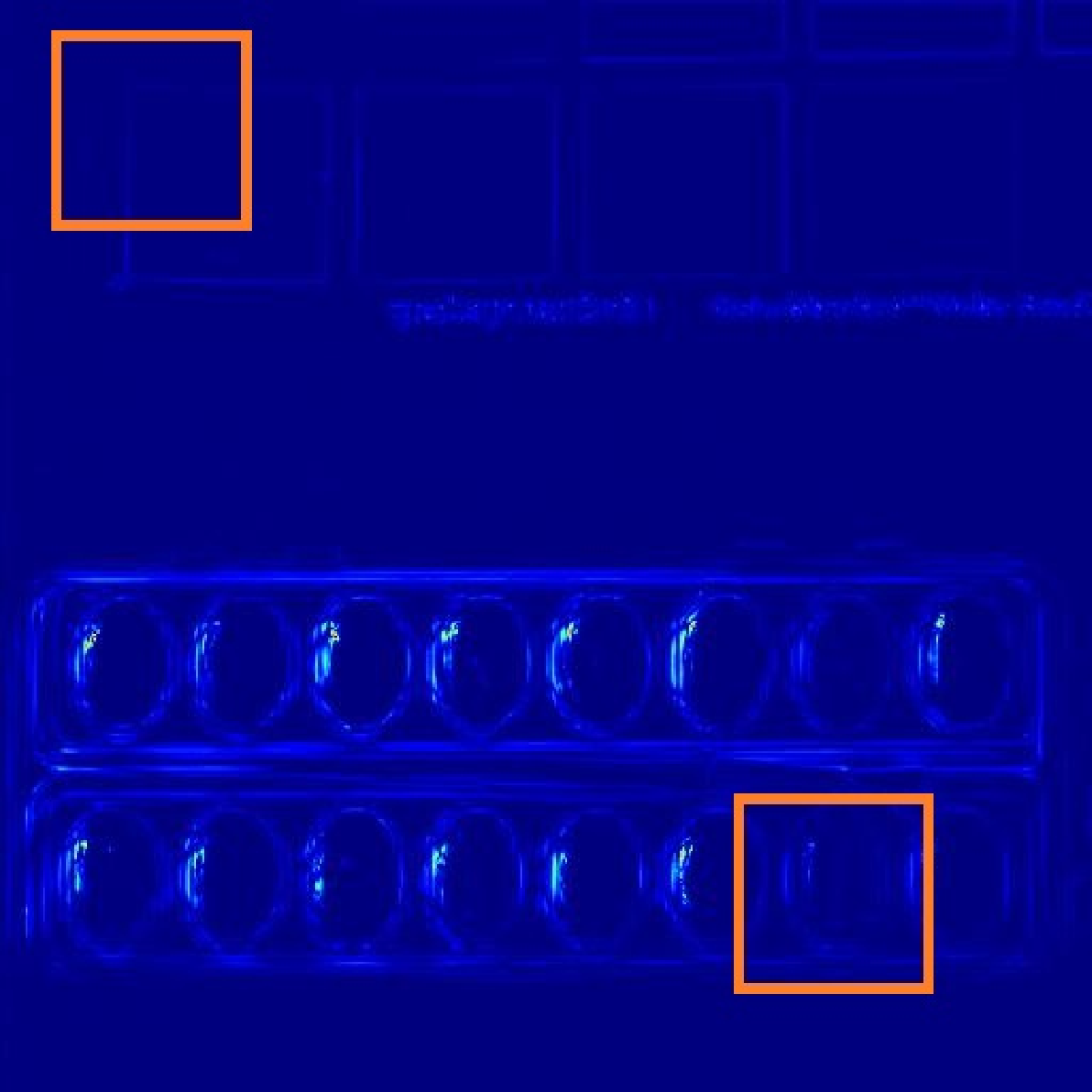}
            \centerline{FRLGN}
        \end{minipage}
    }
\caption{Mean error maps of superballs and paints hyperspectral images from the
CAVE testing dataset with a scale factor of 4}
\label{fig:CAVEResult1}
\end{figure*}

\begin{figure*}[htbp]
    \centering
    \subfigure{
        \begin{minipage}[t]{0.135\textwidth}
            \includegraphics[width=1\textwidth]{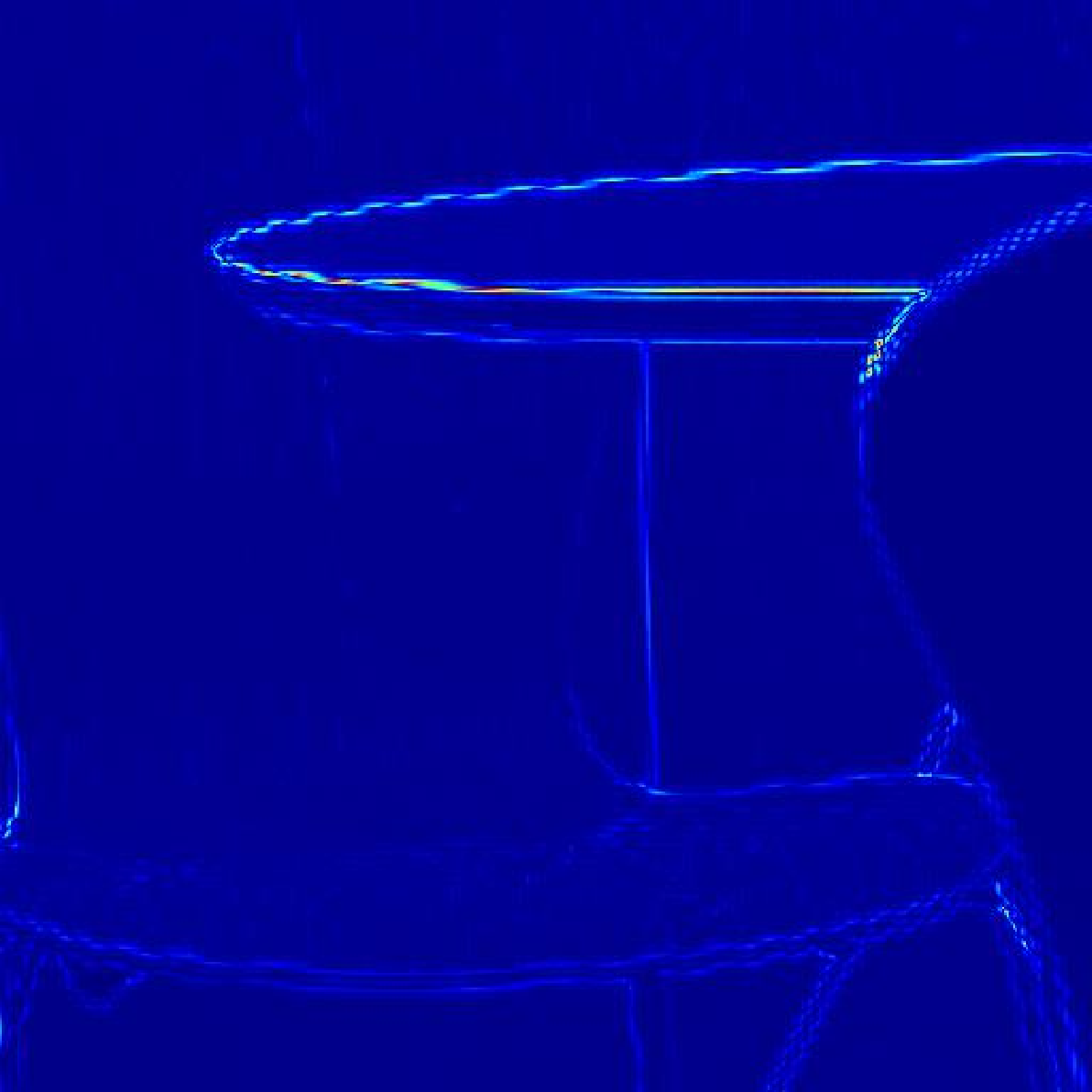}
            \centerline{Bicubic}
        \end{minipage}
    }\hspace{-2.7mm}
    \subfigure{
        \begin{minipage}[t]{0.135\textwidth}
            \includegraphics[width=1\textwidth]{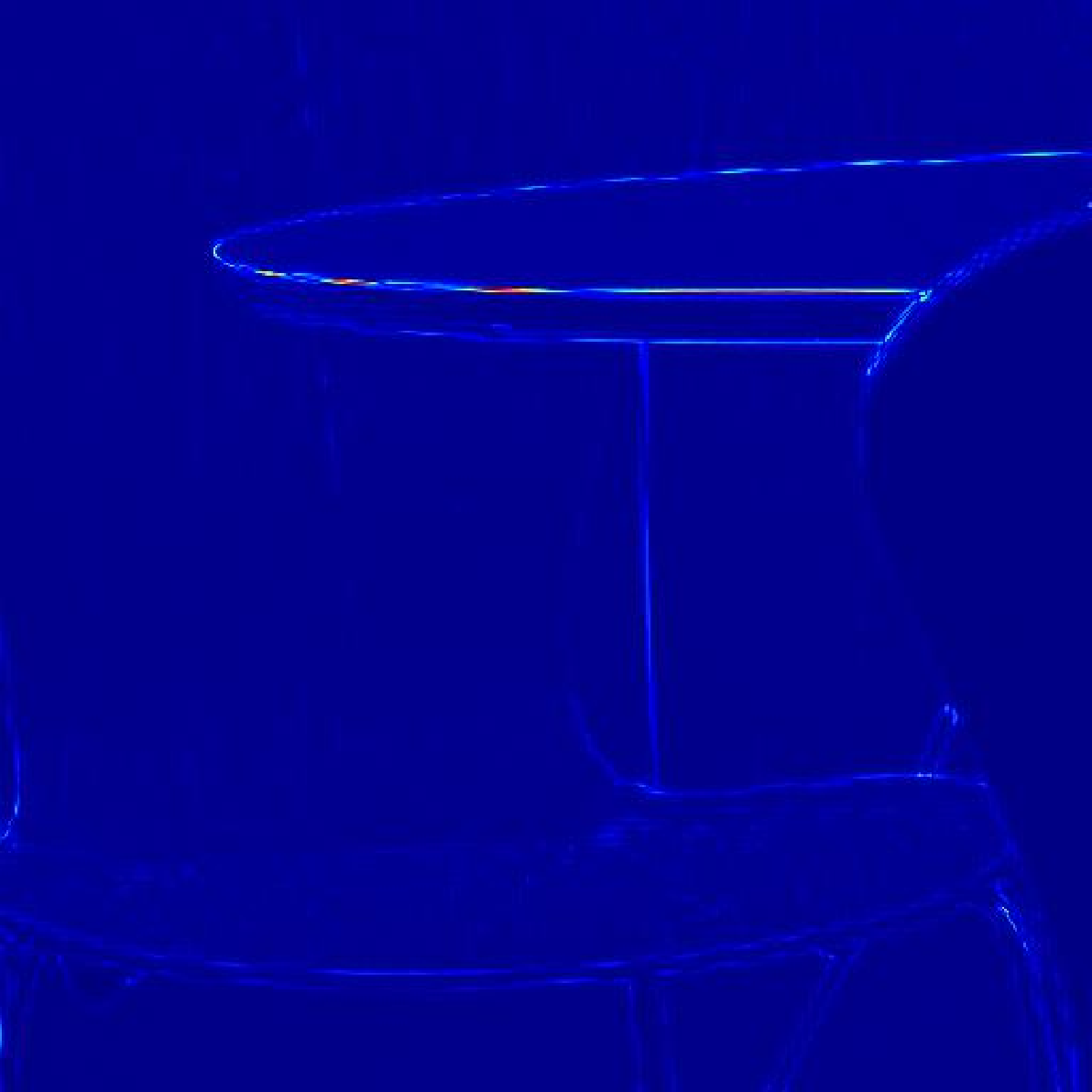}
            \centerline{VDSR}
        \end{minipage}
    }\hspace{-2.7mm}
    \subfigure{
        \begin{minipage}[t]{0.135\textwidth}
            \includegraphics[width=1\textwidth]{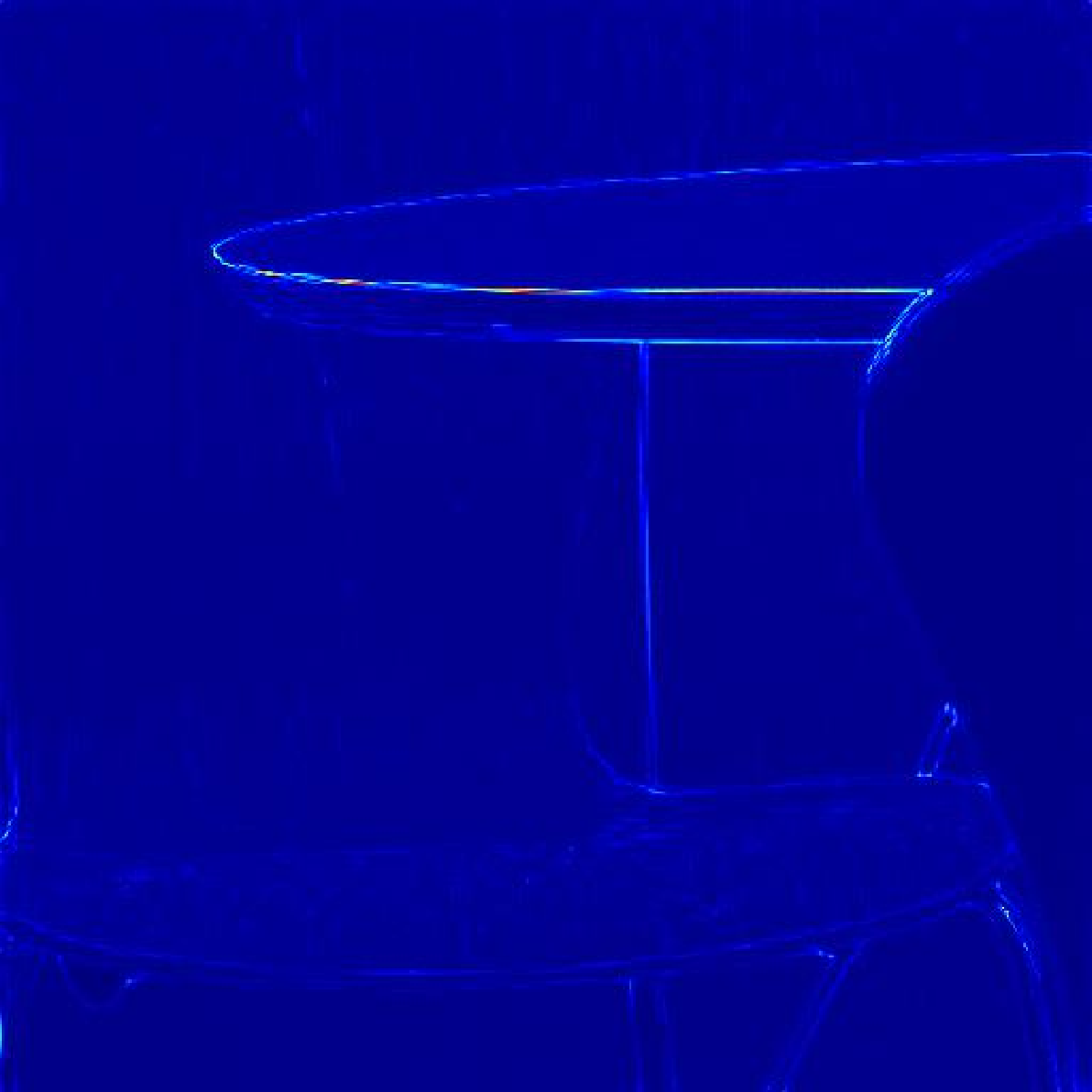}
            \centerline{RCAN}
        \end{minipage}
    }\hspace{-2.7mm}
    \subfigure{
        \begin{minipage}[t]{0.135\textwidth}
            \includegraphics[width=1\textwidth]{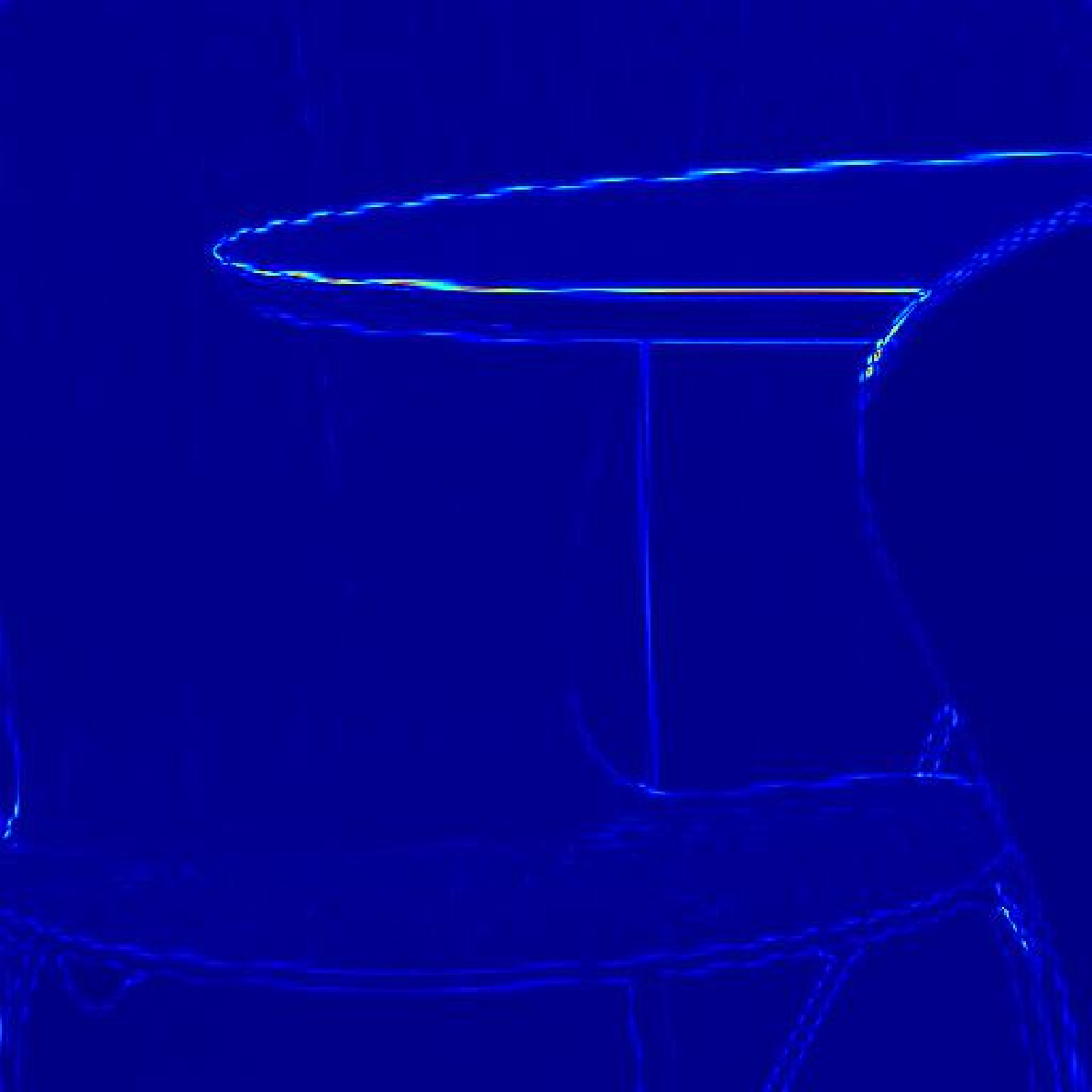}
            \centerline{3DCNN}
        \end{minipage}
    }\hspace{-2.7mm}
    \subfigure{
        \begin{minipage}[t]{0.135\textwidth}
            \includegraphics[width=1\textwidth]{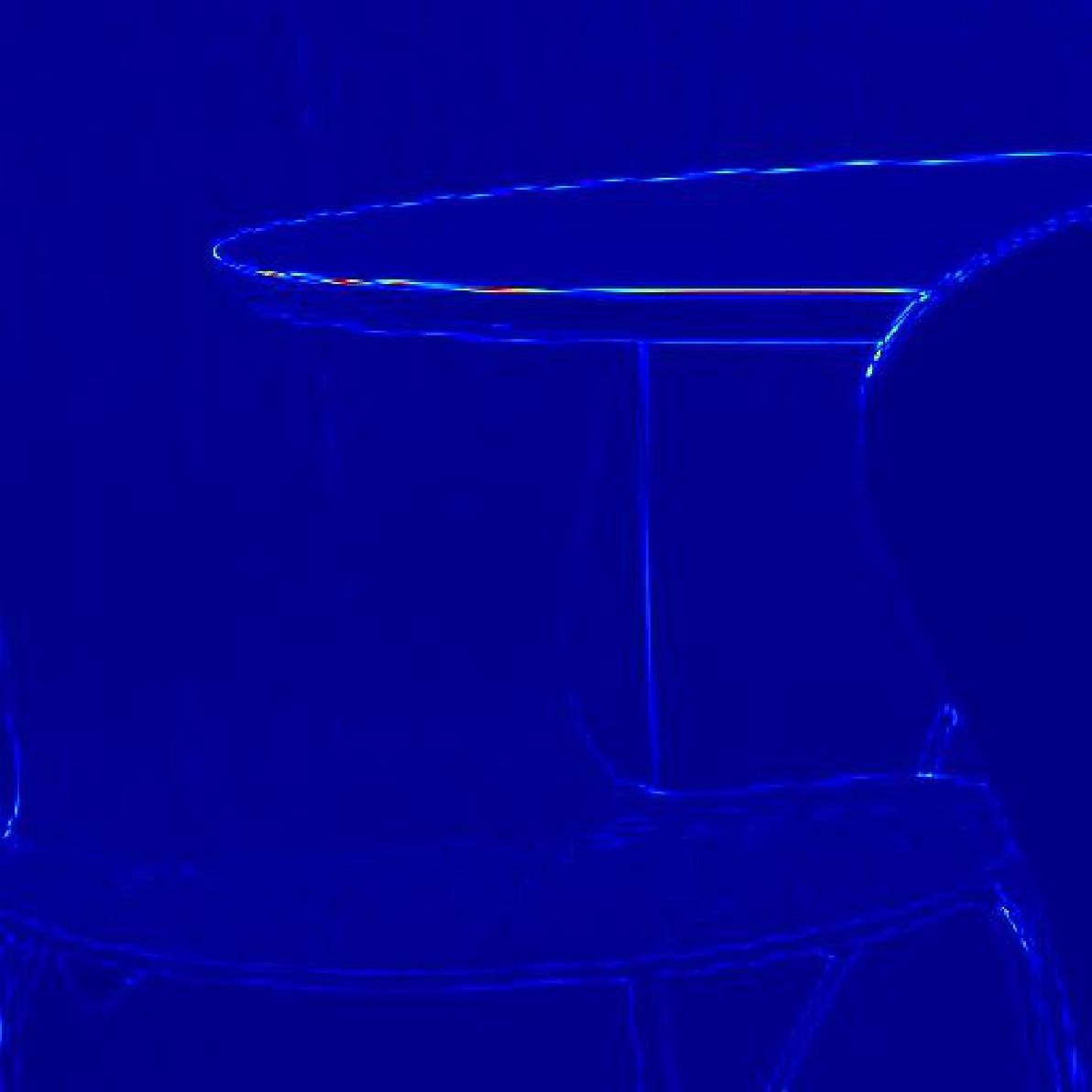}
            \centerline{GDRRN}
        \end{minipage}
    }\hspace{-2.7mm}
    \subfigure{
        \begin{minipage}[t]{0.135\textwidth}
            \includegraphics[width=1\textwidth]{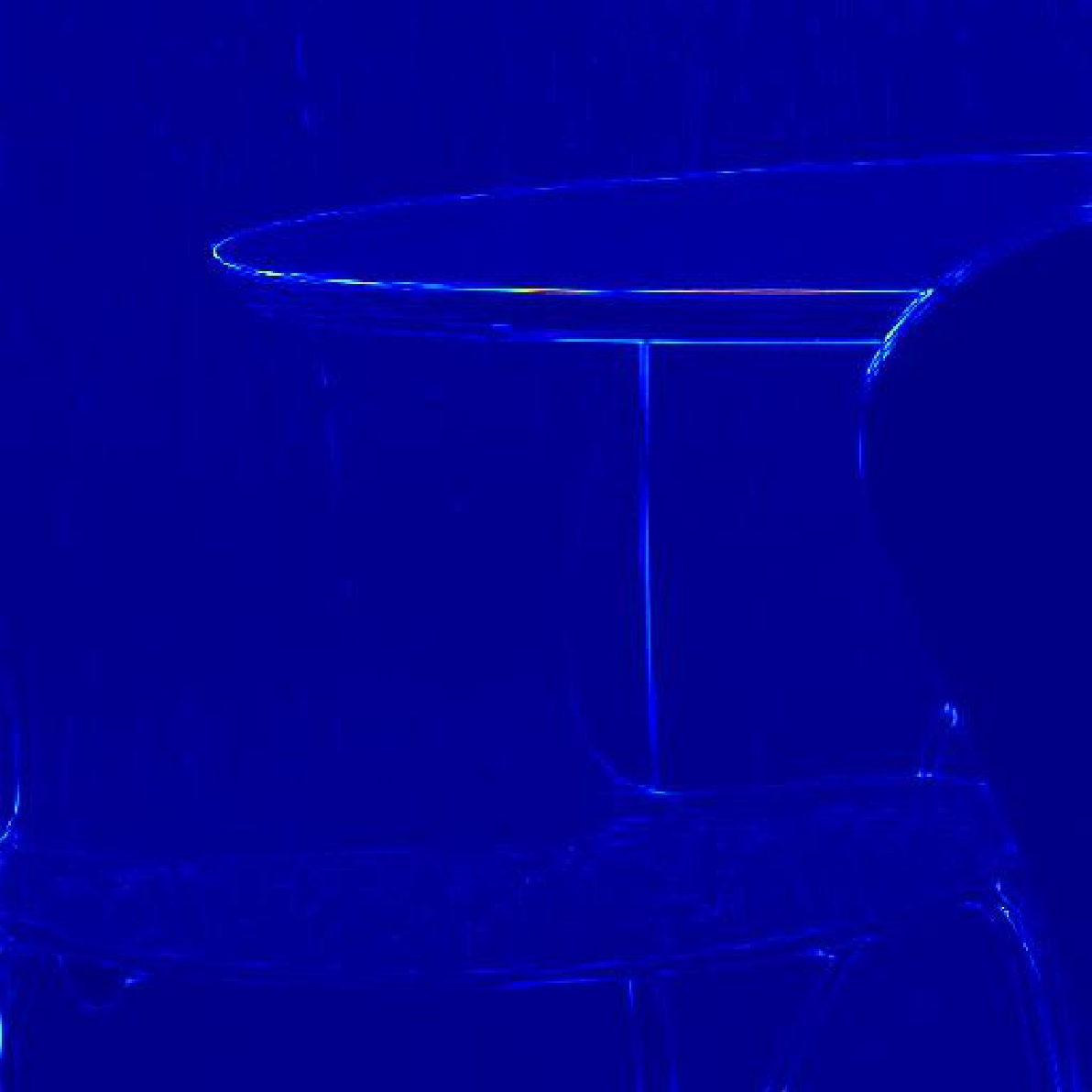}
            \centerline{SSPSR}
        \end{minipage}
    }\hspace{-2.7mm}
    \subfigure{
        \begin{minipage}[t]{0.135\textwidth}
            \includegraphics[width=1\textwidth]{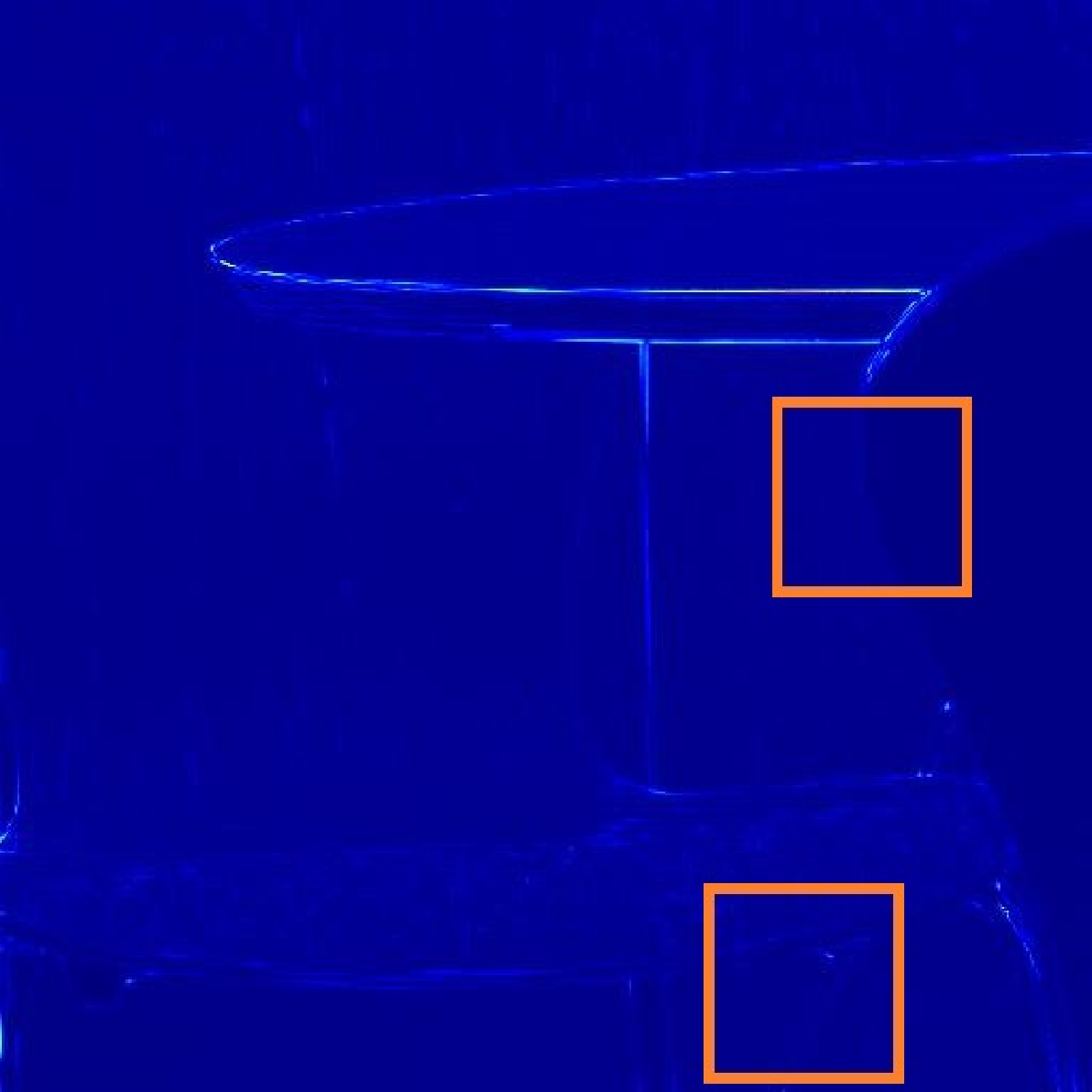}
            \centerline{FRLGN}
        \end{minipage}
    }
    \\
    \subfigure{
        \begin{minipage}[t]{0.135\textwidth}
            \includegraphics[width=1\textwidth]{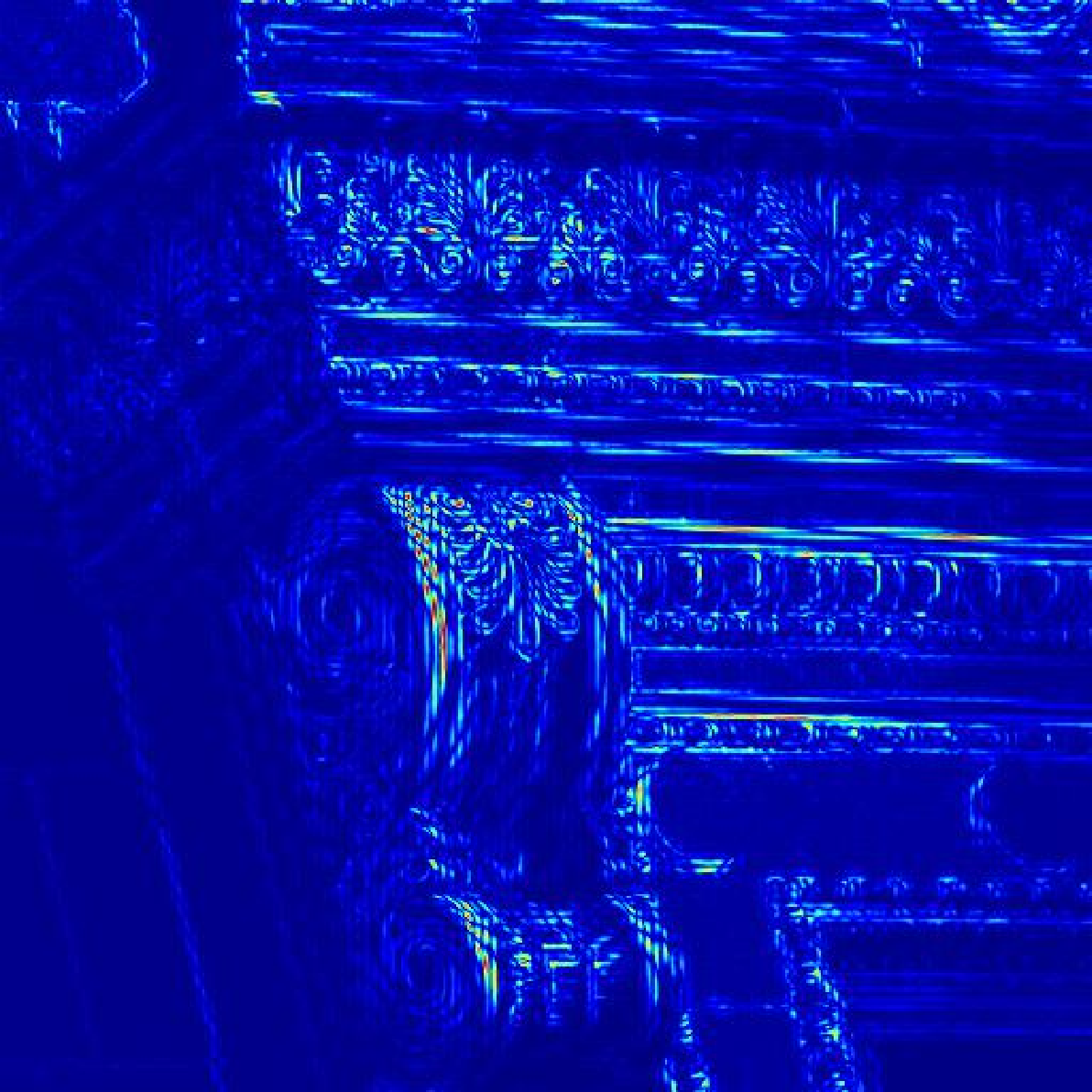}
            \centerline{Bicubic}
        \end{minipage}
    }\hspace{-2.7mm}
    \subfigure{
        \begin{minipage}[t]{0.135\textwidth}
            \includegraphics[width=1\textwidth]{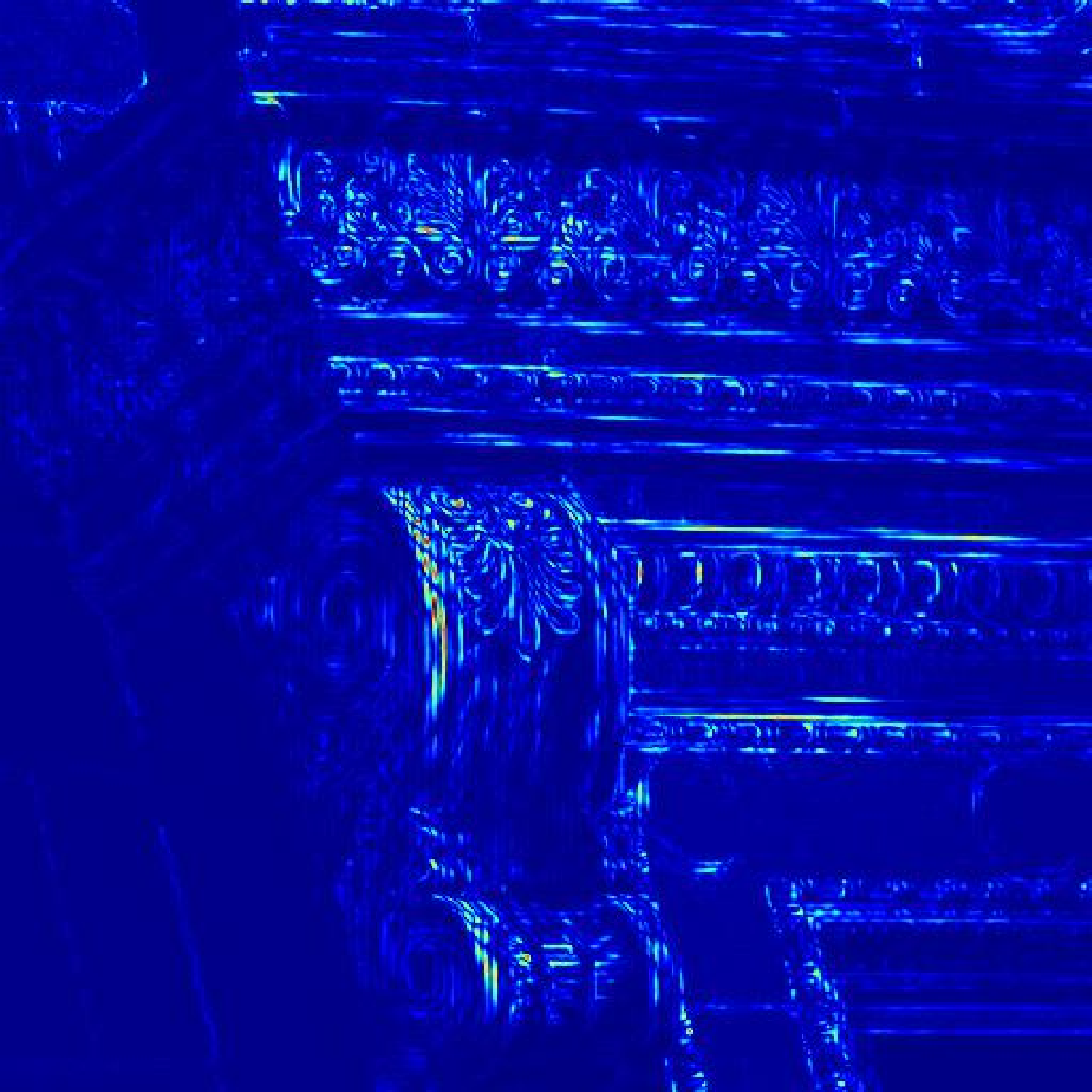}
            \centerline{VDSR}
        \end{minipage}
    }\hspace{-2.7mm}
    \subfigure{
        \begin{minipage}[t]{0.135\textwidth}
            \includegraphics[width=1\textwidth]{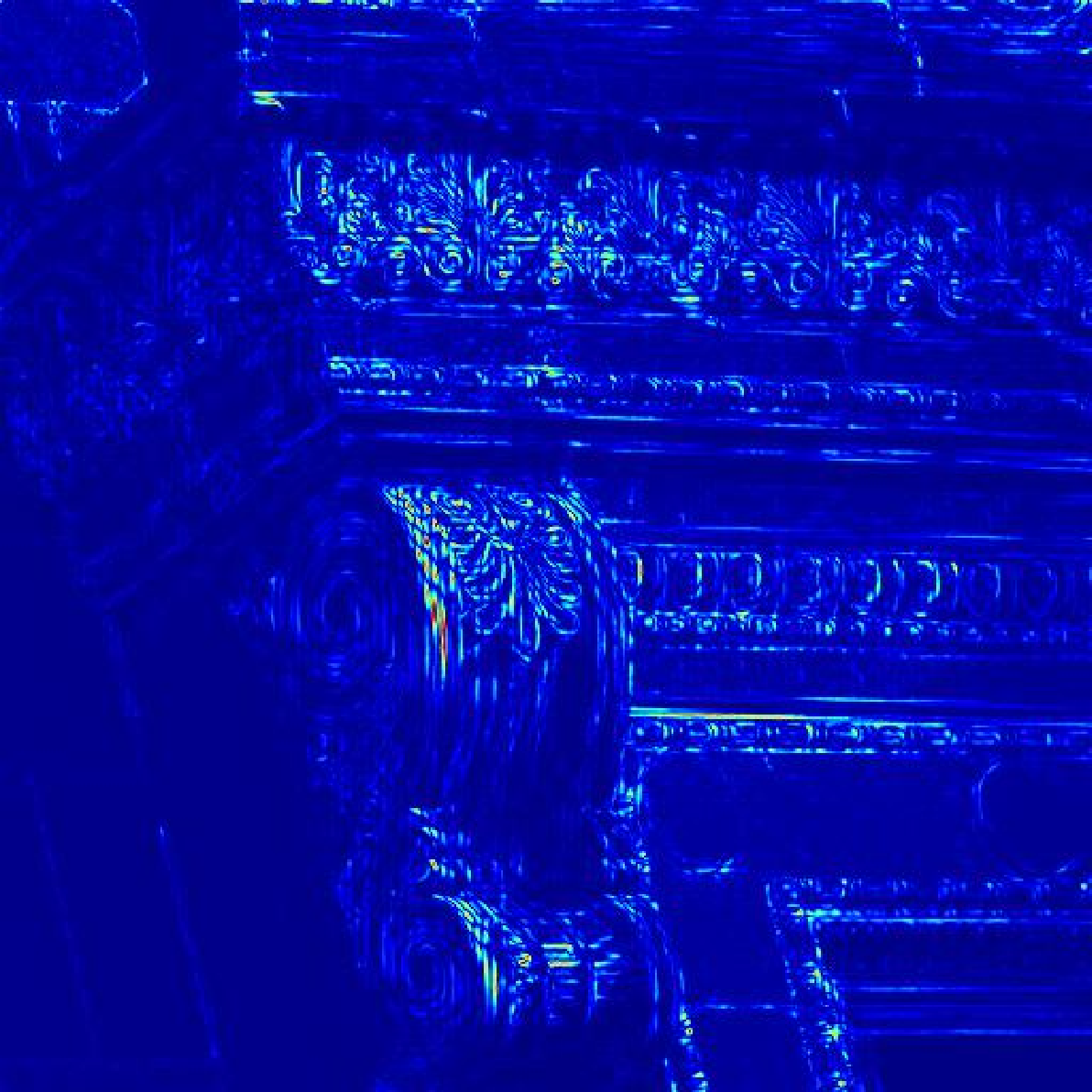}
            \centerline{RCAN}
        \end{minipage}
    }\hspace{-2.7mm}
    \subfigure{
        \begin{minipage}[t]{0.135\textwidth}
            \includegraphics[width=1\textwidth]{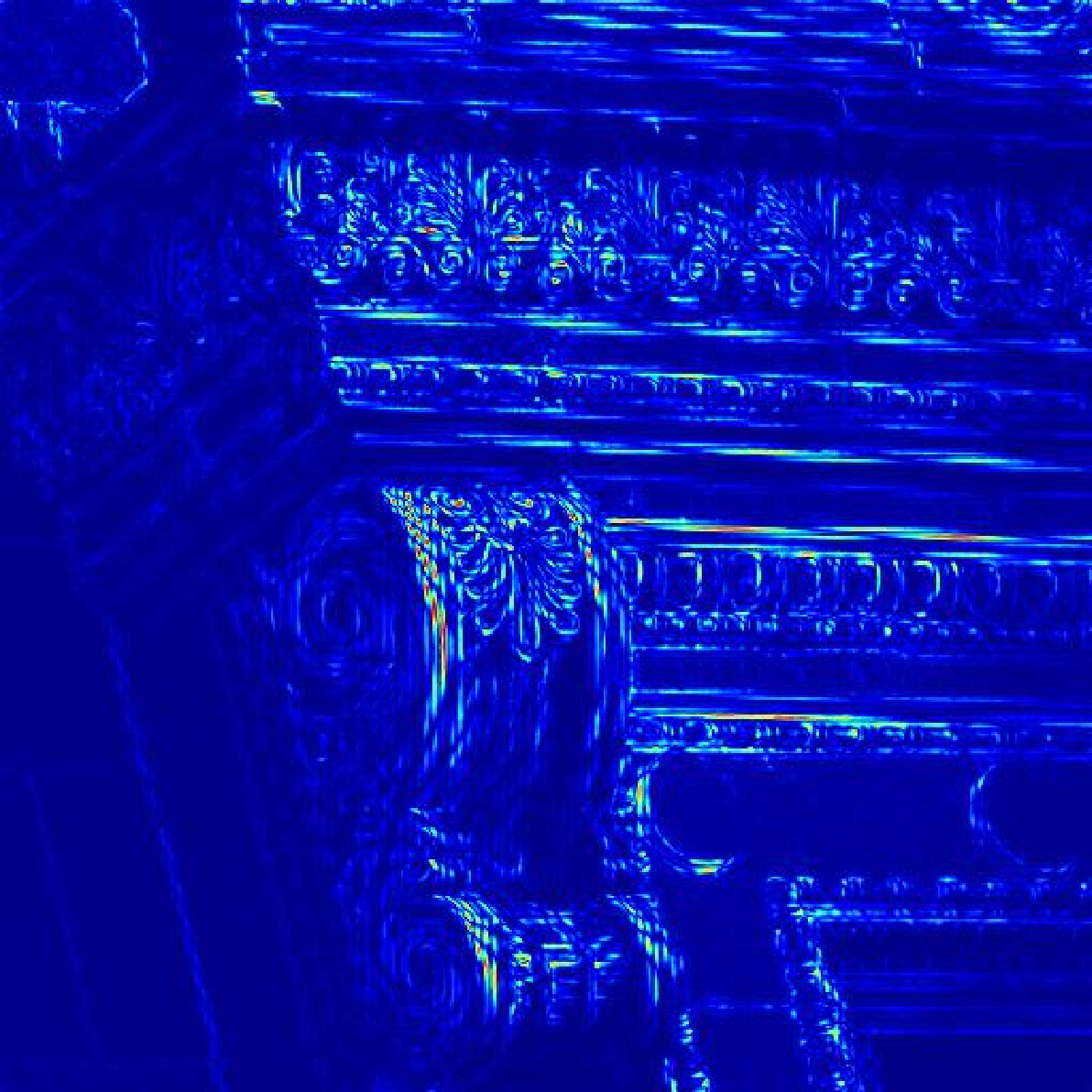}
            \centerline{3DCNN}
        \end{minipage}
    }\hspace{-2.7mm}
    \subfigure{
        \begin{minipage}[t]{0.135\textwidth}
            \includegraphics[width=1\textwidth]{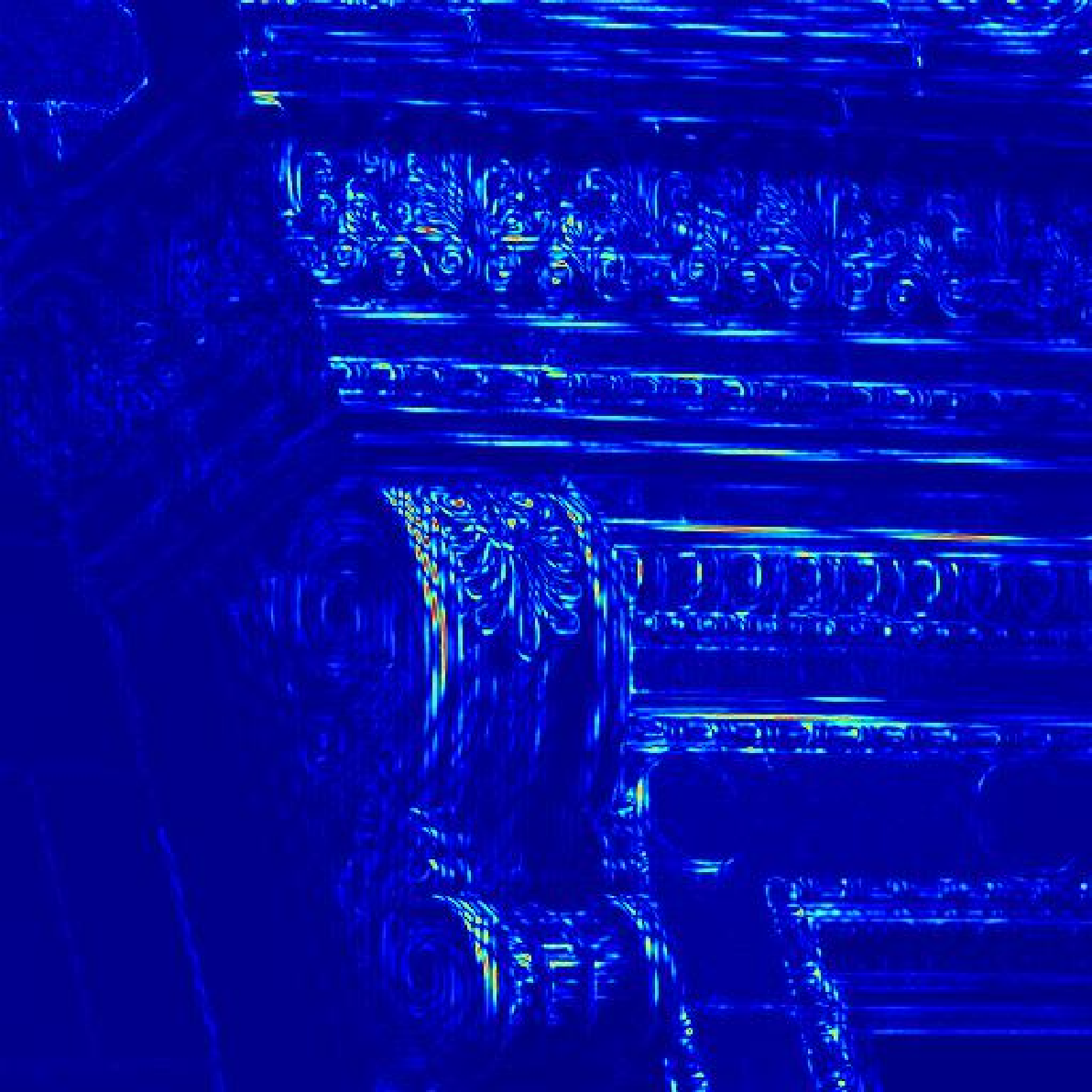}
            \centerline{GDRRN}
        \end{minipage}
    }\hspace{-2.7mm}
    \subfigure{
        \begin{minipage}[t]{0.135\textwidth}
            \includegraphics[width=1\textwidth]{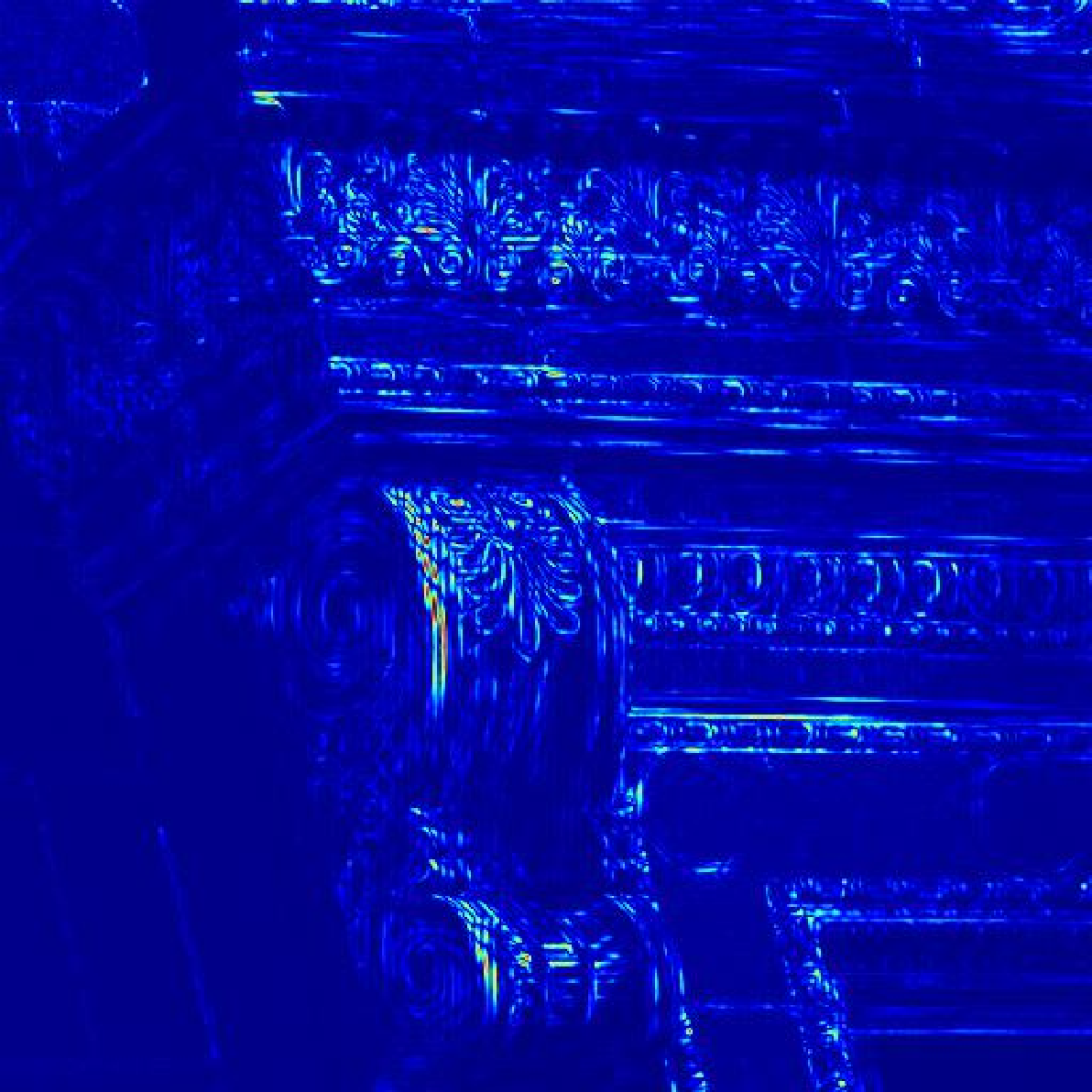}
            \centerline{SSPSR}
        \end{minipage}
    }\hspace{-2.7mm}
    \subfigure{
        \begin{minipage}[t]{0.135\textwidth}
            \includegraphics[width=1\textwidth]{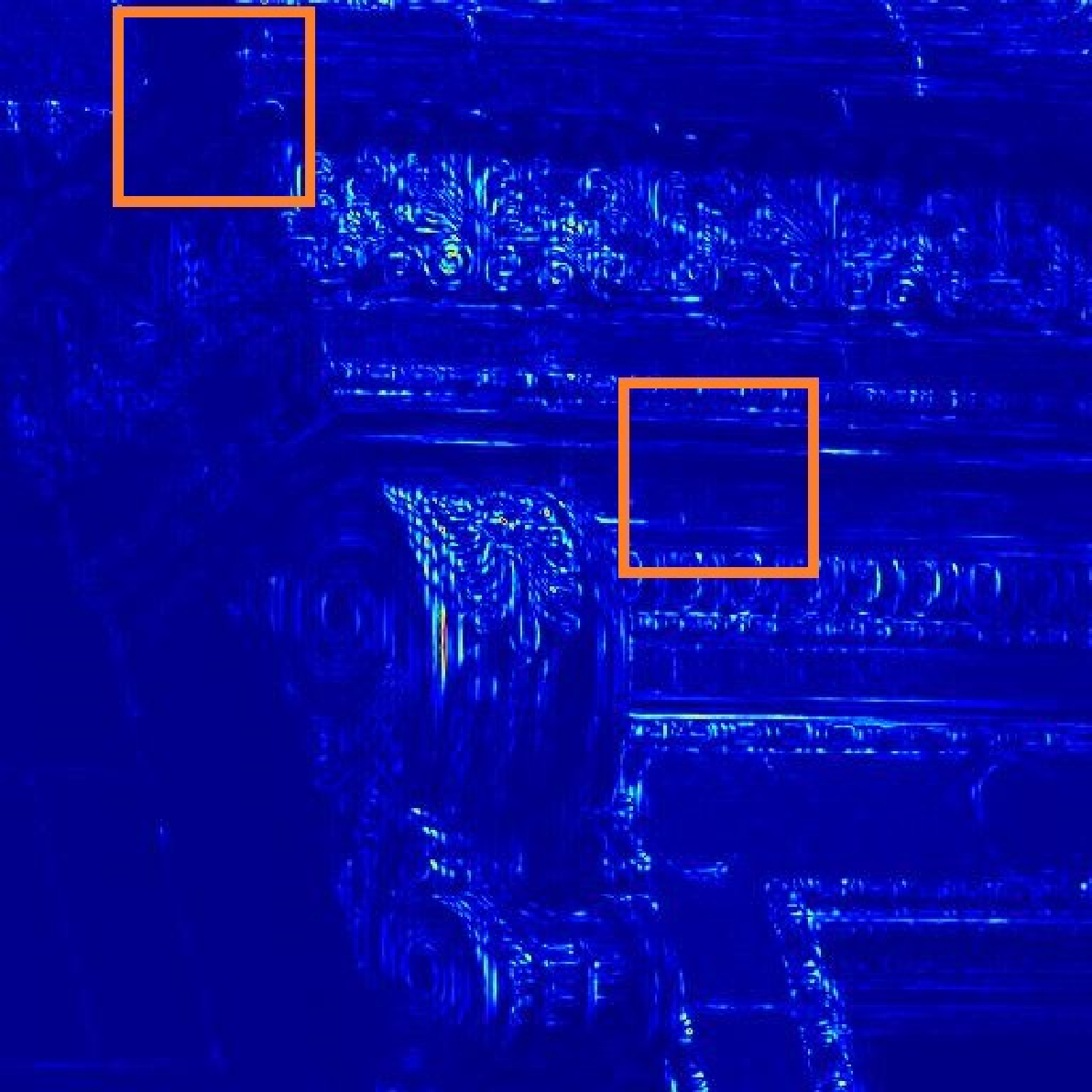}
            \centerline{FRLGN}
        \end{minipage}
    }
\caption{Mean error maps of two hyperspectral images from the
Harvard testing dataset with the scale factor 4.}
\label{fig:HarvardResult1}
\end{figure*}

\begin{figure*}[htbp]
    \centering
    \subfigure{
        \begin{minipage}[t]{0.135\textwidth}
            \includegraphics[width=1\textwidth]{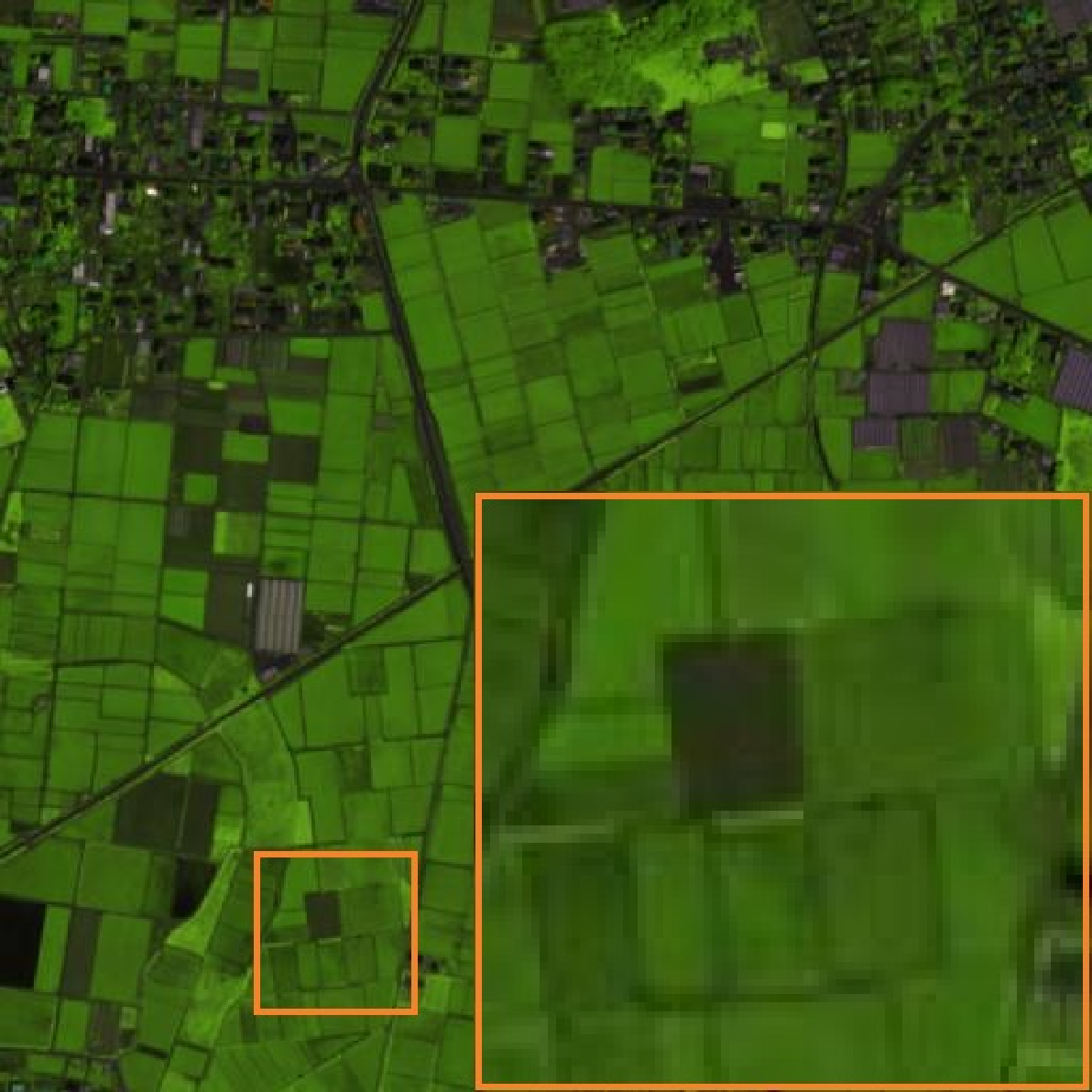}
            \centerline{HR}
        \end{minipage}
    }\hspace{-2.7mm}
    \subfigure{
        \begin{minipage}[t]{0.135\textwidth}
            \includegraphics[width=1\textwidth]{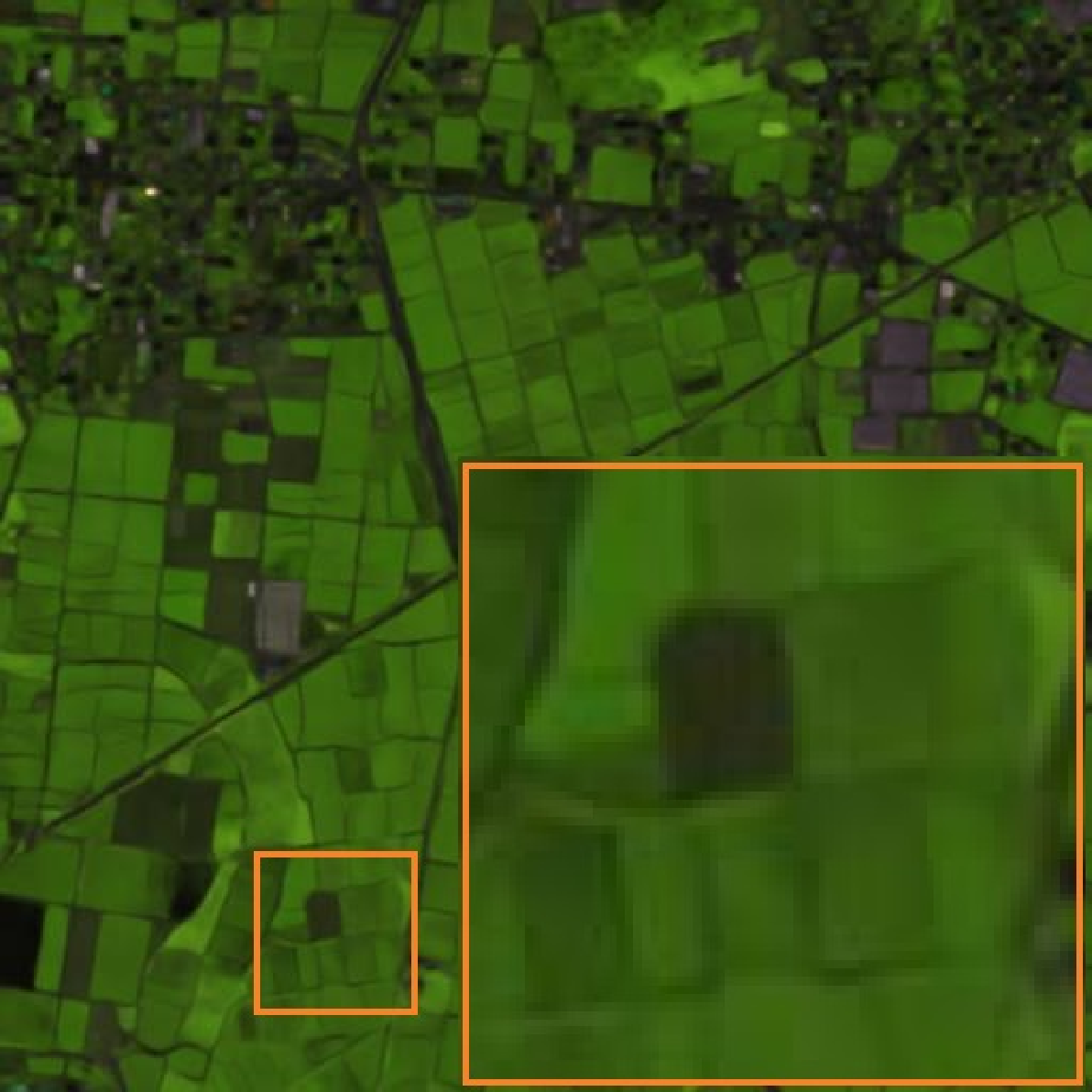}
            \centerline{VDSR}
        \end{minipage}
    }\hspace{-2.7mm}
    \subfigure{
        \begin{minipage}[t]{0.135\textwidth}
            \includegraphics[width=1\textwidth]{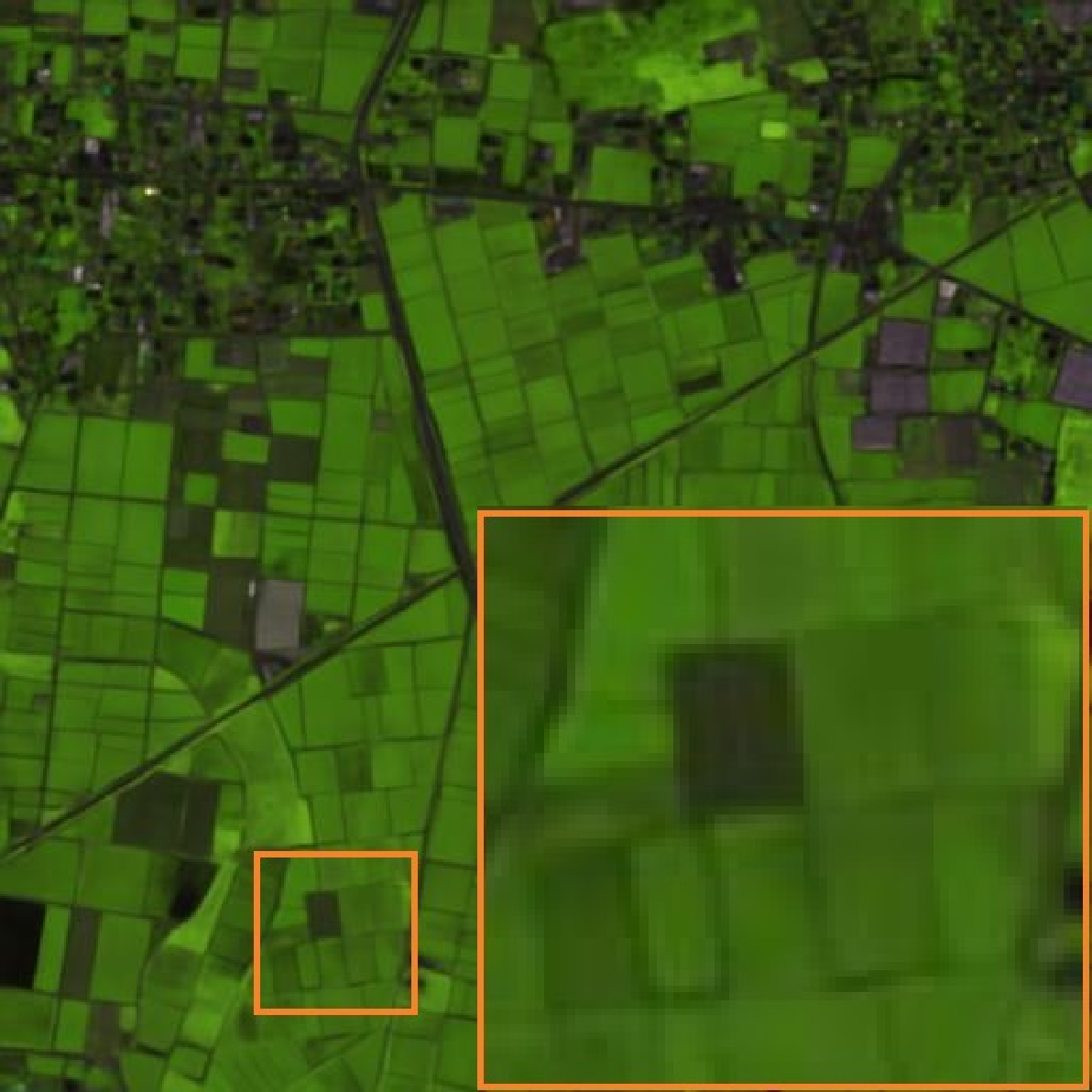}
            \centerline{RCAN}
        \end{minipage}
    }\hspace{-2.7mm}
    \subfigure{
        \begin{minipage}[t]{0.135\textwidth}
            \includegraphics[width=1\textwidth]{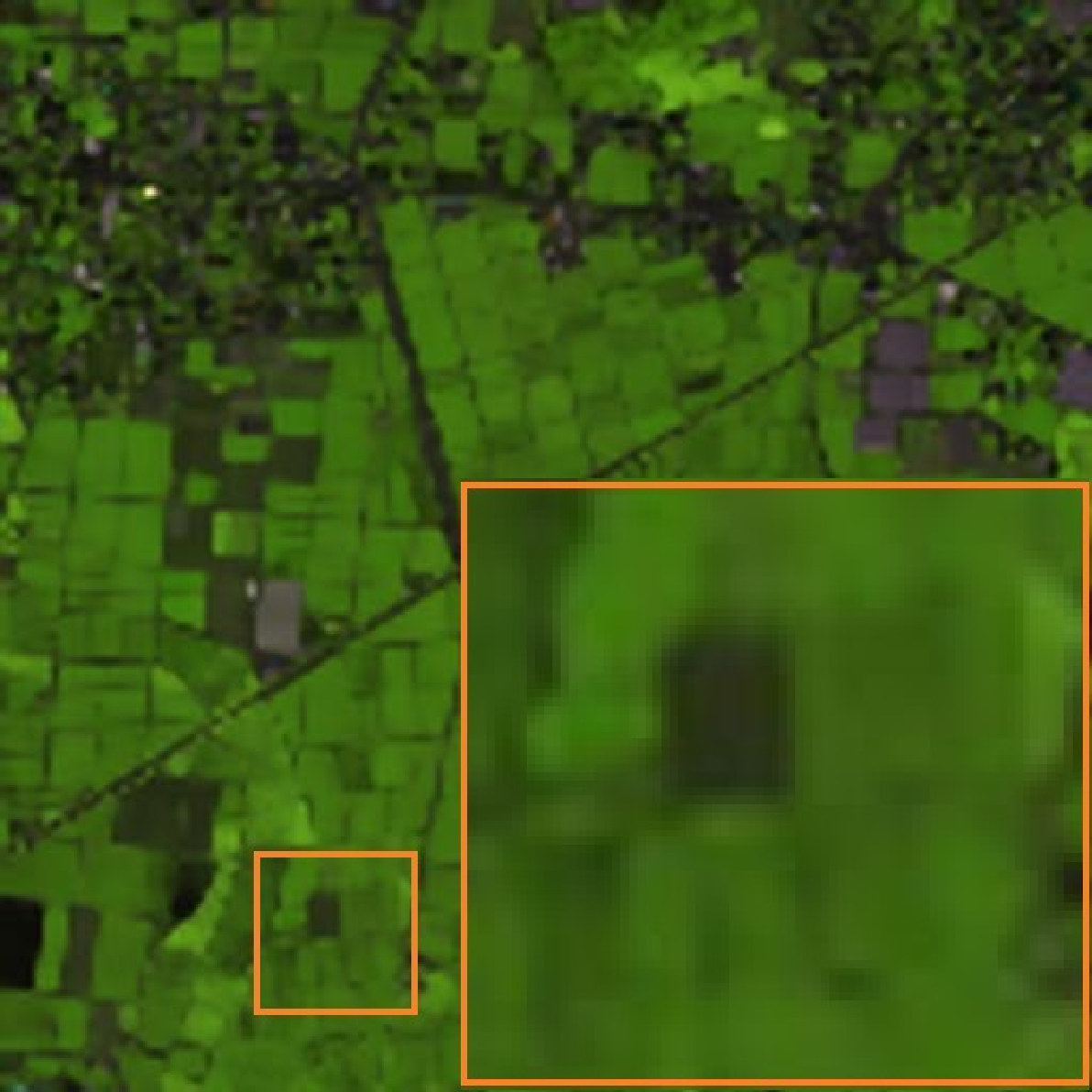}
            \centerline{3DCNN}
        \end{minipage}
    }\hspace{-2.7mm}
    \subfigure{
        \begin{minipage}[t]{0.135\textwidth}
            \includegraphics[width=1\textwidth]{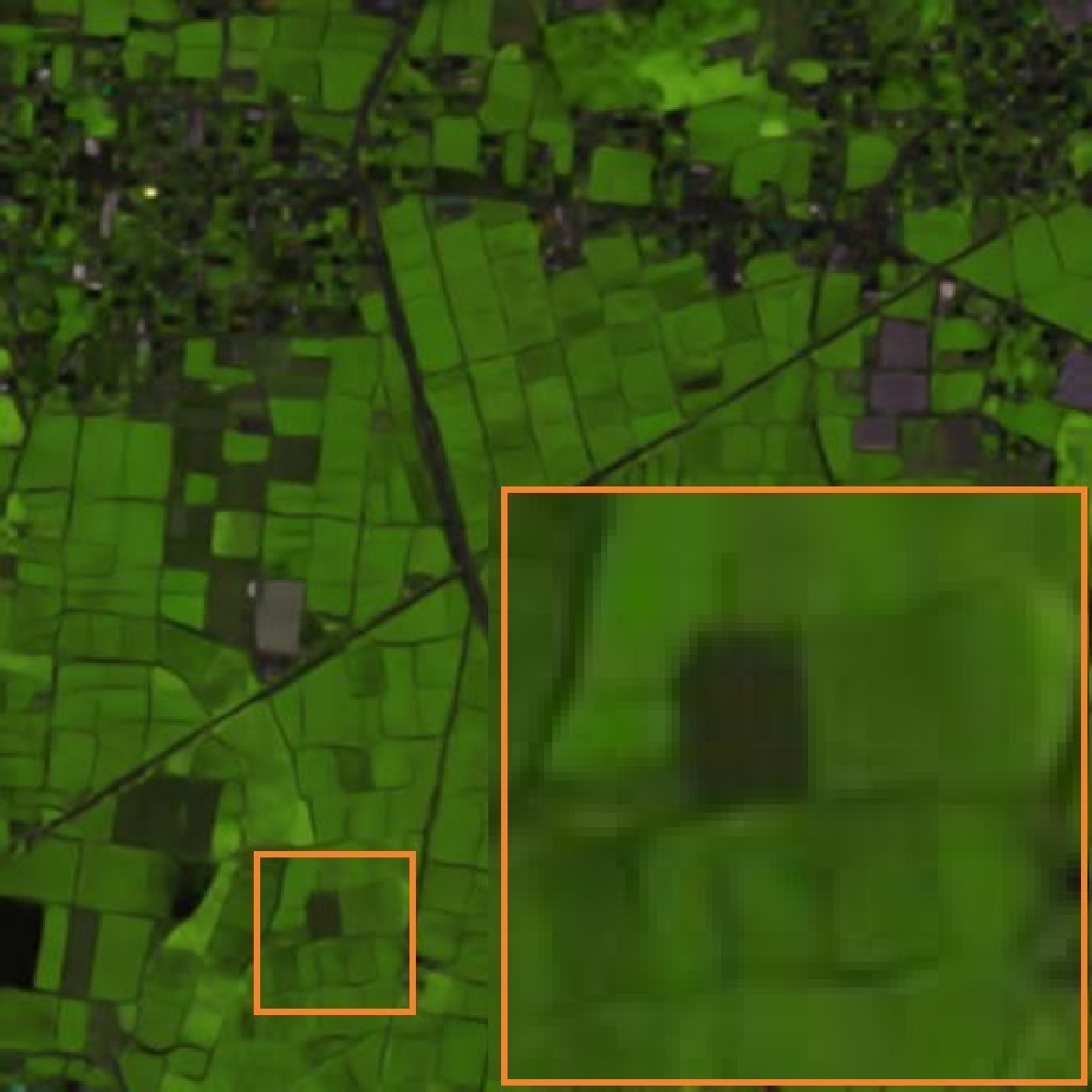}
            \centerline{GDRRN}
        \end{minipage}
    }\hspace{-2.7mm}
    \subfigure{
        \begin{minipage}[t]{0.135\textwidth}
            \includegraphics[width=1\textwidth]{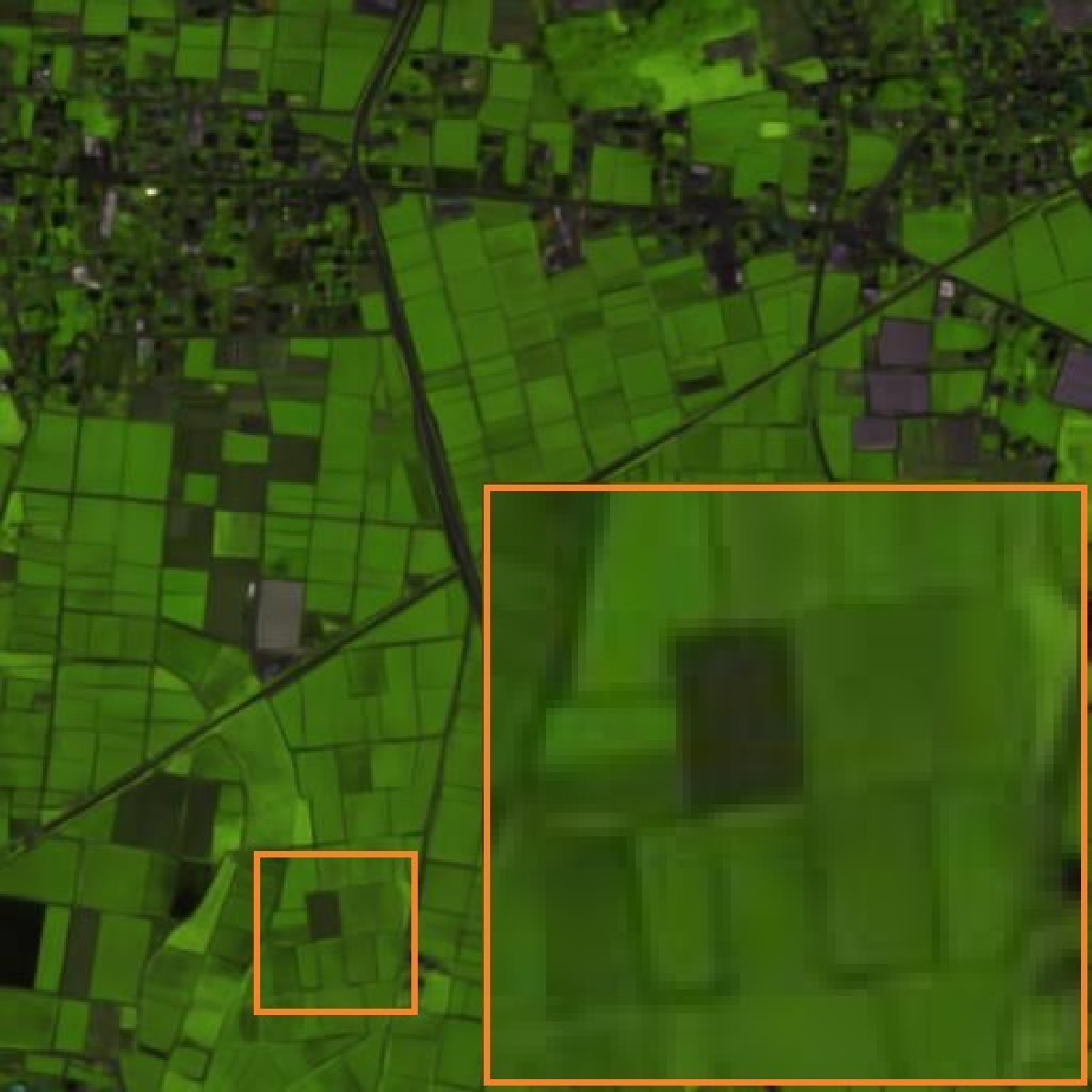}
            \centerline{SSPSR}
        \end{minipage}
    }\hspace{-2.7mm}
    \subfigure{
        \begin{minipage}[t]{0.135\textwidth}
            \includegraphics[width=1\textwidth]{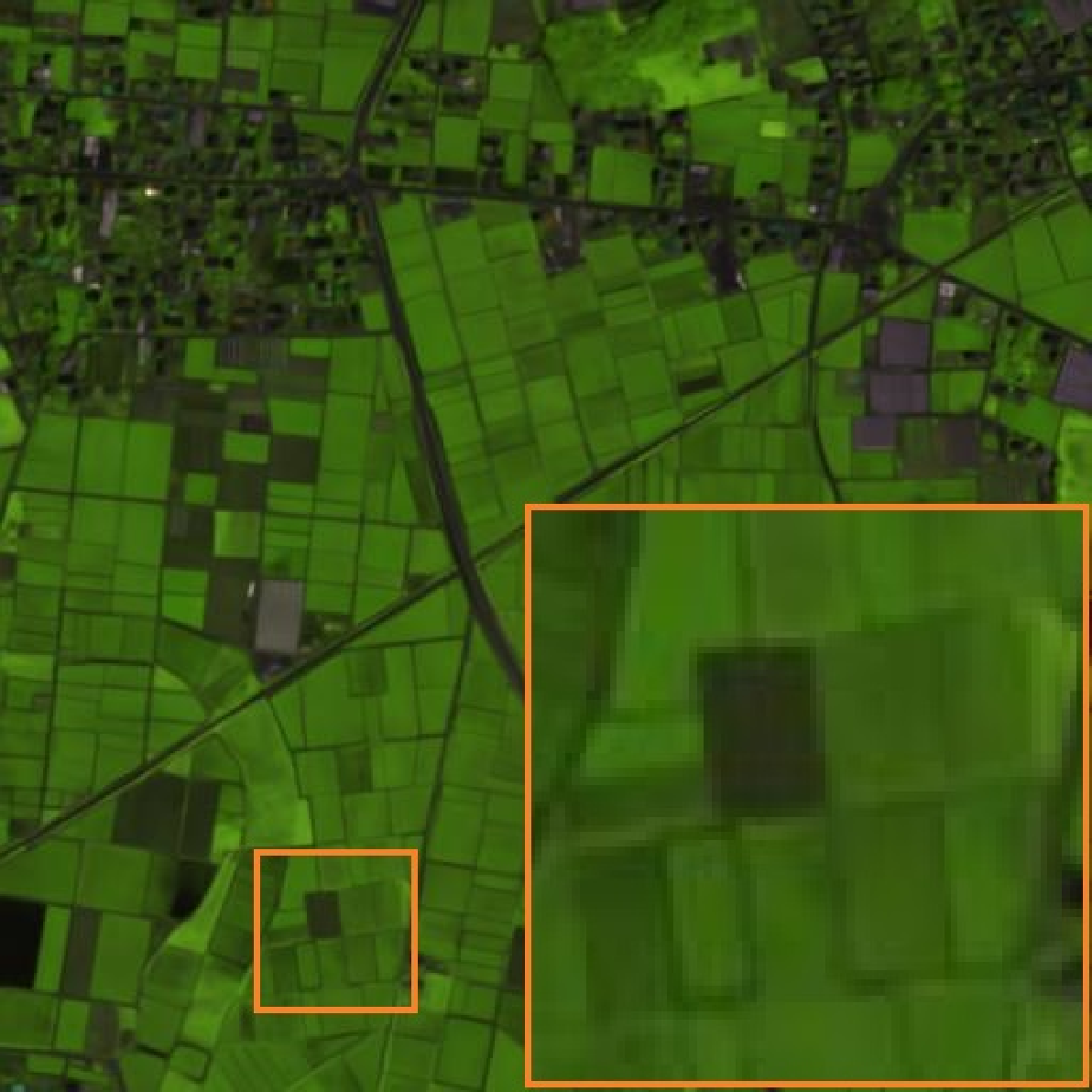}
            \centerline{FRLGN}
        \end{minipage}
    }
    \\
    \subfigure{
        \begin{minipage}[t]{0.135\textwidth}
            \includegraphics[width=1\textwidth]{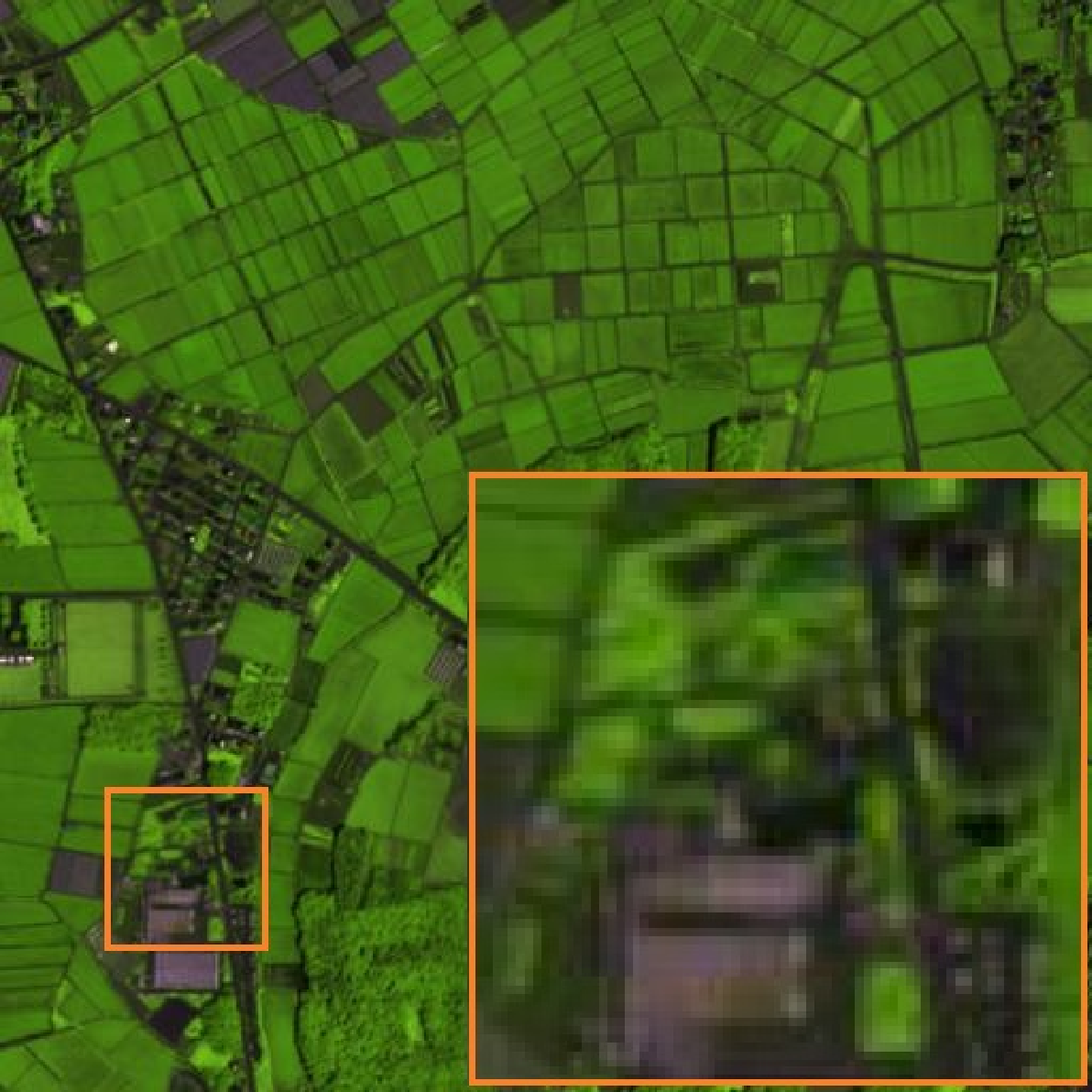}
            \centerline{HR}
        \end{minipage}
    }\hspace{-2.7mm}
    \subfigure{
        \begin{minipage}[t]{0.135\textwidth}
            \includegraphics[width=1\textwidth]{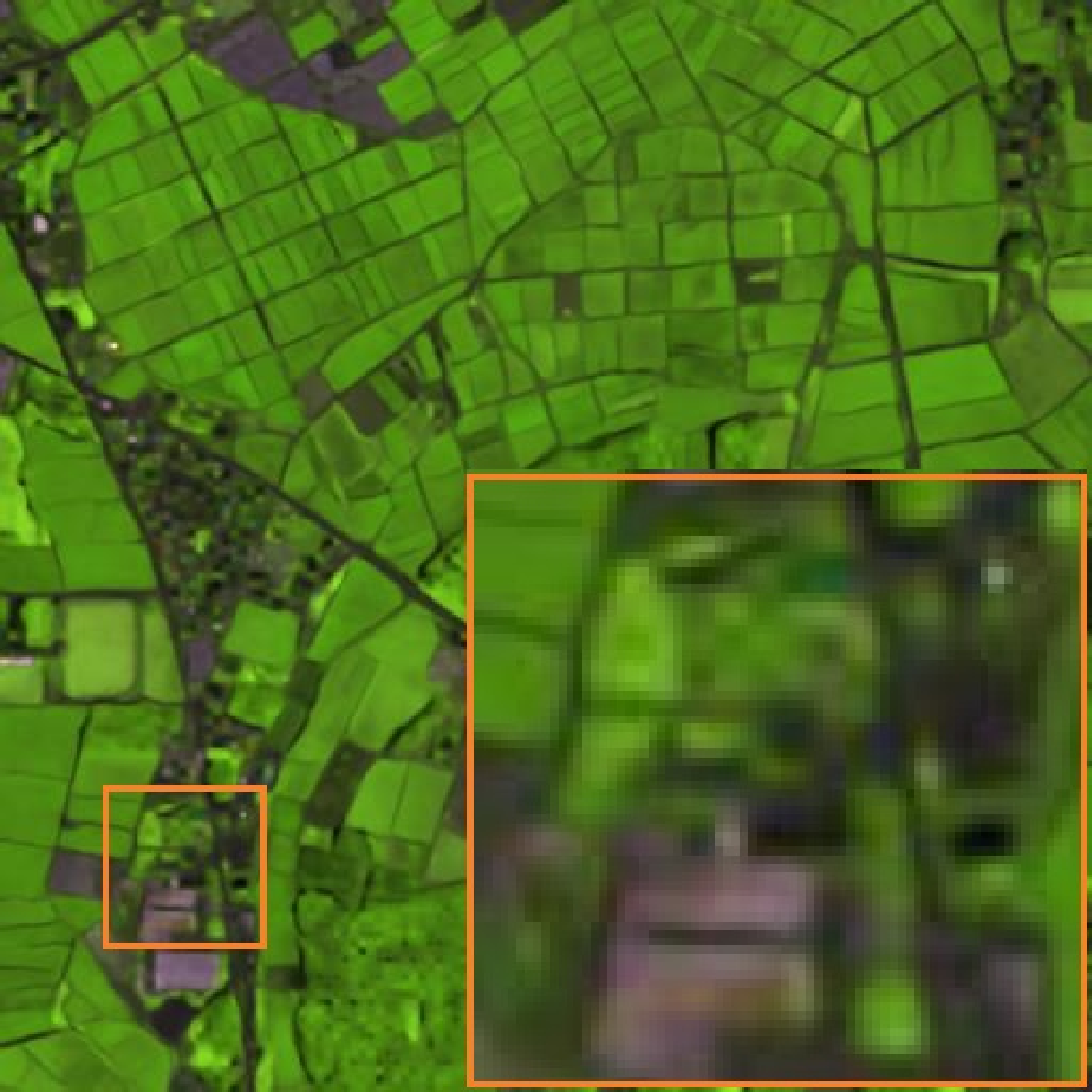}
            \centerline{VDSR}
        \end{minipage}
    }\hspace{-2.7mm}
    \subfigure{
        \begin{minipage}[t]{0.135\textwidth}
            \includegraphics[width=1\textwidth]{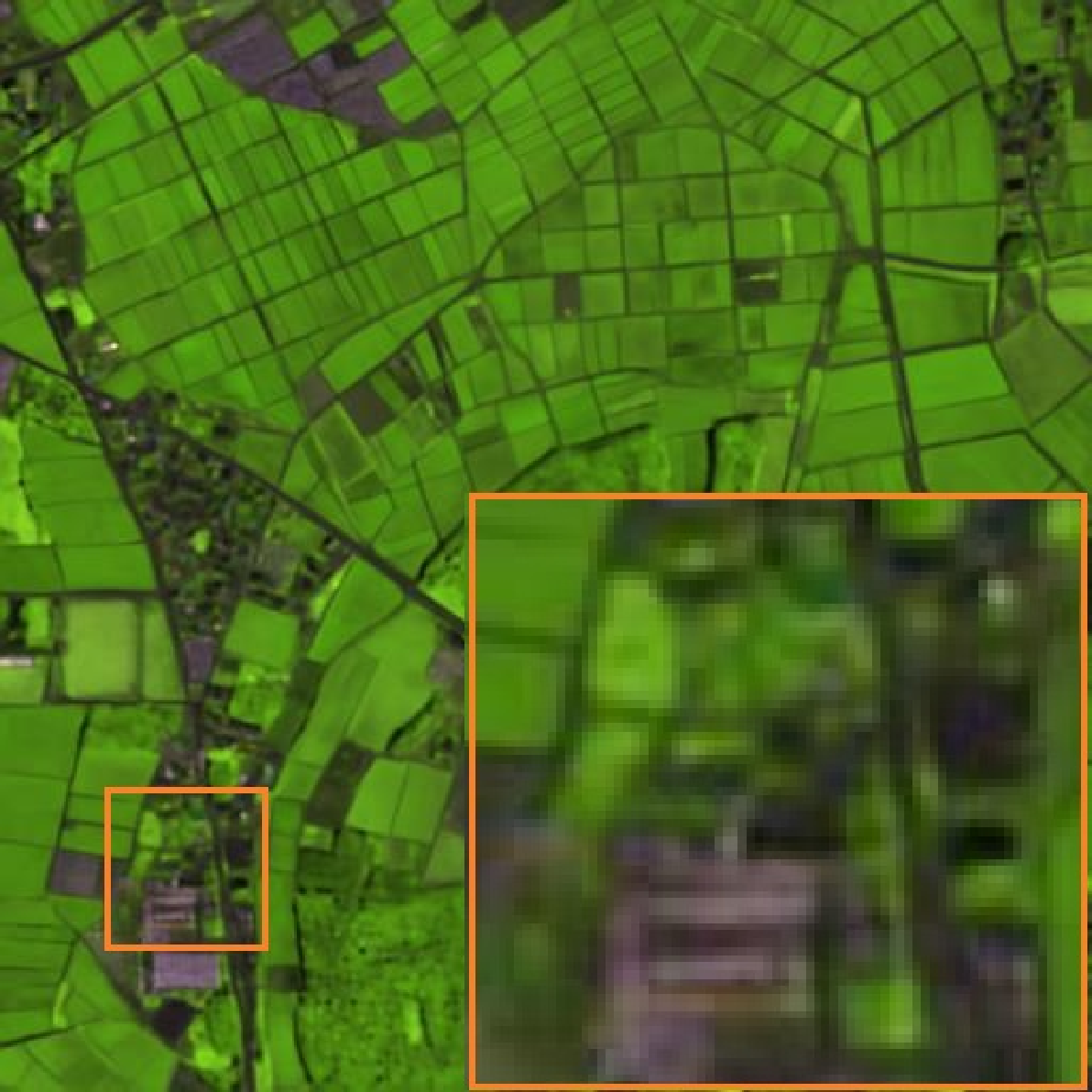}
            \centerline{RCAN}
        \end{minipage}
    }\hspace{-2.7mm}
    \subfigure{
        \begin{minipage}[t]{0.135\textwidth}
            \includegraphics[width=1\textwidth]{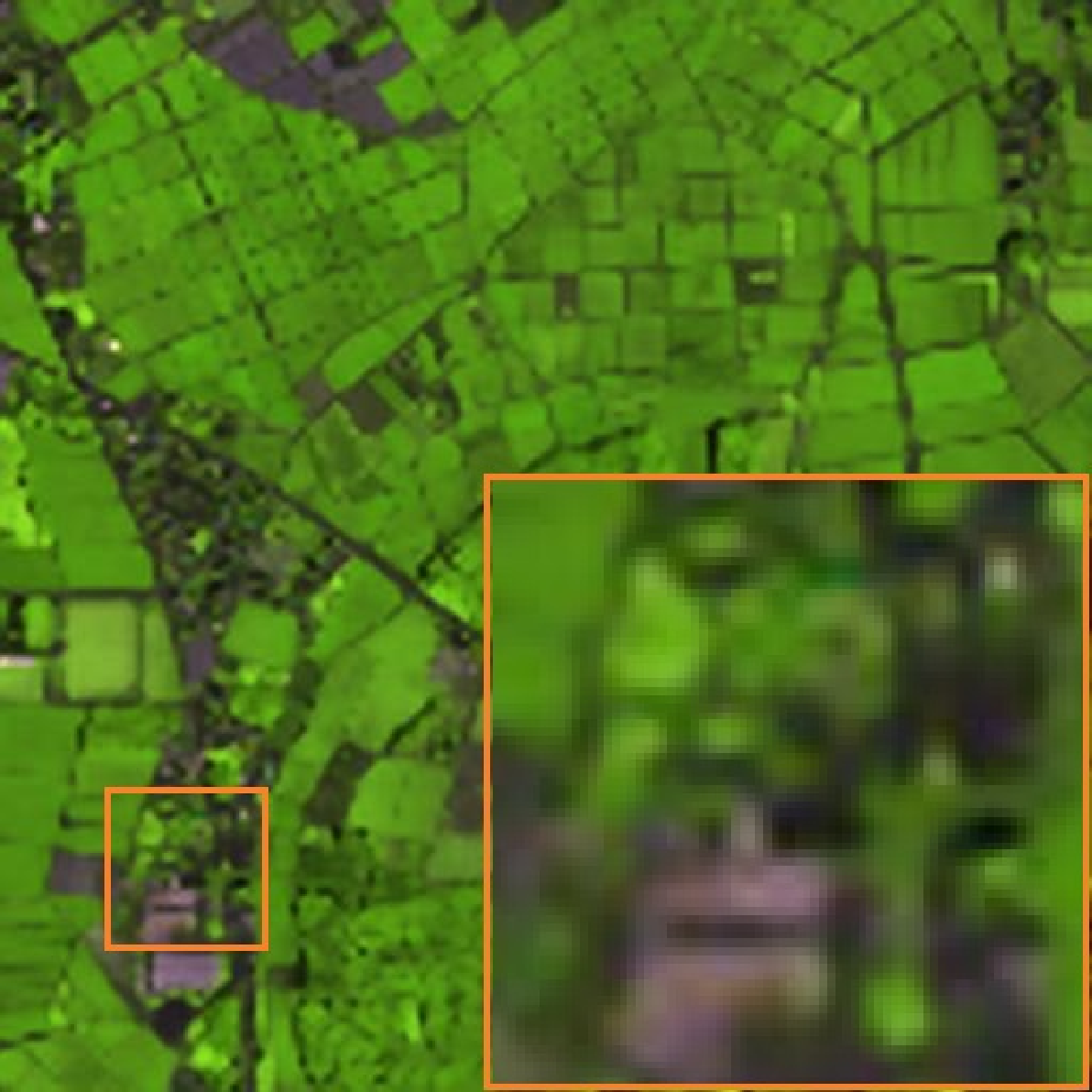}
            \centerline{3DCNN}
        \end{minipage}
    }\hspace{-2.7mm}
    \subfigure{
        \begin{minipage}[t]{0.135\textwidth}
            \includegraphics[width=1\textwidth]{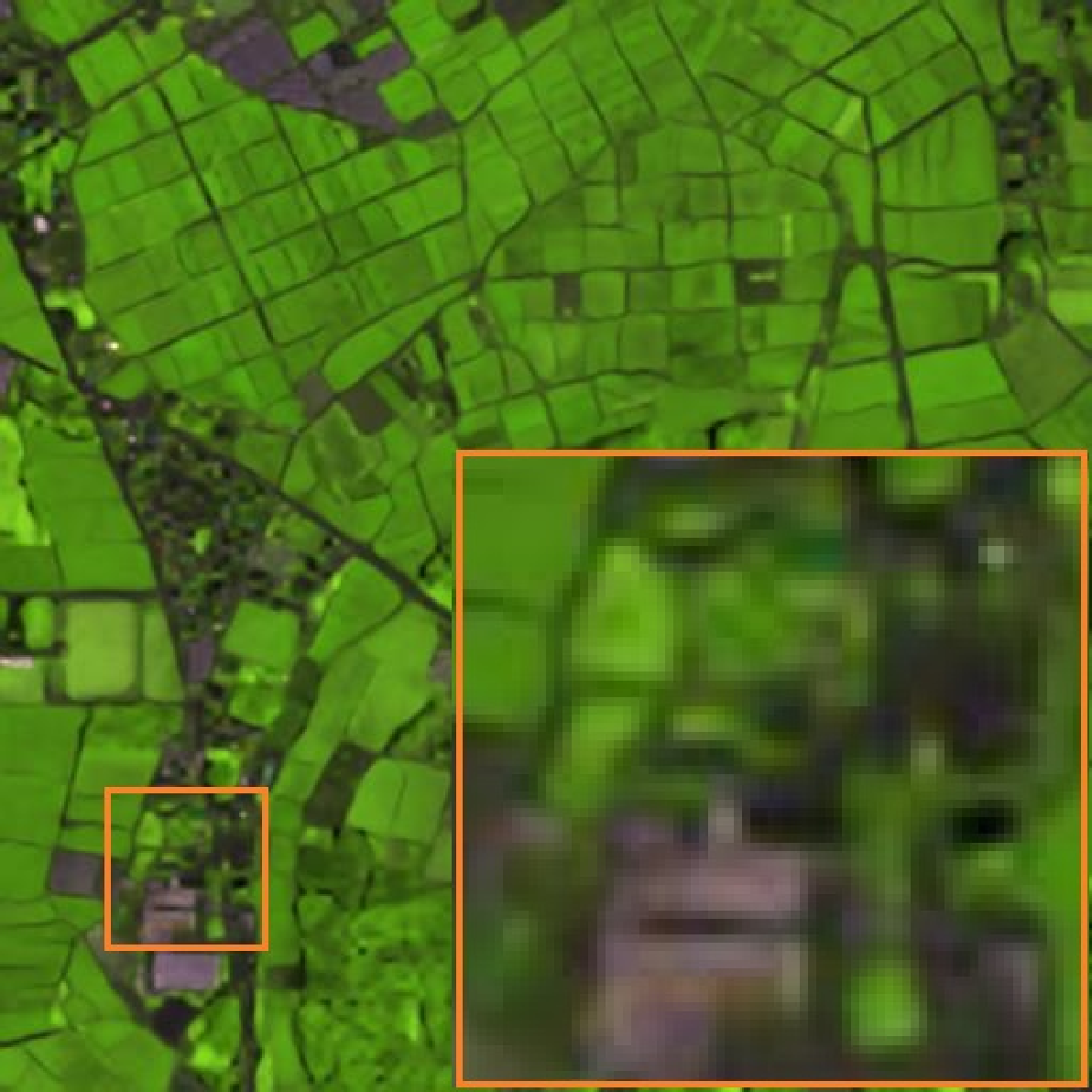}
            \centerline{GDRRN}
        \end{minipage}
    }\hspace{-2.7mm}
    \subfigure{
        \begin{minipage}[t]{0.135\textwidth}
            \includegraphics[width=1\textwidth]{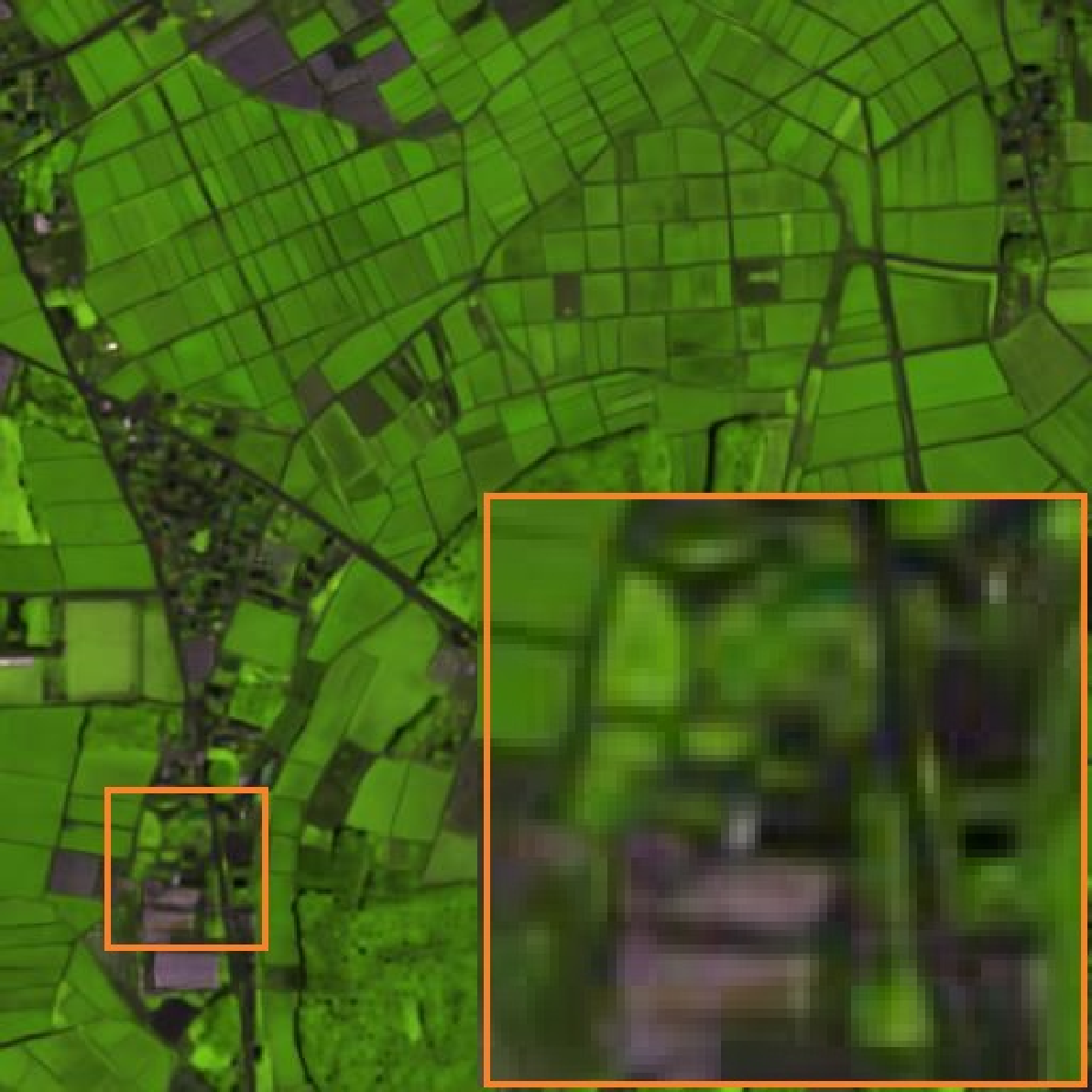}
            \centerline{SSPSR}
        \end{minipage}
    }\hspace{-2.7mm}
    \subfigure{
        \begin{minipage}[t]{0.135\textwidth}
            \includegraphics[width=1\textwidth]{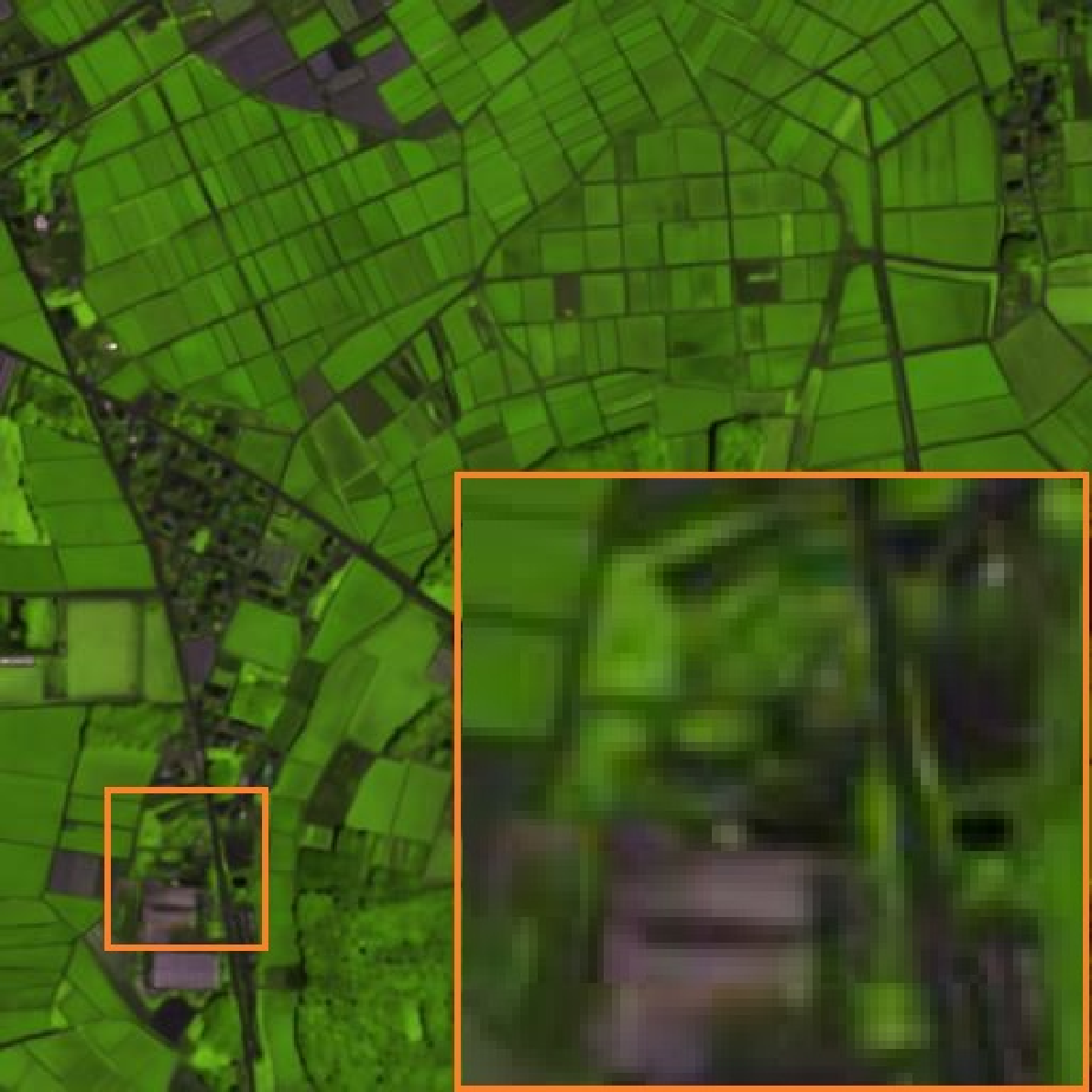}
            \centerline{FRLGN}
        \end{minipage}
    }
\caption{Two reconstructed hyperspectral images from 
the Chikusei testing dataset with the
scale factor 4, in which the bands 70-100-36 is treat as R-G-B.} \label{fig:ChikuseiResult1}
\end{figure*}


\begin{figure*}[htbp]
    \centering
    \subfigure{
        \begin{minipage}[t]{0.32\textwidth}
            \includegraphics[width=1\textwidth]{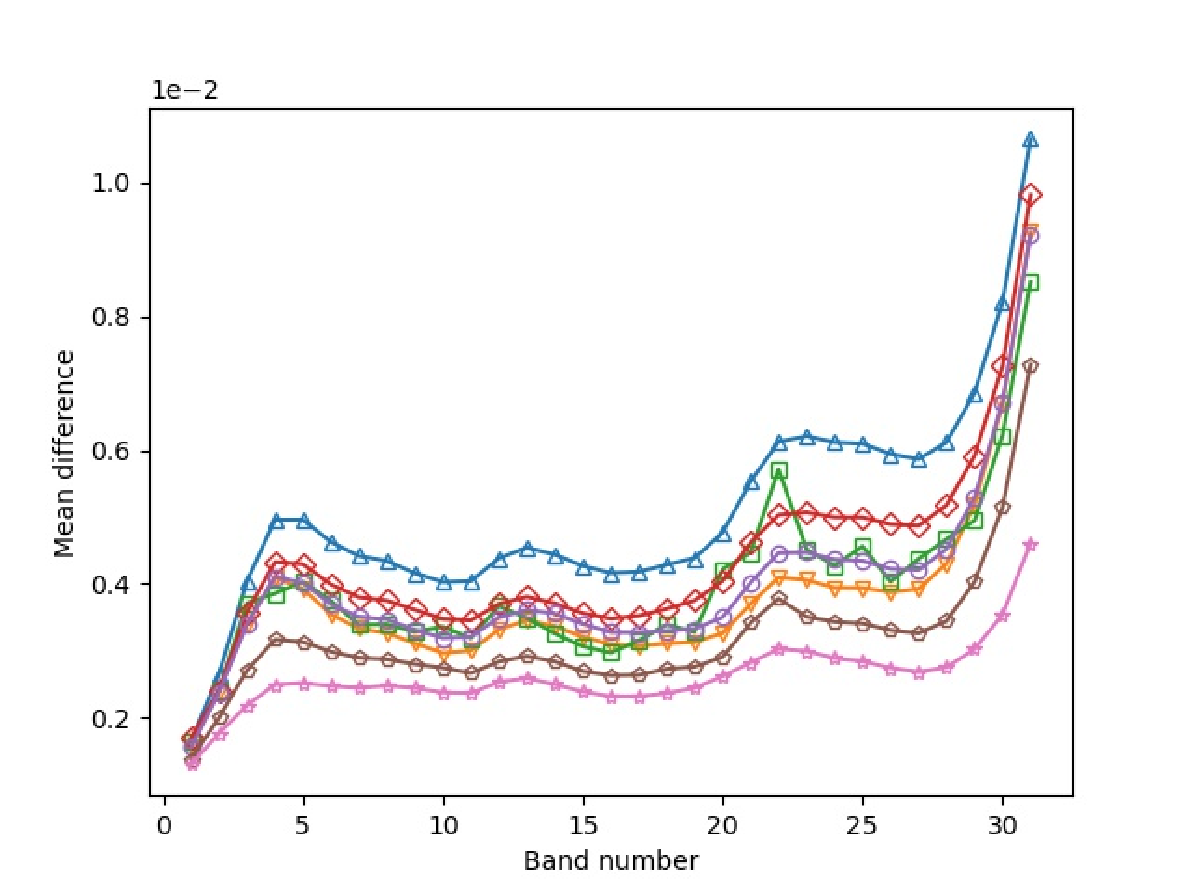}
        \end{minipage}
    }\hspace{-2.7mm}
    \subfigure{
        \begin{minipage}[t]{0.32\textwidth}
            \includegraphics[width=1\textwidth]{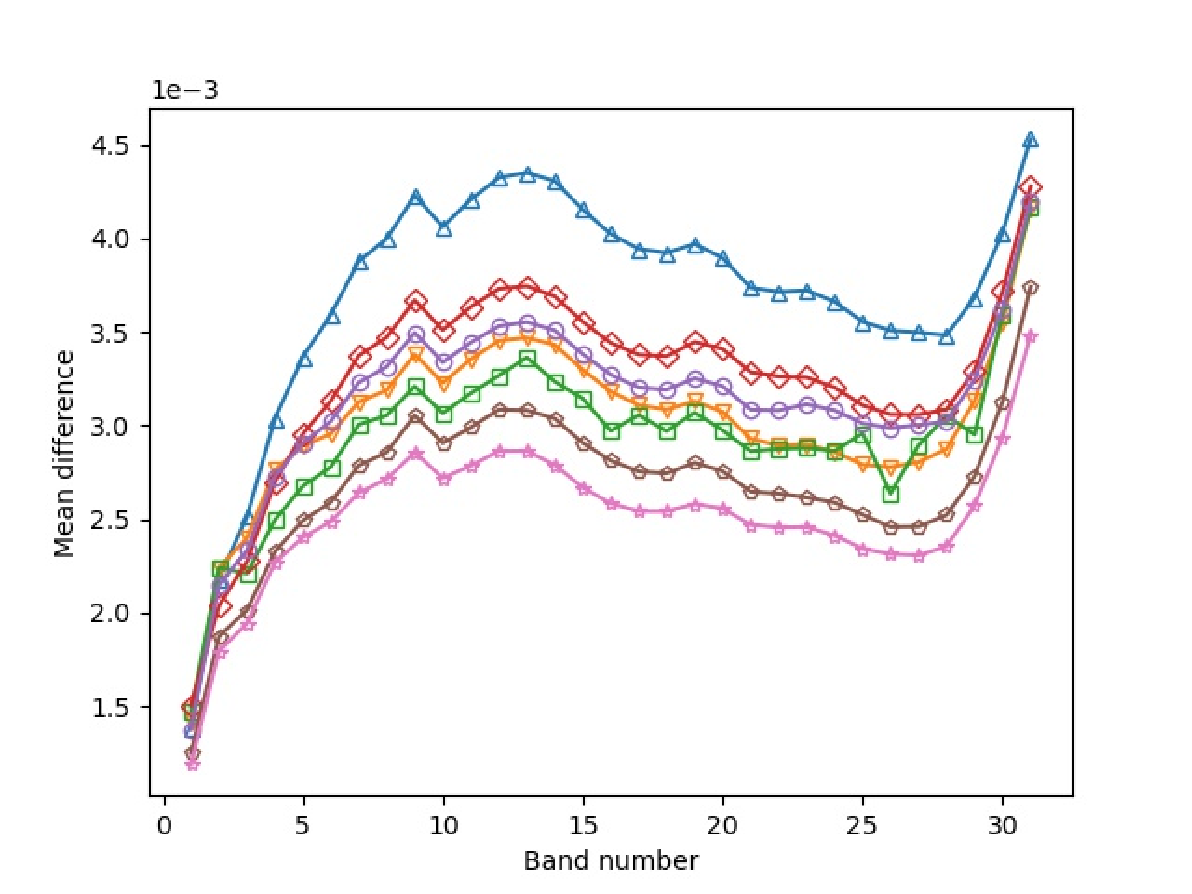}
        \end{minipage}
    }\hspace{-2.7mm}
    \subfigure{
        \begin{minipage}[t]{0.32\textwidth}
            \includegraphics[width=1\textwidth]{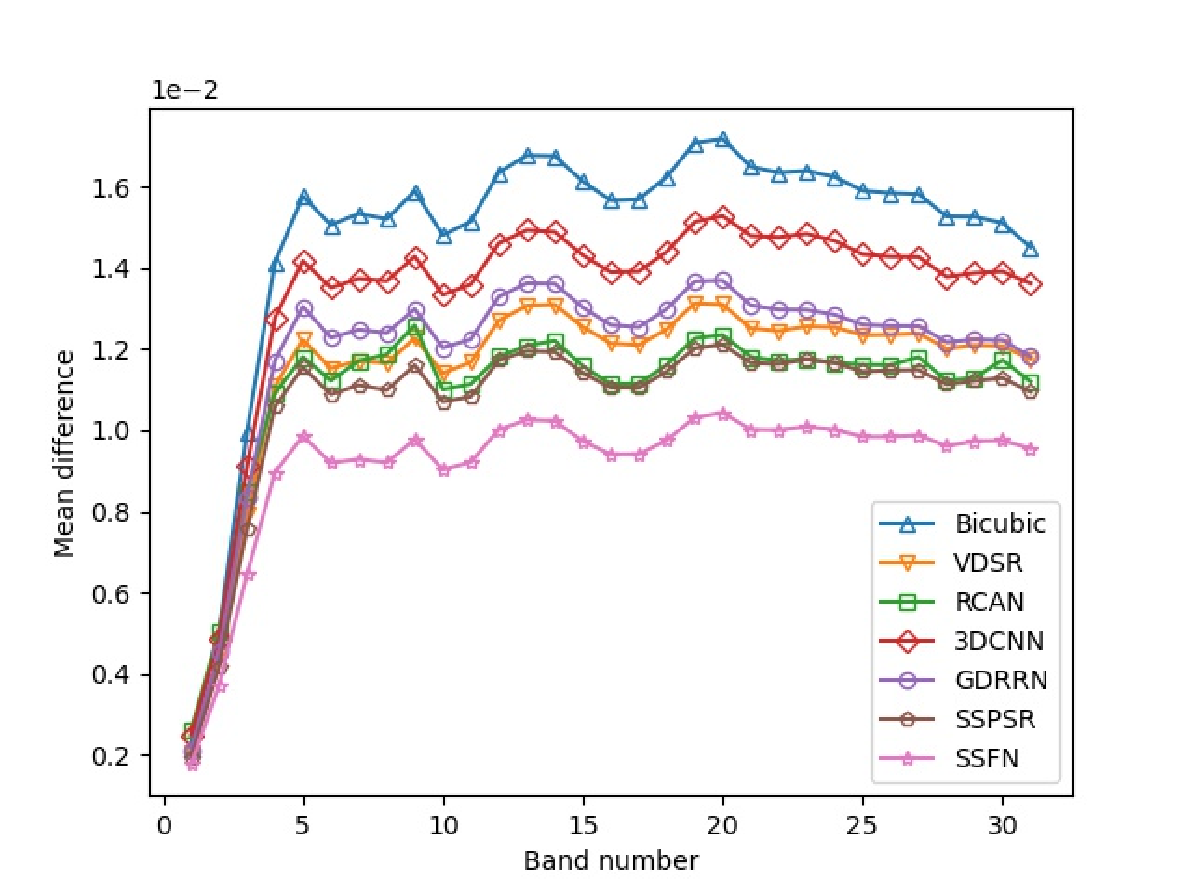}
        \end{minipage}
    }
    \caption{Mean spectral difference curve of three hyperspectral images from the CAVE testing dataset with the scale factor 4:superballs,sushi,chart_and_stuffed.}
    \label{fig:CAVEResult2}
\end{figure*}

\begin{figure*}[htbp]
    \centering
    \subfigure{
        \begin{minipage}[t]{0.32\textwidth}
            \includegraphics[width=1\textwidth]{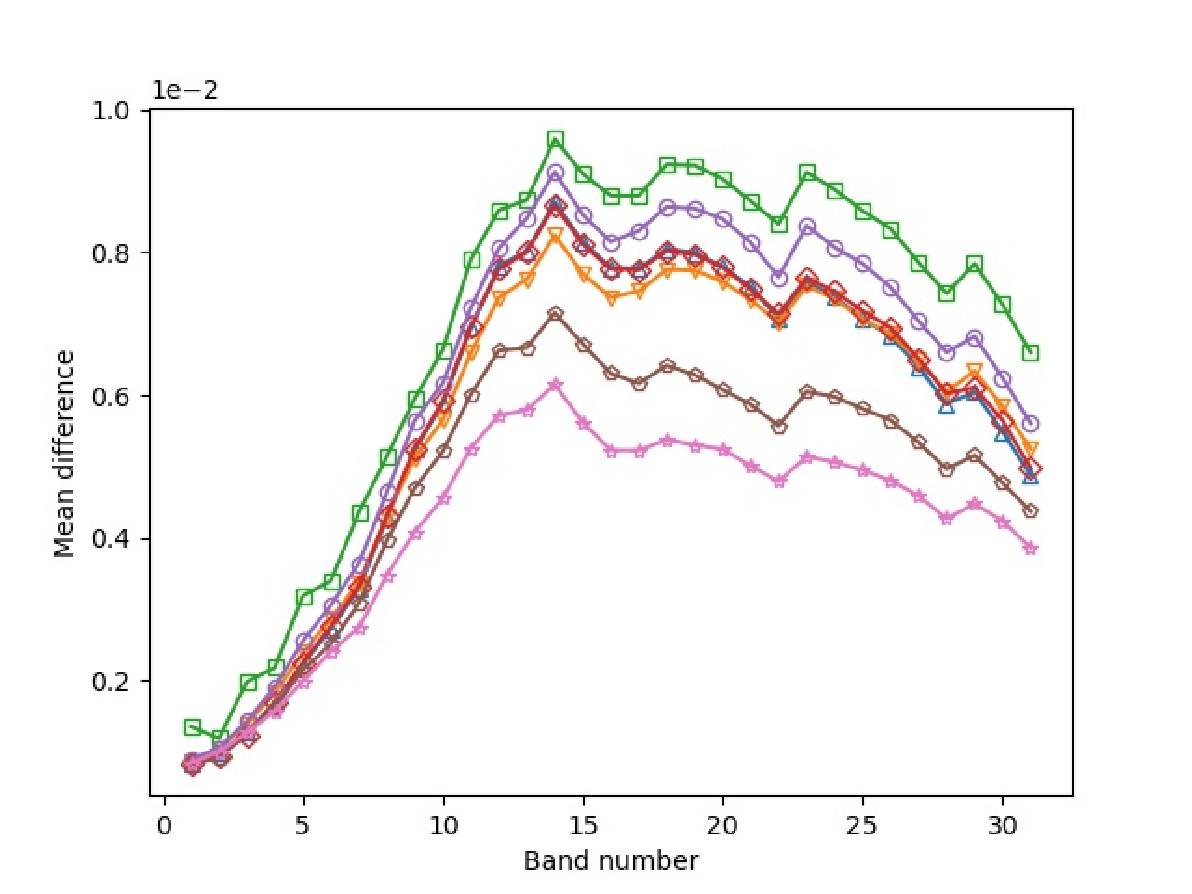}
        \end{minipage}
    }\hspace{-2.7mm}
    \subfigure{
        \begin{minipage}[t]{0.32\textwidth}
            \includegraphics[width=1\textwidth]{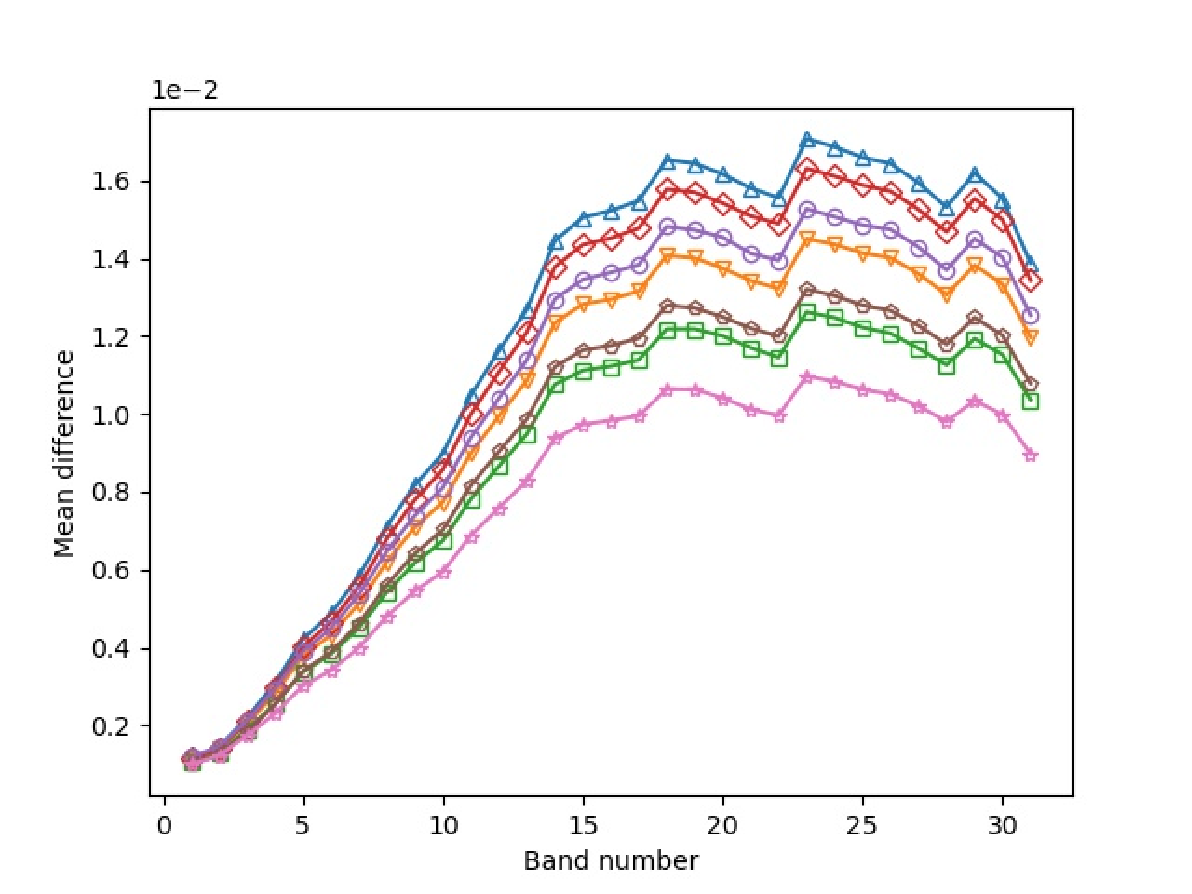}
        \end{minipage}
    }\hspace{-2.7mm}
    \subfigure{
        \begin{minipage}[t]{0.32\textwidth}
            \includegraphics[width=1\textwidth]{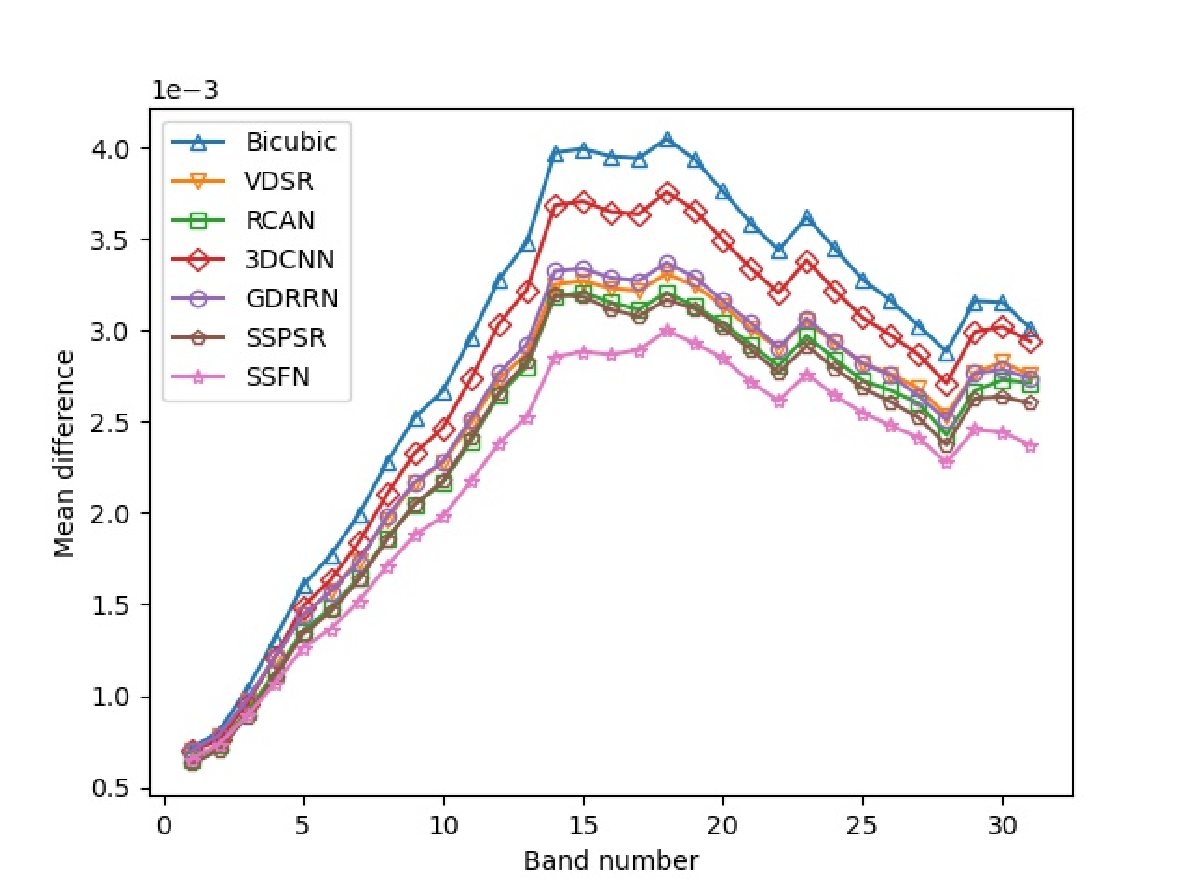}
        \end{minipage}
    }
    \caption{Mean spectral difference curve of three hyperspectral images from the Harvard testing dataset with the scale factor 4.}
    \label{fig:HarvardResult2}
\end{figure*}

\begin{figure*}[htbp]
    \centering
    \subfigure{
        \begin{minipage}[t]{0.32\textwidth}
            \includegraphics[width=1\textwidth]{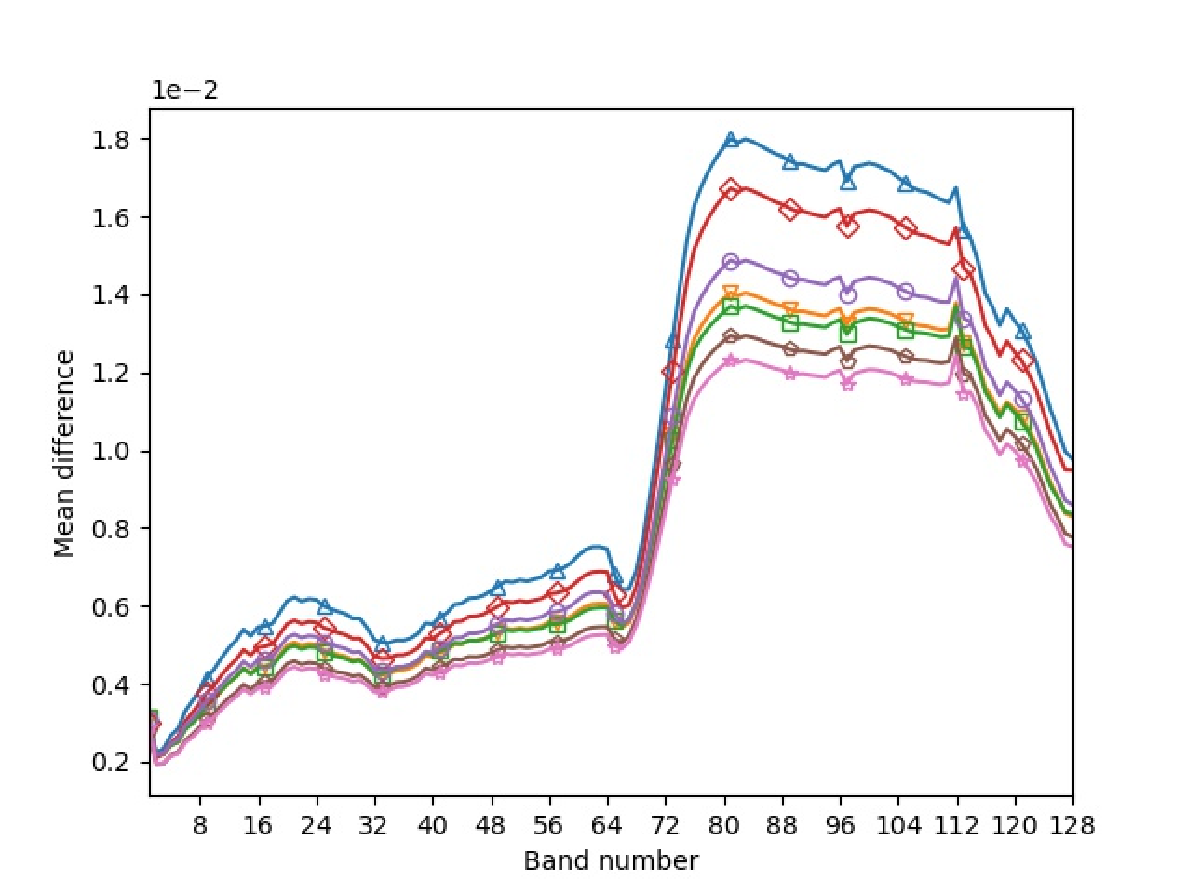}
        \end{minipage}
    }\hspace{-2.7mm}
    \subfigure{
        \begin{minipage}[t]{0.32\textwidth}
            \includegraphics[width=1\textwidth]{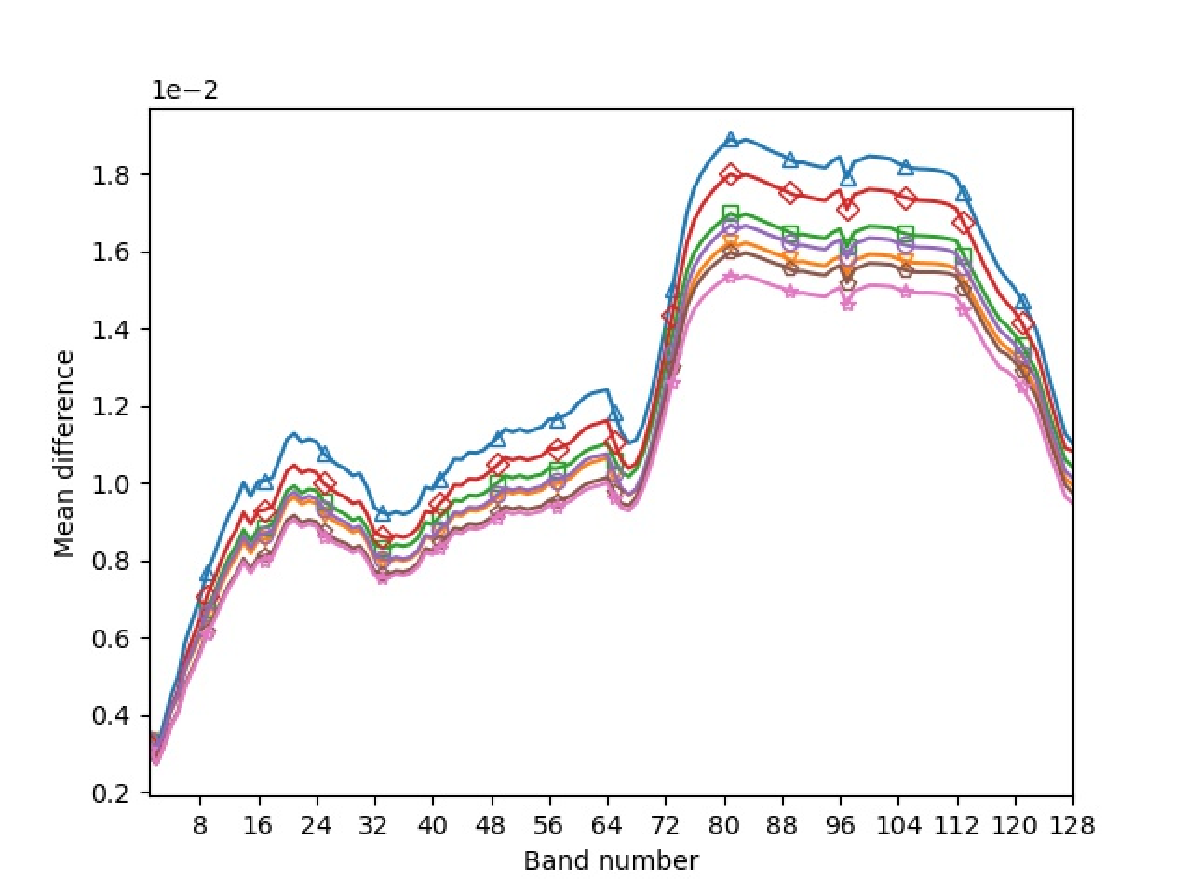}
        \end{minipage}
    }\hspace{-2.7mm}
    \subfigure{
        \begin{minipage}[t]{0.32\textwidth}
            \includegraphics[width=1\textwidth]{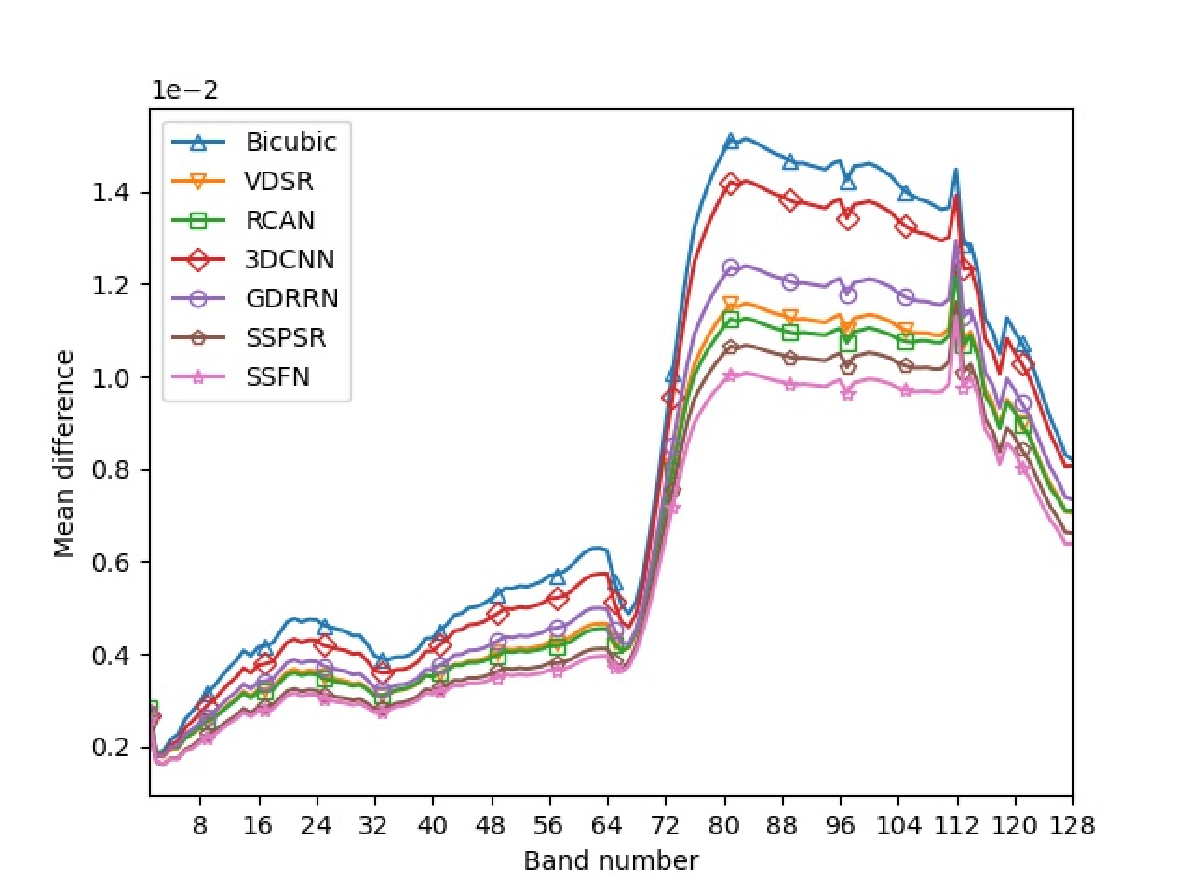}
        \end{minipage}
    }
    \caption{Mean spectral difference curve of three hyperspectral images from the Chikusei testing dataset with the scale factor 4.}
    \label{fig:ChikuseiResult2}
\end{figure*}

\subsection{Comparisons with the State-of-the-Art Methods}
In this section, we evaluate the single image super-resolution effect of FRLGN in detail on
three benchmarks, namely CAVE dataset \cite{yasuma2010generalized}, Harvard dataset
\cite{chakrabarti2011statistics} and Chikusei dataset
\cite{NYokoya2016}. Specifically, we compare the FRLGN with five existing super-resolution approaches, including two advanced deep multispectral image super-resolution methods, VDSR \cite{kim2016accurate}, RCAN \cite{zhang2018image},  and three
representative HSI super-resolution
methods, 3DCNN \cite{mei2017hyperspectral}, GDRRN
\cite{li2018single} and SSPSR \cite{jiang2020learning}. In addition, we carefully tune the hyper-parameters of these super-resolution methods to obtain a good performance. Moreover, the bicubic interpolation is used as our baseline model. Table \ref{tab:CAVE}, \ref{tab:Harvard} and
\ref{tab:Foster} depict the quantitative performance of all super-resolution algorithms over testing images on three datasets, where bold indicates the best results.

Table \ref{tab:CAVE} shows that our FRLGN method outperforms other comparative methods in all objective assessment metrics. Specifically, the baseline approach has the worst performance among these compared algorithms. As the competitive multispectral image super-resolution methods, VDSR and RCAN can generate very satisfactory results. Nonetheless, in comparison with those HSI super-resolution methods, \emph{i.e}, 3DCNN \cite{mei2017hyperspectral} and SSPSR \cite{jiang2020learning}, their spectral reconstruction effect (SAM) is relatively poor. This indicates that the multispectral super-resolution approaches cannot effectively explore the spectral prior information from the hyperspectral data. Similar to our work, SSPSR \cite{jiang2020learning} also
adopts a group strategy but neglects the continuous relationship among band groups. Therefore, it achieves the suboptimal results for the SAM indices. Compared with other comparison SR methods, our
proposed FRLGN can obtain better performance in spectral and spatial dimensions. In term of PSNR,
the FRLGN was 0.8 and 0.6 higher than the suboptimal method for upsampling factors $d$ of 4 and 8, respectively. The table \ref{tab:Harvard} and \ref{tab:Foster} show the similar results. In conclusion, FRLGN has presented advantages on three datasets compared to existing SR methods, especially for PSNR and SSIM.

In order to further prove the effectiveness of FRLGN, Fig.~\ref{fig:CAVEResult1} and \ref{fig:HarvardResult1} display
the mean absolute error maps across all spectral bands of two HSIs with the scale factor $\times 4$ from the CAVE testing dataset and Harvard testing dataset, respectively.
Principally, the bluer the color of the error map, the better the reconstructed HSI. From fig.~\ref{fig:CAVEResult1} and \ref{fig:HarvardResult1}, we can easily discover that the FRLGN method can obtain better reconstruction fidelity when restoring the spatial information of the original HSI. Specifically, in contrast to with the suboptimal SSPSR method, FRLGN performs better in reconstructing textures such as edges and structures. Besides, we also display two reconstructed high-resolution HSIs from Chikusei test dataset with a downsampling factor of 4 in Fig.~\ref{fig:ChikuseiResult1}. As can be seen from Fig.~\ref{fig:ChikuseiResult1}, our FRLGN can restore finer texture details than other comparison methods.

In addition, to prove our advantage in reconstructing spectral information, Fig.~\ref{fig:CAVEResult2}, \ref{fig:HarvardResult2} and \ref{fig:ChikuseiResult2} show the average absolute difference of all comparison methods along the spectral dimension. The average spectral error curve has a better visualization effect than displaying the spectral reflectance of multiple locations. As shown in Fig.~\ref{fig:CAVEResult2},
\ref{fig:HarvardResult2} and \ref{fig:ChikuseiResult2}, our method has the lowest average spectral error curve, which
indicates that FRLGN has better spectral reconstruction ability. This can be attributed to the guidance of the global spectral feedback information to the local spectral band group. Moreover, as iterations increase, the local spectral group information gradually accumulates, leading to better spectral reconstruction performance.

\section{Conclusion}
Considering the difficulty of simultaneously exploring the spatial and spectral information of hyperspectral data, we propose a new approach for the single HSI super-resolution task, called Feedback Refined Local-Global Network. FRLGN can produce a clear high-resolution HSI by introducing a Feedback Structure and a Local-Global Spectral Block. In particular, we construct a recurrent neural network with feedback connections to refine low-level feature representations using feedback global spectral high-level semantic information. 
Furthermore, taking advantage of the feedback high-level semantic information, we carefully design a Local-Global Spectral Block to guide the extraction process of low-level representations between local spectral bands using the feedback information, and then generate a more powerful high-level feature among global spectral bands. With the increasing number of iterations, the spatial-spectral prior gradually accumulates, leading to better HSI reconstruction performance. The comprehensive experimental results and visual data analysis show the effectiveness of the proposed FRLGN.


%

\ifCLASSOPTIONcaptionsoff
  \newpage
\fi



%
\bibliographystyle{IEEEtran}
\bibliography{ref}

%








\end{document}